\documentclass{article} 
\usepackage{algorithm}%
\usepackage{algorithmicx}%
\usepackage{algpseudocode}%
\usepackage{dsfont}
\usepackage{mathrsfs}

\usepackage{wrapfig}
\usepackage{enumerate}
\usepackage{amsmath,amssymb}
\usepackage{amsthm}
\usepackage{mathrsfs}
\usepackage{natbib}
\usepackage{xcolor}%
\usepackage{mathtools}

\usepackage{color}
\usepackage{graphicx}
\usepackage{caption}
\usepackage{subcaption}
\usepackage{multirow}
\usepackage[colorlinks,
            linkcolor=red,
            anchorcolor=blue,
            citecolor=blue
            ]{hyperref}
\usepackage{cleveref}

\usepackage{fullpage}

\newtheoremstyle{thmstyleone}
{18pt plus2pt minus1pt}
{18pt plus2pt minus1pt}
{\small\itshape}
{0pt}
{\small\bfseries}
{}
{.5em}
{\thmname{#1}\thmnumber{\@ifnotempty{#1}{ }\@upn{#2}}%
  \thmnote{ {\the\thm@notefont(#3)}}}

\newtheoremstyle{thmstyletwo}
{18pt plus2pt minus1pt}
{18pt plus2pt minus1pt}
{\small\normalfont}
{0pt}
{\small\itshape}
{}
{.5em}
{\thmname{#1}\thmnumber{\@ifnotempty{#1}{ }{#2}}%
  \thmnote{ {\the\thm@notefont(#3)}}}
\newtheoremstyle{thmstylethree}
{18pt plus2pt minus1pt}
{18pt plus2pt minus1pt}
{\small\normalfont}
{0pt}
{\small\bfseries}
{}
{.5em}
{\thmname{#1}\thmnumber{\@ifnotempty{#1}{ }\@upn{#2}}%
  \thmnote{ {\the\thm@notefont(#3)}}}
\theoremstyle{thmstyleone}%
\newtheorem{theorem}{Theorem}
\newtheorem{proposition}[theorem]{Proposition}%

\theoremstyle{thmstyletwo}%

\theoremstyle{thmstylethree}%

\begin{document}

\title{Learning Roles with Emergent Social Value Orientations}

\author{\normalsize
    Wenhao Li\thanks{The Chinese University of Hong Kong, Shenzhen; \texttt{liwenhao@cuhk.edu.cn}} \qquad
    Xiangfeng Wang\thanks{East China Normal University; \texttt{xfwang@cs.ecnu.edu.cn}} \qquad
    Bo Jin\thanks{East China Normal University; \texttt{bjin@cs.ecnu.edu.cn}} \qquad
    Jingyi Lu\thanks{East China Normal University; \texttt{jylu@psy.ecnu.edu.cn}}\qquad
    Hongyuan Zha\thanks{The Chinese University of Hong Kong, Shenzhen; \texttt{zhahy@cuhk.edu.cn}}
}

\maketitle

\begin{abstract}
Social dilemmas can be considered situations where individual rationality leads to collective irrationality.
The multi-agent reinforcement learning community has leveraged ideas from social science, such as social value orientations (SVO), to solve social dilemmas in complex cooperative tasks.
In this paper, by first introducing the typical ``division of labor or roles" mechanism in human society, we provide a promising solution for intertemporal social dilemmas (ISD) with SVOs.
A novel learning framework, called Learning \textbf{R}oles with \textbf{E}mergent \textbf{SVO}s (\textbf{RESVO}), is proposed to transform the learning of roles into the social value orientation emergence, which is symmetrically solved by endowing agents with altruism to share rewards with other agents.
An SVO-based role embedding space is then constructed by individual conditioning policies on roles with a novel rank regularizer and mutual information maximizer.
Experiments show that RESVO achieves a stable division of labor and cooperation in ISDs with different complexity.
\end{abstract}

\section{Introduction}\label{sec:intro}

The continuity of human civilization and the prosperity of the race depends on our ability to cooperate.
From evolutionary biology to social psychology and economics, cooperation in human populations has been regarded as a paradox and a challenge~\citep{fehr2003nature,elizabeth2009origin,santos2021complexity}.
Cooperation issues vary in scale and are widespread in daily human life, ranging from assembly line operations in factories and scheduling of seminars to peace summits between significant powers, business development, and pandemic control~\citep{dafoe2020open}.

Although cooperation can benefit all parties, it might be costly.
Thus, the temptation to evade any cost (i.e., the free-riding) becomes a tempting strategy, which leads to cooperation collapsing, or the multi-person social dilemma~\citep{rapoport1965prisoner,xu2019cooperation}.
That is, ``individually reasonable behavior leads to a situation in which everyone is worse off than they might have been otherwise''~\citep{kollock1998social}.
Just as cooperation widely exists in human social, economic, and political activities, most thorny problems we face, from the interpersonal to the international, are at their core social dilemmas.
This article presents two cases in recent years closely related to the future economic and political decisions of countries, namely autonomous driving and carbon trading, and the role of social dilemmas in them.

Autonomous driving (AV), which promises world-changing benefits by increasing traffic efficiency~\citep{van2006impact}, reducing pollution~\citep{spieser2014toward}, and eliminating up to $90\%$ of traffic accidents~\citep{gao2014road}, is a very complex systems engineering. 
Existing work mainly focuses on accomplishing generic tasks, such as following a planned path while obeying traffic rules. 
However, there are many driving scenarios in practice, most of which have social dilemmas. 
Examples include lane changing~\citep{dafoe2020open}, meeting, parking~\citep{li2022cooperation}, and even ethical aspects of aggressive versus conservative driving behavior choices~\citep{bonnefon2016social}.
Therefore, the practicality of AV depends on the efficient solution to social dilemmas.

Carbon trading is a greenhouse gas emission right (emission reduction) transaction based on the United Nations Framework Convention on Climate Change established by the Kyoto Protocol to promote the reduction of greenhouse gas emissions, using a market mechanism~\citep{grimeaud2001overview}.
Carbon emission is a representative social dilemma in which countries' direct gas emissions for the sake of economic development undermine collective interests.
The typical mechanisms in carbon trading, such as \textit{distribution of allowances}~\citep{fullerton1997environmental}, \textit{joint implementation}~\citep{grimeaud2001overview}, etc., have obvious correspondences with the \textit{boundaries}~\citep{ibrahim2020reward, Ibrahim2020RewardRM} and \textit{institutions}~\citep{koster2020silly,lupu2020gifting} used to solve social dilemmas in economics and social psychology.

The social dilemma has been comprehensively studied in economics, social psychology, and evolutionary biology in the past few decades. 
This paper focuses on the \textit{public good dilemma} in the intertemporal social dilemma (ISD).
A public good is a resource from which all may benefit, regardless of whether they have helped provide the good (producer)~\citep{kollock1998social}.
This is to say that public goods are \textit{non-excludable}.
As a result, there is the temptation to enjoy the good (consumer) without contributing to its creation or maintenance.
Those who do so are termed \textit{free-riders}, and while it is individually rational to free-ride if all do so, the public good is not provided, and all are worse off.

Artificial intelligence (AI) advances pose increasing opportunities for AI research to promote human cooperation and enable new tools for facilitating cooperation~\citep{dafoe2020open}.
Recently, multi-agent reinforcement learning (MARL) has been utilized as a powerful toolset to study human cooperative behavior with great success~\citep{Lowe2017MultiAgentAF,Silver2018AGR,jaderberg2019human,Liao2020IterativelyRefinedI3,Li2021StructuredCR}.
We believe it is reasonable to use MARL as a first step in exploring the use of AI tools to study multi-person social dilemmas.
The current model for reinforcement learning suggests that reward maximization is sufficient to drive behavior that exhibits abilities studied in the human cooperation and social dilemmas, including ``knowledge, learning, perception, social intelligence, language, generalization and imitation''~\citep{yang2021many,silver2021reward,vamplew2022scalar}.
The justification for this claim is deeply rooted in the \textit{von Neumann Morgenstern utility theory}~\citep{von2007theory}, which is the basis for the well-known \textit{expected utility theory}~\citep{schoemaker2013experiments} and essentially states that it is safe to assume an intelligent entity will always make decisions according to the highest expected utility in any complex scenarios\footnote{Although follow-up works have shown that some of the assumptions on rationality could be violated by real decision-makers in practice~\citep{gigerenzer2002bounded}, those conditions are rather taken as the “axioms” of rational decision making~\citep{yang2021many}.}~\citep{yang2021many}.

In MARL, the critical issue of multi-person social dilemma can be formalized as an ISD~\citep{leibo2017multi,Hughes2018InequityAI}, and most MARL methods have introduced ideas from social psychology and economics more or less.
These methods could be divided into three categories, \textit{strategic} solutions, \textit{structural} solutions, and \textit{motivational} solutions, based on whether the solutions assume egoistic agents and whether the structure of the game can be changed~\citep{kollock1998social} according to the taxonomy of social science.

\textit{Structural} solutions reduce the difficulty of the original social dilemma by changing the game's rules or completely avoiding the occurrence of the social dilemma.
The mechanisms introduced into MARL mainly include boundaries and sanctions~\citep{ostrom1990governing}.
\citet{ibrahim2020reward} indirectly sets boundaries for resources by introducing a shared periodic signal and a conditional policy based on this signal, allowing agents to access shared resources in a fixed order.
\citet{Ibrahim2020RewardRM} achieves resource boundarization by introducing a centralized government module through taxation and wealth redistribution..
\citet{koster2020silly,lupu2020gifting} introduce a centralized module and use rules and learning methods to punish the free-riding agent separately. 
LIO~\citep{yang2020learning} enables each agent to punish, thereby implementing the sanction mechanism in a decentralized manner.
\citet{vinitsky2021learning} adopts a combination of centralized and decentralized modules and judges the decentralized sanctioning behavior of the agent through the centralized module, thereby encouraging appropriate sanctioning behaviors and avoiding unreasonable behaviors.
Furthermore, \citet{dong2021birds} introduces homophily into the MARL to solve the second-order social dilemma caused by sanctions.

\textit{Strategic} solutions assume that all individuals in the group are egoists and that the algorithm does not change the game's structure. 
Such methods rely on an individual's ability to shape other individuals' payoffs, thereby directly influencing the behavior of others.
Direct and indirect reciprocity is the main mechanisms introduced into MARL.
\citet{eccles2019learning} introduces the classic direct reciprocity algorithm tit-for-tat~\citep{axelrod1981evolution} into the solution of ISD.
In order to realize the ``imitation" at the core of tit-for-tat and the definition of the binary action (cooperate and defect) in ISD, \citet{eccles2019learning}~divides the agents into innovators and imitators and introduces the niceness function based on the deep advantage function.
\citet{anastassacos2021cooperation} introduces two core concepts of indirect reciprocity, reputation and social norm~\citep{santos2021complexity} into MARL and uses them as fixed rules to construct the agent's action space.

\textit{Motivational} solutions assume agents are not entirely egoistic and so give some attention (passively or actively) to the outcomes of their partners.
One of the typical mechanisms is communication.
Across a wide variety of economics and social psychology studies, when individuals are given a chance to talk with each other, cooperation increases significantly~\citep{orbell1988explaining,orbell1990limits}.
Although there are many works~\citep{sheng2020learning,ahilan2020correcting} on communication learning in MARL, little attention has been paid to the role of communication in solving ISD.
\citet{pretorius2020game} first uses empirical game-theoretic analysis~\citep{tuyls2018generalised} to study existing communication learning methods in ISD and to verify the effects of these methods experimentally.
Another typical mechanism is social value orientation.
Social value orientations (SVOs), or heterogeneous distributive preferences~\citep{batson2012history,cooper20164,eckel1996altruism,rushton1981altruistic,simon1993altruism}, are widely recognized in social psychology and economics as an effective mechanism for promoting the emergence of human cooperative behavior in different social dilemmas~\citep{mckee2020social}.

\begin{figure}[htb!]
    \centering
    \includegraphics[width=\textwidth]{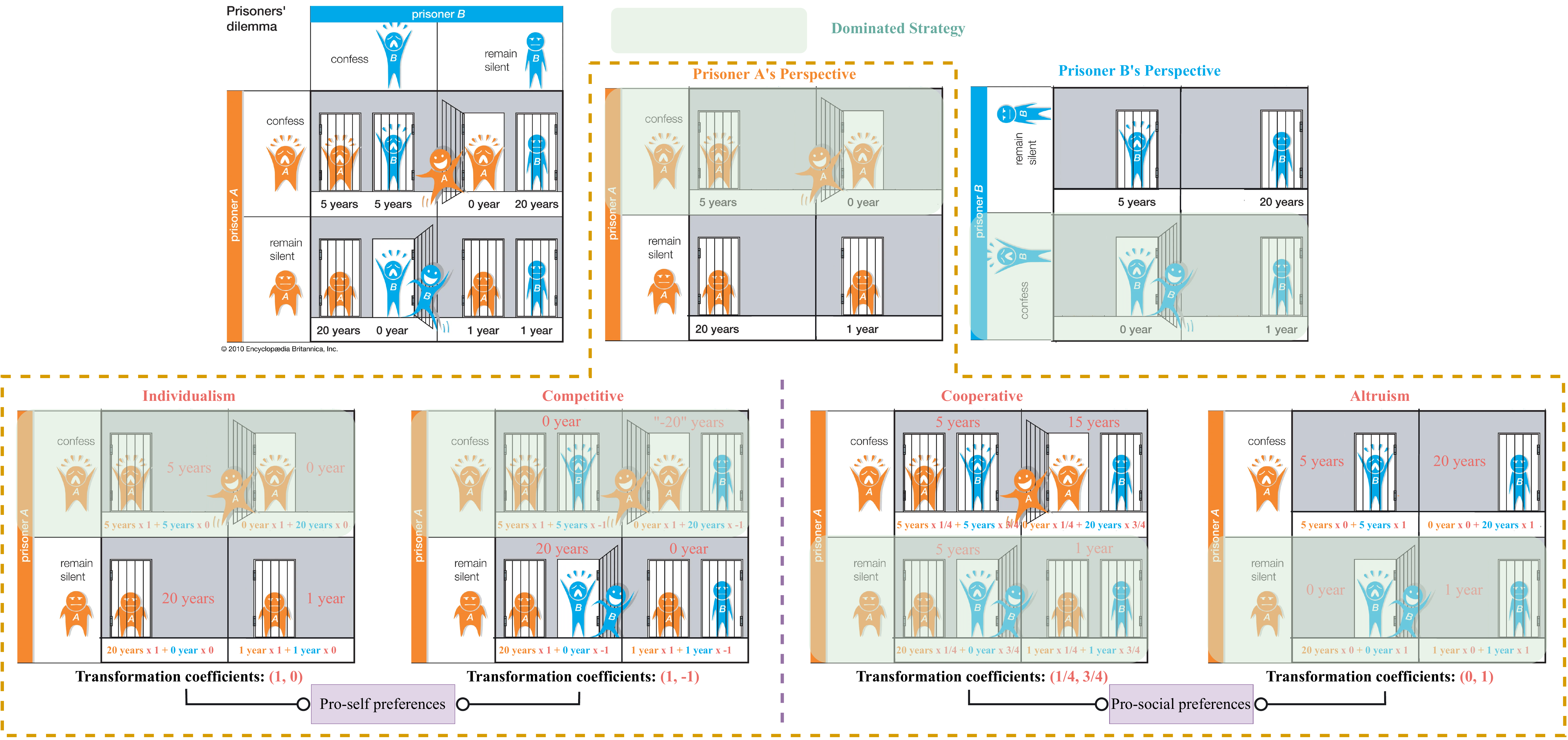}
    \caption{Interdependence theory in the prisoner's dilemma~\citep{pd2022}: the four pathways depict transformation processes for a row player who has individualistic, competitive, cooperative, and altruistic preferences, respectively; four resulting transformations suggest different dominant strategies (highlighted in green).}
    \label{fig:interdependence}
\end{figure}

The above three types of methods mainly make breakthroughs in methodology and are accompanied by simulation experiments to verify the correctness of the conclusions.
Considering the completeness of the theory and the feasibility of convergence analysis, this paper mainly focuses on solving intertemporal or public good social dilemmas based on social value orientations.
The aforementioned mainstream conclusions about SVO from social psychology and economics are mainly supported by interdependence theory~\citep{hansen1982interpersonal}.
In social psychology and economics games, classical game theory does not accurately predict human behavior. 
This is because, in these human-involved games, each player does not rely on the given payoff matrix to make decisions but on their own ``effective'' payoff matrix~\citep{hansen1982interpersonal,mckee2020social}.
The effective payoff matrix is constructed by redistributing payoffs for the given payoff matrix based on the players' respective SVOs.
As seen from Figure~\ref{fig:interdependence}, different SVO will make players choose different dominant strategies when facing the prisoner's dilemma, thus affecting the emergence of cooperation.
Many different social value orientations are theoretically possible, but most work has concentrated on various linear combinations of individuals' concern for the rewards for themselves and their partners.

Inspired by the interdependence theory, many previous works have introduced the SVO into MARL to solve the ISD~\citep{peysakhovich2018prosocial,hughes2018inequity,zhang2019sa,wang2019evolving,baker2020emergent,gemp2020d3c,yi2021learning,ivanov2021balancing,Schmid2021StochasticMG}.
\citet{peysakhovich2018prosocial} introduces the SVO into MARL for the first time and proposes the concept of prosocial, that is, cooperative orientation agents. 
The reward function of a prosocial agent is shaped as a fixed linear combination of its reward and the others.
\citet{hughes2018inequity} introduces an inequity aversion model in ISD, namely equality orientation, which promotes cooperation by minimizing the gap between one's return and that of other individuals.
The latter work is no longer satisfied with a fixed linear combination and begins to introduce trainable weight parameters. 
\citet{baker2020emergent} first attempts to randomize the linear weights of one's and others' rewards to observe whether cooperative behavior emerges. 
Since the linear weights are always greater than $0$, all agents can be roughly classified into three categories: cooperative-oriented, altruistic-oriented, or individual-oriented.
Going a step further, D3C~\citep{gemp2020d3c} optimizes the linear combination weights by using the ratio of the worst equilibrium to the optimal solution (Price of Anarchy, PoA) that measures the quality of the equilibrium points.
Concurrent work LToS~\citep{yi2021learning} models the optimization problem of linearly transforming weights as a bi-level problem and uses an end-to-end approach to train weights and policies jointly.
Considering the noise or privacy issues that instantaneous rewards for SVO modeling may introduce, some recent works shape the agents' reward in other ways.
\citet{Schmid2021StochasticMG} realizes the conditional linear combination of agent rewards by introducing the idea of the market economy.
\citet{zhang2019sa} and~\citet{ivanov2021balancing} use state-value and action-value functions to implement SVO modeling.
\citet{wang2019evolving} directly uses reward-to-go and reward-to-come, combined with evolutionary algorithms, to optimize the weights of nonlinear (MLP-based) combinations.
However, these methods cannot stably and efficiently converge to mutual cooperation under complex ISDs, which are further verified in our numerical experiments in Section~\ref{sec:exps}.

\begin{figure}[htb!]
    \centering
    \includegraphics[width=\textwidth]{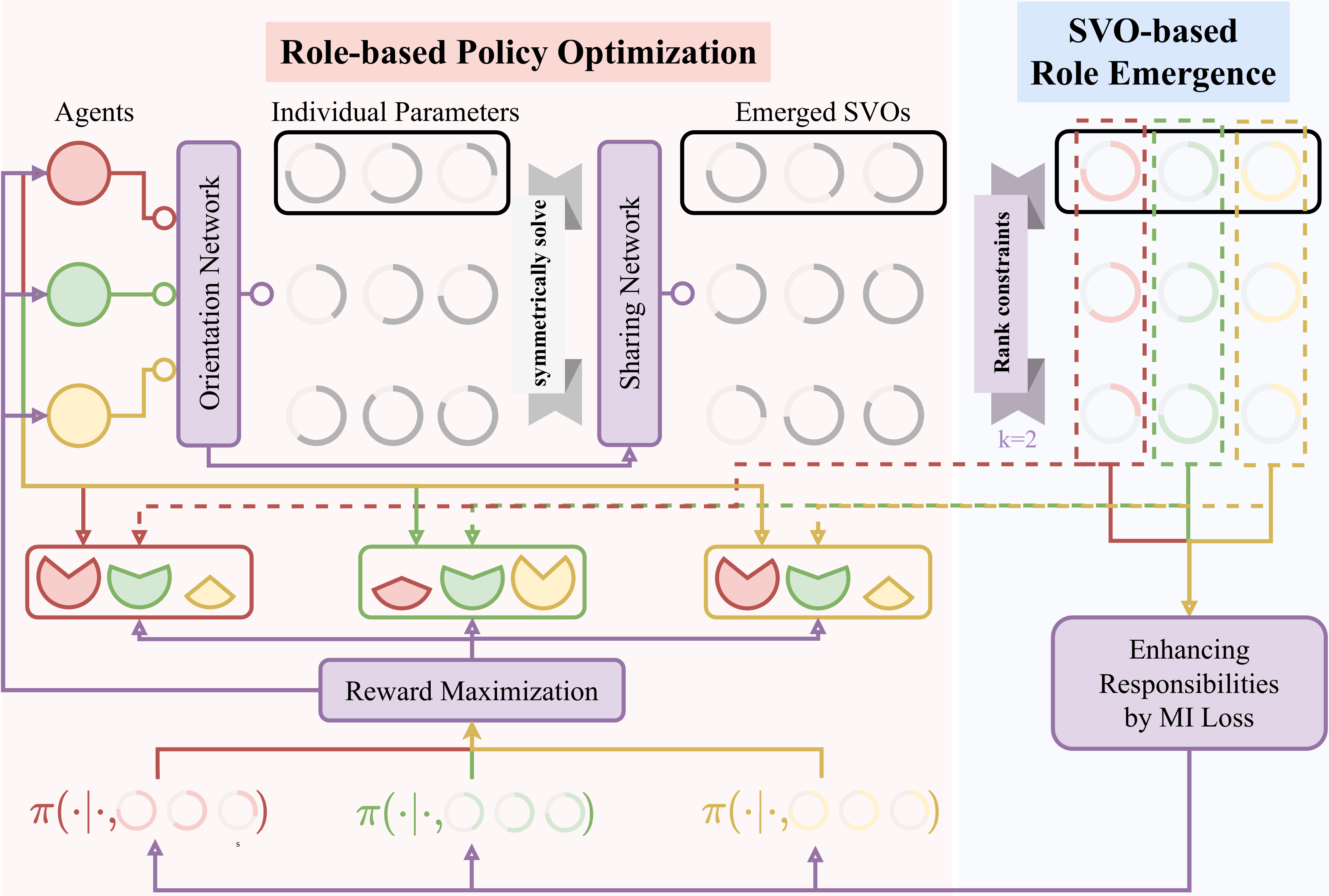}
    \caption{
    The conceptual diagram of the proposed RESVO, which is based on the social value orientations combined with a typical ``division of labor or roles'' mechanism, can benefit from providing a promising solution for the intertemporal social dilemma.
    RESVO is divided into two training phases of joint optimization and interleaved update: SVO-based role or division of labor emergence and role or division-based policy optimization. 
    In the first phase, RESVO transforms the learning of roles into a social value orientation emergence problem, which is symmetrically solved by endowing agents with altruism to learn to share rewards with other agents.
    An SVO-based role embedding space is then constructed by conditioning individual policies on roles with a novel rank regularizer and mutual information maximizer.
    Moreover, RESVO optimizes the policies based on the multi-agent policy gradient theorem in the second phase by maximizing the shaped rewards of all agents with different emerged social value orientations.
    }
    \label{fig:framework}
\end{figure}

\begin{figure}[htb!]
    \centering
    \begin{subfigure}[b]{0.24\textwidth}
        \includegraphics[width=\textwidth]{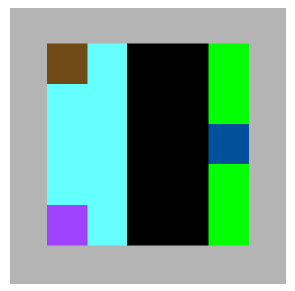}
        \caption{}
        \label{fig:roleinsaction-1}
    \end{subfigure}
    \begin{subfigure}[b]{0.24\textwidth}
        \includegraphics[width=\textwidth]{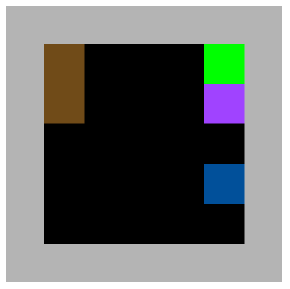}
        \caption{}
        \label{fig:noroleinsaction-1}
    \end{subfigure}
    \begin{subfigure}[b]{0.24\textwidth}
        \includegraphics[width=\textwidth]{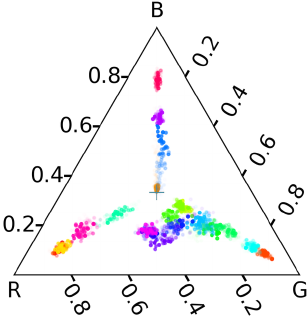}
        \caption{}
        \label{fig:roleinsaction-2}
    \end{subfigure}
    \begin{subfigure}[b]{0.24\textwidth}
        \includegraphics[width=\textwidth]{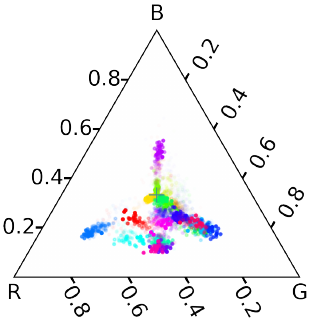}
        \caption{}
        \label{fig:noroleinsaction-2}
    \end{subfigure}
    \caption{(a) is a snapshot of the division of labor found by~\citet{yang2020learning} in \textit{Cleanup} task, where the blue agent picks apples, and the purple one stays on the riverside to clean waste. 
    In contrast, (b) shows a jointly suboptimal division where two failure agents compete for apples. 
    (c-d) \citet{vinitsky2021learning} shows similar results in \textit{Allelopathic Harvest} task.}
    \label{fig:roleffective}
\end{figure}

The conceptual diagram of our solution is shown in Figure~\ref{fig:framework}.
Specifically, we find that a typical mechanism of human society, i.e., division of labor or roles, can benefit from providing a promising solution for the ISD combined with SVOs.
The effectiveness of the division of labor in solving the ISD has emerged in existing MARL works but is still underexplored.
The numerical results from sanction-based methods~\citet{yang2020learning,vinitsky2021learning} on the typical ISD task \textit{Cleanup}~\citep{Hughes2018InequityAI} and \textit{Allelopathic Harvest}~\citep{koster2020model} show that policies solving ISDs effectively exhibit a clear division of labor (Figure~\ref{fig:roleffective}).
Many natural systems feature emergent division of labor, such as ants~\citep{gordon1996organization}, bees~\citep{jeanson2005emergence}, and humans~\citep{butler2012condensed}.
In these systems, the division of labor is closely related to the roles and is critical to labor efficiency.
The division of labor, or the role theory, has been widely studied in sociology and economics~\citep{Smith2013OnTW}.
A role is a comprehensive pattern of behavior, and agents with different roles will show different behaviors. 
Thus the overall performance can be improved by learning from others' strengths~\citep{Wang2020ROMAMR}.
These benefits inspired multi-agent system designers, who try to reduce the design complexity by decomposing the task and specializing agents with the same role to certain sub-tasks~\citep{wooldridge2000gaia,omicini2000soda,padgham2002prometheus,pavon2003agent,cossentino2005passi,zhu2008role,spanoudakis2010using,deloach2010mase,bonjean2014adelfe}.
However, roles and the associated responsibilities (or subtask-specific rewards~\citet{sun2020reinforcement}) are predefined using prior knowledge in this systems~\citep {Lhaksmana2018RoleBasedMF}.
Although pre-definition can be efficient in tasks with a clear structure, such as software engineering~\citep{bresciani2004tropos}, it hurts generalization and requires prior knowledge that may not be available in practice.
To solve this problem, \citet{wilson2010bayesian} uses Bayesian inference to learn a set of roles, and ROMA~\citep{Wang2020ROMAMR} designs a specialization objective to encourage the emergence of roles.
\citet{Wang2021RODELR} improves the learning efficiency in hard-exploration tasks by first decomposing joint action spaces according to action effects, which makes role discovery much more effortless. 
Unfortunately, none of these methods considers the intertemporal social dilemma.

Drawing the insight from studies in social psychology that characteristics of laborers, or roles, influence the SVOs reciprocally~\citep{Sutin2010ReciprocalIO, Holman2021TransactionsBB}, this paper uses the agent's SVO to represent the role of each agent, transforming the role learning problem into the emergence of the agent's SVO, thereby naturally constructing a role-based framework in MARL to solve ISD.
Specifically, we use the SVOs, i.e., the coefficients in the transformation matrices from independence theory, to represent each agent's role, e.g., $(0,1)$ in individualistic preference and $(0.5, 0.5)$ in cooperative preference.
This method assumes that all agents will have real-time access to one another's rewards while learning.
However, making reward data unrestrictedly accessible is undesirable for several reasons.
For example, agent designers want to imperceptibly modify the agent's reward function or prohibit from sharing their agents' reward function~\citep{yang2021advances}.
This makes the emergence of social value orientation unfeasible, making it impossible to promote the division of labor based on SVOs.

\begin{figure}[htb!]
    \centering
    \includegraphics[width=0.6\linewidth]{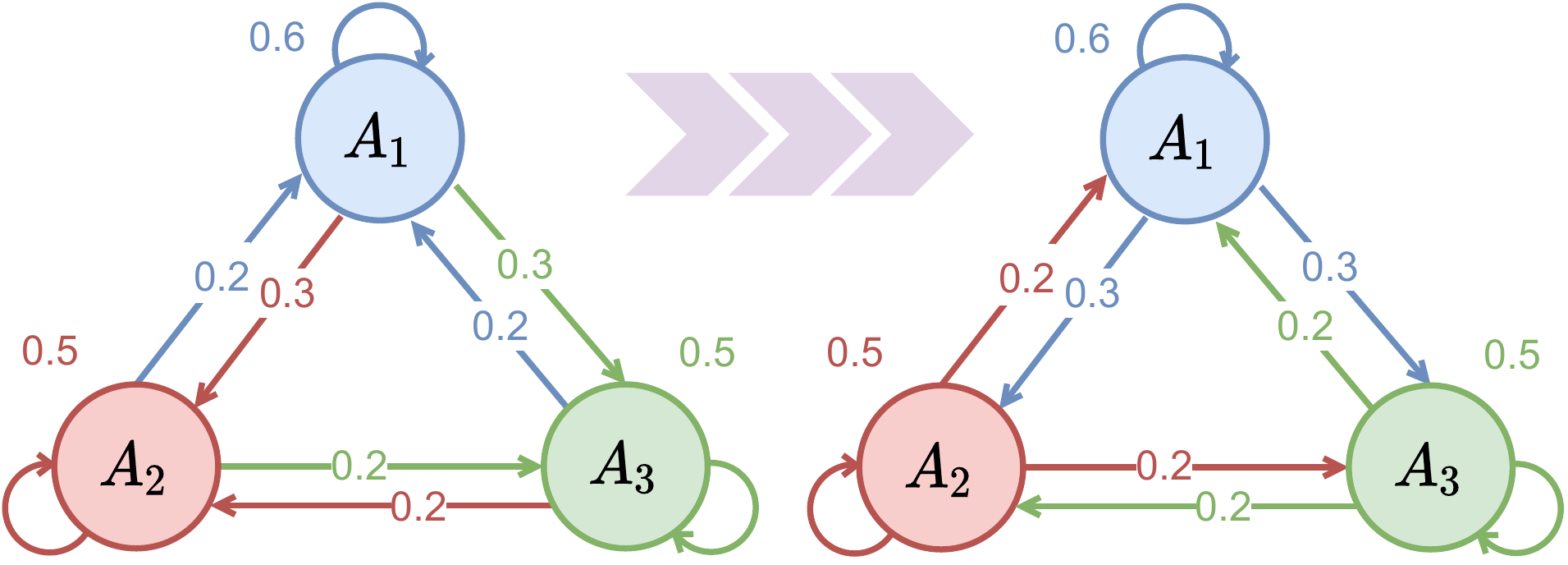}
    \caption{Symmetrically converting (left) the social value orientation learning problem to (right) the learning to share problem in a three-agents environment. The circles of different colors represent different agents, and the numbers of different colors represent the parameters that each agent needs to learn. Assuming that $r_1$, $r_2$, and $r_3$ represent the extrinsic reward of each agent, the shaped reward of $A_1$ is $0.6r_1+0.2r_2+0.2r_3$. The shaped rewards of other agents can also be computed similarly.}
    \label{fig:l2s}
\end{figure}

Inspired by the fact that \textit{altruism} plays a crucial role in human's solution to social dilemmas~\citep{kollock1998social,eisenberg1989roots}, that is, consumers altruistic share a part of their profits with producers, this paper proposes a novel algorithm framework, called Learning Role with Emergent SVO (RESVO), that establish a symmetric relationship between SVO emergence and learning to share.
RESVO encourages agents to learn to dynamically share the reward with other agents (see Figure~\ref{fig:l2s}).
In this learning paradigm, the learnable parameters, or the SVO of each agent, are the proportions\footnote{The reward-receiver agent cannot access these proportions, so it is impossible to reverse infer the reward giver's reward.} of rewards it receives from other agents to the extrinsic rewards of these reward giver.
We take these emergent SVOs as the role representations of each agent.
RESVO then imposes a novel low-rank constraint on the SVO matrix of all agents to effectively represent the different roles of agents and uses projected gradient descent to solve constrained optimization problems.
Furthermore, to establish the connection between roles and decentralized policies, RESVO conditions agents' policies on individual emergent SVOs by explicitly feeding agents' SVO-based role embeddings into their local policies correspondingly.
Furthermore, to associate roles with responsibilities, we propose to learn SVOs that are identifiable by agents' long-term behaviors by maximizing the conditional mutual information between the individual trajectory and the emergent SVO given the current observation and other agents' actions, which is similar as~\citep{Wang2020ROMAMR}.

\section{Preliminaries}\label{sec:pre}

Although studies on social dilemmas have contributed significantly to the research of cooperation emergence for decades~\citep{axelrod1981evolution,peysakhovich2018consequentialist,anastassacos2020partner}, they focus on matrix games and fixed binary policies.
To be more realistic, as in real-world situations, the MARL community considers the intertemporal social dilemmas (ISDs,~\citet{leibo2017multi, Hughes2018InequityAI}).
Before conducting numerical experiments, we first give the formal definition of ISD as follows.
An ISD can be modeled as a partially observable general-sum Markov game~\citep{hansen2004dynamic},
$$
\mathcal{M}=
\left\langle \mathcal{I}, \mathcal{S}, \left\{ \mathcal{A}_i \right\}_{i=1}^{N}, \left\{ \mathcal{O}_i \right\}_{i=1}^{N}, \mathcal{P}, \mathcal{E}, \left\{ \mathcal{R}_i \right\}_{i=1}^{N} \right\rangle,
$$ where $\mathcal{I}$ represents the $N$-agent space.
$s\in\mathcal{S}$ represents the true state of the environment.
We consider partially observable settings, where agent $i$ is only accessible to a local observation $o_i\in\mathcal{O}_i$ according to the emission function $\mathcal{E}(o_i\mid s)$.
At each timestep, each agent $i$ selects an action according to a policy $a_{i} \in \pi_{i}\left(a \mid o_{i}\right)$, forming a joint action $\boldsymbol{a}=\left\langle a_{1}, \ldots, a_{N}\right\rangle \in \times\mathcal{A}_i$, results in the next state $s^{\prime}$ according to the transition function $P\left(s^{\prime} \mid s, \boldsymbol{a}\right)$ and a reward $r_i=\mathcal{R}_i(s, \boldsymbol{a})$.
In ISDs, agents must learn cooperation or defection policies consisting of potentially long sequences of environmental actions instead of taking atomic cooperation or defection actions.
In this paper, we focus on the episodic game with horizon $T$, and the goal of each agent is to maximize the \textit{local} expected return, i.e.,
$$Q^{\pi}_{i}(s, \boldsymbol{a})=\mathbb{E}_{s_{0: T}, \boldsymbol{a}_{0: T}\sim \pi, P}\left[\sum_{t=0}^{T} \gamma^{t} \mathcal{R}_{i}\left(s_{t}, \boldsymbol{a}_{t}\right) \mid s_{0}=s, \boldsymbol{a}_{0}=\boldsymbol{a}\right].$$

\section{Methods}\label{sec:method}

This section proposes a novel learning framework, RESVO, that transforms role-based learning into an SVO emergence problem to solve the ISD.
Because consumers altruistically share a part of their profits with producers and making reward data unrestrictedly accessible is undesirable for several reasons, the proposed RESVO achieves SVO emergence by endowing agents with altruism to learn to share rewards with different weights to other agents.
An SVO-based role embedding space is then constructed by introducing a novel low-rank constraint and conditioning individual policies on roles to ensure that the emergent SVO can effectively represent the different roles of agents and associate roles with responsibilities.
Therefore, the following content in this section will be expanded from two aspects: SVO-based role emergence and role-based policy optimization.

\subsection{SVO-based Role Emergence}\label{sec:svo-based}

As mentioned in Section~\ref{sec:intro}, to consider the fact that consumers altruistically share a part of their profits with producers and avoid the realistic constraint (making reward data unrestrictedly accessible) imposed by directly learning the SVO of each agent according to the independence theory, RESVO enables agents to learn to dynamically share the reward with other agents, as shown in Figure~\ref{fig:l2s}.
Specifically, the SVO-based role emergence mechanism learns an orientation function for each agent by explicitly accounting for its impact on recipients' behavior and, through them, the impact on its extrinsic objective.
Each agent gives rewards using its orientation function and learns an SVO-conditioned policy with all received rewards.
For clarity, we use index $i$ when referring to the reward-sharing part of an agent, and we use $j$ for the part that learns from the received reward, which is similar with~\citet{yang2020learning}.

A reward-sharing agent $i$ learns a individual orientation function, $w^{i}_{\eta_i}:\mathcal{O}_{i}\times\mathcal{A}_{-i}\mapsto\mathbb{R}^{N}$, parameterized by $\eta_i$, that maps its own observation $o_i$ and all other agents' actions $a_{-i}$ to a vector of reward-sharing ratios for all $N$ agents.
Unlike the existing methods based on SVO~\citep{peysakhovich2018prosocial,baker2020emergent,gemp2020d3c,yi2021learning} or saction~\citep{koster2020silly,lupu2020gifting,yang2020learning,vinitsky2021learning,dong2021birds} mechanism, the orientation function in RESVO
\textbf{(1)} allows agents to reward itself
\footnote{But, this does not mean that the orientation function will converge to some trivial function, such as the agent giving itself an infinite reward. Because the final reward received by the agent is its reward multiplied by the sharing ratio, and the sharing ratio is in the closed range of $0$ to $1$.}, and 
\textbf{(2)} the sum of all sharing ratios does not need to be equal to $1$.
This is one of the reasons why reward sharing, a mechanism used by existing work, can encourage the division of labor and solve ISD.
The intuition behind this lies in the particular properties of the public good dilemma.
If there is a good division of labor among agents, the reward (punishment) of the agent does not come entirely from its behavior but partly from the producers (the consumers). 
Therefore, the agent that gets the reward needs to share a part with other agents and only gets a part of it (corresponding to the first point);
Moreover, in a multi-agent scenario, a reward may come from the behavior of multiple producers, so it needs to share the same reward with multiple agents (corresponding to the second point).

Similar with~\citet{lupu2020gifting,yang2020learning}, $w_{\eta_i}$ is separate from the agent's conventional policy and is learned via direct gradient descent on the agent's extrinsic objective to reduce the learning difficulty.
Specifically, at each timestep $t$, each recipient $j$ receives a total reward
\begin{equation}
    r_j(\boldsymbol{\eta}, \boldsymbol{r}):=w^{j}_{\eta_j}[j] \cdot r_{j} + \sum_{i \neq j} w^{i}_{\eta_i}[j] \cdot r_{i},
\end{equation}
where $w^{j}_{\eta_j}[j]$ and $w^{i}_{\eta_i}[j]$ denotes the $j$-th elements of $w^{j}_{\eta_j}$ and $w^{i}_{\eta_i}$ respectively, $\boldsymbol{r}:=[r_0, \cdots, r_N], \boldsymbol{\eta}=[\eta_1, \cdots, \eta_N]$.
Although the sharers' rewards appear in Equation 1, the recipients can only see the sharers' discounted rewards when implemented.
Each agent $j$ learns a SVO-based role conditioned policy $\pi_j(\cdot \mid o_j,e_j(\boldsymbol{\eta}))$ parameterized by $\theta_j$, where $e_j(\cdot)$ is the SVO-based role embedding of agent $j$.
After each agent has updated its policy to $\hat{\pi}_j$, parameterized by new $\hat{\theta}_j$, with role-based policy optimization (Section~\ref{sec:role-policy}) via trajectories $\tau_i$ sampled by joint policies $\{\pi_j\}$, we sample a set of new trajectories with new joint policy $\{\hat{\pi}_j\}$.
Using these trajectories, each agent $i$ updates the individual orientation parameters $\eta_i$ to maximize the following objective
\begin{equation}\label{eq:svo-rank}
        \max_{\eta_i} J^{\mathrm{svo}}(\hat{\tau}_i,\tau_i,\hat{\boldsymbol{\theta}},\boldsymbol{\eta}):=\mathbb{E}_{\hat{\boldsymbol{\pi}}}\left[\sum_{t=0}^{T}\gamma^t \hat{r}_i^t\right],
        \quad\mathrm{s.t.}\;\mathrm{rank}(W^t_{\boldsymbol{\eta}}) = k, \forall t\in[0, T),
\end{equation}
where $\hat{r}_i^t$ is the newly sampled extrinsic reward in $\hat{\theta}_i$, $W^t_{\boldsymbol{\eta}}=\{w^{i,t}_{\eta_i}\}_{i=1}^{N}$ is the matrix composed of the reward sharing ratios of all agents at timestep $t$, and $k \leq N$ is a hyperparameter.
To ensure that the emergent SVO can effectively represent the different roles of agents, RESVO introduces a novel rank constraint on the SVO matrix $W^t_{\boldsymbol{\eta}}$ of all agents, and $k$ can be regarded as the theoretical optimal number of roles.
To be able to optimize (\ref{eq:svo-rank}) with an automatic differentiation toolkit in an end-to-end manner, we transform (\ref{eq:svo-rank}) into the following unconstrained optimization problem based on projected gradient descent by introducing an intrinsic reward
\begin{equation}\label{eq:svo-rank-u}
    \max_{\eta_i} J^{\mathrm{svo}}:=\mathbb{E}_{\hat{\boldsymbol{\pi}}}\left[\sum_{t=0}^{T}\gamma^t \left(\hat{r}_i^t-\alpha\|W^{i,t}_{\eta_i}-W_k^{i,t}\|^{2}_{2}\right)\right],
\end{equation}
where $W_k^t$ is the $k$-rank approximation of $W^t_{\eta}$ obtained with SVD algorithm and $\alpha^t$ is another hyperparameter, and superscription $i$ denotes the $i$-th column of the matrix.
In practice, following the derivation process of~\citep{yang2020learning}, one can define the loss as
\begin{equation}\label{eq:svo-loss}
    \begin{aligned}
        -\sum_{t=0}^{T} &\sum_{j=1}^{N} \log \pi^{j}_{\hat{\theta}^{j}}\left(\hat{a}^{t}_{j} \mid \hat{o}^{t}_{j}, \hat{e}^{t}_{j}(\boldsymbol{\eta})\right) \cdot \\
        \quad &\sum_{\ell=t}^{T} \gamma^{\ell-t} \left(\hat{r}_i^{\ell}-\alpha\|\Delta^{i,\ell}(W,k)\|^{2}_{2}\right)-2\alpha\nabla_{\eta_i}W^{i,t}_{\eta_i}\Delta^{i,t}(W,k),
    \end{aligned}
\end{equation}
and directly minimize it via automatic differentiation, where $\Delta^{i,\ell}(W,k):=W^{i,\ell}_{\eta_i}-W_k^{i,\ell}$ and $\Delta^{i,t}(W,k):=W^{i,t}_{\eta_i}-W_k^{i,t}$.
Crucially, $\hat{\theta}^{j}$ must preserve the functional dependence of the policy update step~(\ref{eq:policy-update}) on $\eta_i$ within the same computation graph.

It is worth noting that the agent's role representation $e_i(\boldsymbol{\eta})$ is \textbf{NOT} composed of its reward \textit{sharing} ratios $w^{i}_{\eta_i}$. 
However, its reward \textit{recipient} ratios, i.e., we take the $i$-th row in the matrix $W^t_{\boldsymbol{\eta}}$ as the role representation of agent $i$ at timestep $t$, as shown by interdependence theory in Figure~\ref{fig:interdependence}. 
Intuitively, agents with similar role representations have similar divisions of labor and thus receive similar rewards.
In other words, RESVO decomposes the extrinsic reward received by all agents into several parts according to the functional composition of the agent during the SVO emergence learning stage.
The difference in rewards will directly lead to differences in the agents' policies, thereby encouraging the formation of the division of labor among the agents.

\subsection{Role-based Policy Optimization}\label{sec:role-policy}

Introducing SVO-based role embedding and conditioning individual policies on this embedding explicitly establishes the connection between the role and the individual policies to encourage the division of labor through the diversity of roles. 
However, this does not enable the role to constrain the agent's long-term behavior.
Intuitively, conditioning roles on local observations and actions\footnote{The orientation function $w^{i}_{\eta_i}$ maps agent $i$'s observation $o_i$ and all other agents' actions $a_{-i}$ to a vector of reward-sharing ratios for all $N$ agents.} enables roles to be responsive to the changes in the environment but may cause roles to change quickly.
Thus roles and responsibilities cannot be effectively associated.
To address this problem, we expect SVO-based roles to be temporally stable.

Drawing inspiration from \citet{eysenbach2018diversity, Wang2020ROMAMR}, we propose to learn SVO-based roles that are identifiable by agents' long-term behaviors, which can be achieved by maximizing $I(\tau_i;e_i \mid \boldsymbol{o},\boldsymbol{a})$, the conditional mutual information between the individual trajectory and the role given the current \textit{joint} observation and \textit{joint} action.
The conditional mutual information calculated here is based on the joint actions and observations of all agents because $e_i$ cannot be generated by local observations $o_i$.
The role representation of agent $i$, $e_i$, is the $i$-th \textbf{row} in the matrix $W_{\boldsymbol{\eta}}$.
Based on variational inference, a variational posterior estimator can be proposed to derive a tractable lower bound for the mutual information objective
\begin{equation}\label{eq:lowerbound}
\max_{\boldsymbol{\eta},\phi} \;J_{i}^{\mathrm{mi}}:=I(e_{i}^{t};\tau_{i}^{t-1} \mid \boldsymbol{o}^{t},\boldsymbol{a}^{t}) \geq 
\mathbb{E}_{e_{i}^{t}, \tau_{i}^{t-1}, \boldsymbol{o}^{t},\boldsymbol{a}^{t}}\left[\log \frac{q_{\phi}\left(e_{i}^{t} \mid \tau_{i}^{t-1}, \boldsymbol{o}^{t},\boldsymbol{a}^{t}\right)}{W_{\boldsymbol{\eta}}\left(e_{i}^{t} \mid \boldsymbol{o}^{t},\boldsymbol{a}^{t}\right)}\right],
\end{equation}
where ``$\mathrm{mi}$'' stands for ``mutual information'', $\tau_{i}^{t-1}=\left(o_{i}^{0}, a_{i}^{0}, \cdots, o_{i}^{t-1}, a_{i}^{t-1}\right)$\footnote{There's a bit of misuse of notation here, and compared to Section~\ref{sec:svo-based}, no rewards are included in the trajectory here.} and $q_{\phi}$ is the variational estimator parameterized by $\phi$.
Here we use $W_{\boldsymbol{\eta}}$ to represent orientation function $w^{i}_{\eta_i}$ of all agent $i$.
For $q_{\phi}$, we use a causal transformer~\citep{chen2021decision} (shared by all agents) to encode an agent's history of observations and actions.
The lower bound in (\ref{eq:lowerbound}) can be further rewritten as a loss function to be minimized via automatic differentiation
\begin{equation}
\begin{aligned}
\operatorname{L}^{\mathrm{mi}}\left(\boldsymbol{\tau};\boldsymbol{\eta},\phi\right) &= \frac{1}{n}\sum_{i=1}^{N}\operatorname{L}_{i}^{\mathrm{mi}}\left(\tau_i;\boldsymbol{\eta},\phi\right) \\
&= \mathbb{E}\left[D_{\mathrm{KL}}\left[W_{\boldsymbol{\eta}}\left(e_{i}^{t}\mid \boldsymbol{o}^{t},\boldsymbol{a}^{t}\right) \| q_{\phi}\left(e_{i}^{t} \mid \tau_{i}^{t-1}, \boldsymbol{o}^{t},\boldsymbol{a}^{t}\right)\right]\right],
\end{aligned}
\end{equation}
where $D_{\mathrm{KL}}[\cdot\|\cdot]$ is the KL-divergence operator, and the conditioned terms in $p,q_{\phi}$ are omitted for convenience.
See the Appendix for the detailed derivation.

In addition, to further improve learning stability, the individual policies are no longer based on the current role embedding but on the latest $m$ role embeddings to make decisions.
That is, each agent $j$ learns a SVO-based role conditioned policy $\pi_j(a^t_j \mid o^t_j,\{e^l_j(\boldsymbol{\eta})\}_{l=t-m+1}^{t})$ parameterized by $\theta_j$.
Our experiments found that for complex ISDs with long episodes, mutual information constraints and decision-making based on historical role embeddings can be crucial in improving performance.
Thus, the objective of role-based policy optimization for each agent $j$ is shown as follows based on the multi-agent policy gradient theorem~\citep{Lowe2017MultiAgentAF}
\begin{equation}\label{eq:policy-update}
    \max_{\theta_j} J^{\mathrm{policy}}(\theta_j,\boldsymbol{\eta}):=\mathbb{E}_{\boldsymbol{\pi}(\cdot \mid \cdot,e_j(\boldsymbol{\eta}))}\left[\sum_{t=0}^{T}\gamma^t r_j^t(\boldsymbol{\eta}, \boldsymbol{r}^{t})\right].
\end{equation}

\subsection{Algorithm Summary}

The learning process of RESVO consists of two main steps, namely, SVO-based role emergence (Eq.~\ref{eq:svo-rank}) and role-based policy optimization (Eq.~\ref{eq:policy-update}). 
To improve the expressiveness and responsibility of role representation, we additionally impose rank constraints (Eq.~\ref{eq:svo-rank}) and mutual information regularization (Eq.~\ref{eq:lowerbound}) to the objectives of role emergence and policy optimization.
Therefore, the final learning objective of RESVO for each agent $i$ is
\begin{equation}
    \max_{\theta_i,\eta_i,\phi} J := \overbrace{\lambda_{\mathrm{svo}}J^{\mathrm{svo}} + \lambda_{\mathrm{mi}}J^{\mathrm{mi}}}^{\textcolor{red}{\mathrm{role\;emergence}}} + \underbrace{\lambda_{\mathrm{p}}J^{\mathrm{policy}}}_{\textcolor{blue}{\mathrm{policy\;optimization}}},
\end{equation}
where $\lambda_{\mathrm{svo}},\lambda_{\mathrm{mi}},\lambda_{\mathrm{p}}$ are scaling factors.
Furthermore, the pseudo-code of the proposed RESVO is shown in Algorithm~\ref{alg:resvo}.

\begin{algorithm}[H]
\caption{Leanring Role with Emergent SVO (RESVO)}
\label{alg:resvo}
\begin{algorithmic}[1]
\Require Initialize parameters of policies $\theta_i$, orientation function $\eta_i$, variational estimator $\phi$;
\For {each iteration}
    \State Collect interaction $\tau_i$ with $\pi_{\psi_i}$ and update replay buffer $\mathcal{D} \leftarrow \{\tau_i\}$;
    \State \textcolor{blue}{Generate trajectories $\{\tau_i\}$ using $\boldsymbol{\theta}$ and $\boldsymbol{\eta}$, and for all reward receivers $j$, update $\hat{\theta}_j$ via (\ref{eq:policy-update})};
    \State \textcolor{red}{Generate new trajectories $\{\hat{\tau}_i\}$ using new $\hat{\boldsymbol{\theta}}$ and for reward sharers $i$, compute $\hat{\eta}_i$ via (\ref{eq:svo-rank}),(\ref{eq:lowerbound})};
    \State \textcolor{red}{For variational estimator, compute $\hat{\phi}$ via (\ref{eq:lowerbound})}, $\textcolor{blue}{\theta_i \leftarrow \hat{\theta}_i}$, $\textcolor{red}{\eta_i \leftarrow \hat{\eta}_i}$ for all $i$, $\textcolor{red}{\phi \leftarrow \hat{\phi}}$.
\EndFor
\end{algorithmic}
\end{algorithm}

\section{Experiments}\label{sec:exps}

Although studies on social dilemmas have contributed significantly to the research of cooperation emergence for decades~\citep{axelrod1981evolution,peysakhovich2018consequentialist,anastassacos2020partner}, they focus on matrix games and fixed binary policies.
To be more realistic as in real-world situations, the MARL community considers the intertemporal social dilemmas (ISDs,~\citet{leibo2017multi, Hughes2018InequityAI}, see Appendix \ref{sec:pre} for the formal definition) which can be modeled as a partially observable general-sum Markov game~\citep{hansen2004dynamic}.

\begin{figure}[htb!]
    \centering
    \includegraphics[width=\textwidth]{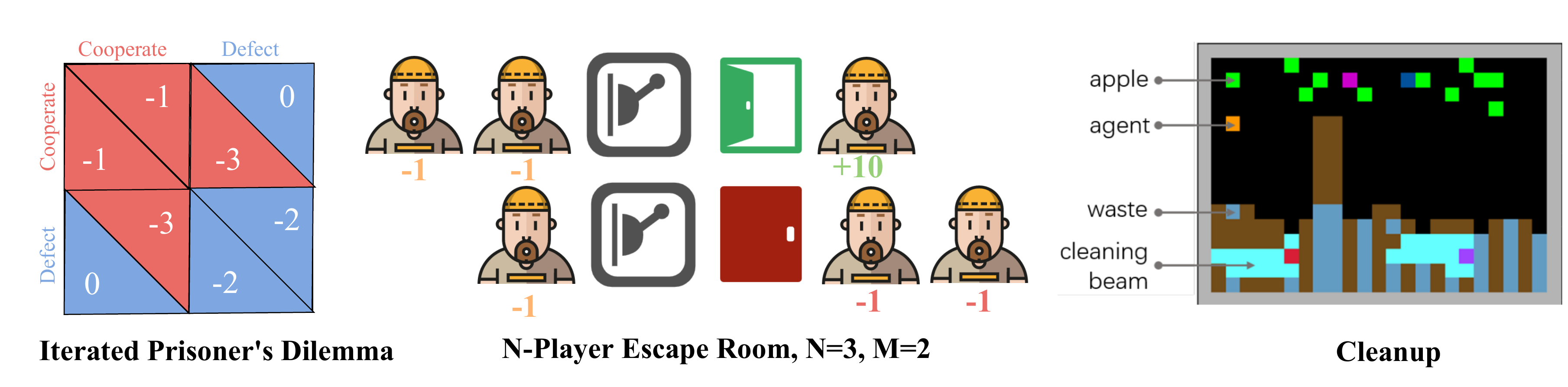}
    \caption{Three different environments with increasing complexity.}
    \label{fig:envs}
\end{figure}

Our experiments demonstrate that the division of labor can effectively solve the ISDs of different difficulties. 
Moreover, compared with other mechanisms, the division of labor can converge to mutual cooperation faster and more stably.
Our method is tested in the following three typical public good dilemmas with increasing complexity (see Figure~\ref{fig:envs}) against several baselines.
\begin{itemize}
    \item The first one, Iterated Prisoner's Dilemma (IPD)~\citep{foerster2018learning}, is often regarded as the canonical and most difficult of $2$-by-$2$ game-theoretic cooperation problems.
    The prisoner setting may seem contrived, but there are, in fact, many examples of human interaction as well as interactions in nature that have the same payoff matrix.
    Many natural processes have been abstracted into models in which living beings are engaged in endless games of prisoner's dilemma, e.g., the climate-change politics in environmental studies~\citep{rehmeyer2012game}, the reciprocal food exchange of vampire bats~\citep{davis2017selfish}, the doping in sport~\citep{schneier2012lance} and the coherence of strategic realism in the international political~\citep{majeski1984arms}.
    This broad applicability of the PD gives the game substantial importance.
    In an IPD, agents observe the joint action taken in the previous round and receive rewards in Figure~\ref{fig:envs}.
    Although, by definition, a public good dilemma needs to contain more than $2$ agents~\citep{kollock1998social}, the prisoner's dilemma can be viewed as a simplified version. 
    The agent that selects the ``cooperate'' action corresponds to the ``producer'', and the agent that selects the ``defect'' corresponds to the ``consumer''. 
    In this simplified version of the public good dilemma, defect or free-riding is the dominant strategy.
    \item The second one, $N$-Player Escape Room (ER) (\citet[Figure 1]{yang2020learning}), is a discrete $N$-player Markov game with parameter $M<N$.
    ER is a more complex public good dilemma. 
    Since $M=2$, $N=3$, the optimal joint policy of the game has an asymmetric division of labor, which is also widespread in the economic activities of human society.
    Specifically, there are $3$ discrete states in this game, i.e., ``start'', ``door'' and ``lever''.
    An agent gets $+10$ extrinsic reward for exiting a ``door'' and ending the game, but the ``door'' can only be opened when $M$ other agents cooperate to pull the ``lever''. 
    However, an extrinsic penalty of $-1$ for any movement discourages all agents from taking this cooperative action.
    \item In the third one, Cleanup~\citep{leibo2017multi, Hughes2018InequityAI} is a high-dimensional grid-world intertemporal social dilemma that serves as a problematic benchmark for independent learning agents.
    Moreover, the Cleanup can also be considered as a simplified version of the tax simulator, \textit{Gather-Trade-Build (GTB)}~\citep{zheng2022aieconomist}, but since there is no significant division of labor in the tasks involved in the latter, it is not considered in this paper.
    In the Cleanup, agents get $+1$ (for the small $10$ by $10$ map) or $+0.25$ (for the big $48$ by $18$ map) reward by collecting apples, which spawn on the map at a linear decay rate as the amount of waste approaches a depletion threshold.
    Each episode starts with a waste level above the threshold and no apple present.
    An agent can contribute to the public good by firing a cleaning beam to clear waste (no reward).
    This would enable other agents to be free riders, resulting in a problematic public good dilemma.
\end{itemize}

We selected the aforementioned role learning algorithm, \textbf{ROMA}~\citep{Wang2020ROMAMR}, and sanction-based algorithm, \textbf{LIO}~\citep{yang2020learning}, as the baselines because our RESVO draws on the core ideas of them in the SVO-based role emergence and the role-based policy optimization\footnote{See Method (Section~\ref{sec:method}) for more details.} respectively.
In addition, we selected several aforementioned SVO-based algorithms, including \textbf{LToS}~\citep{yi2021learning} and \textbf{D3C}~\citep{gemp2020d3c} as baselines. 
Below we will briefly introduce the core idea of each algorithm again.
ROMA designs an end-to-end specialization learning objective to encourage the emergence of roles in general MARL tasks for better cooperation and generalization and avoid the requirement of prior or expert knowledge.
LIO enables each agent to punish, thereby implementing the sanction mechanism in a decentralized manner.
LToS uses a bi-level optimization scheme similar to LIO but uses meta-gradients instead of data sampled by other agents' updated policies for the learning of SVOs; 
D3C uses the price of anarchy instead of the joint expected cumulative reward in LToS as the optimization objective of SVOs.
Meanwhile, the SVO-based role emergence mechanism in RESVO can be introduced into other SVO-based methods as a plug-and-play module.
In order to verify the general promotion effect of the division of labor on SVO-based methods, we added the role learning mechanism in RESVO to two SVO-based works, denoted as \textbf{LToS+r} and \textbf{D3C+r} respectively.

\subsection{Role Emergence in the Classic Tasks}

\begin{figure}[htb!]
    \centering
    \begin{subfigure}[b]{0.45\textwidth}
        \includegraphics[width=\textwidth]{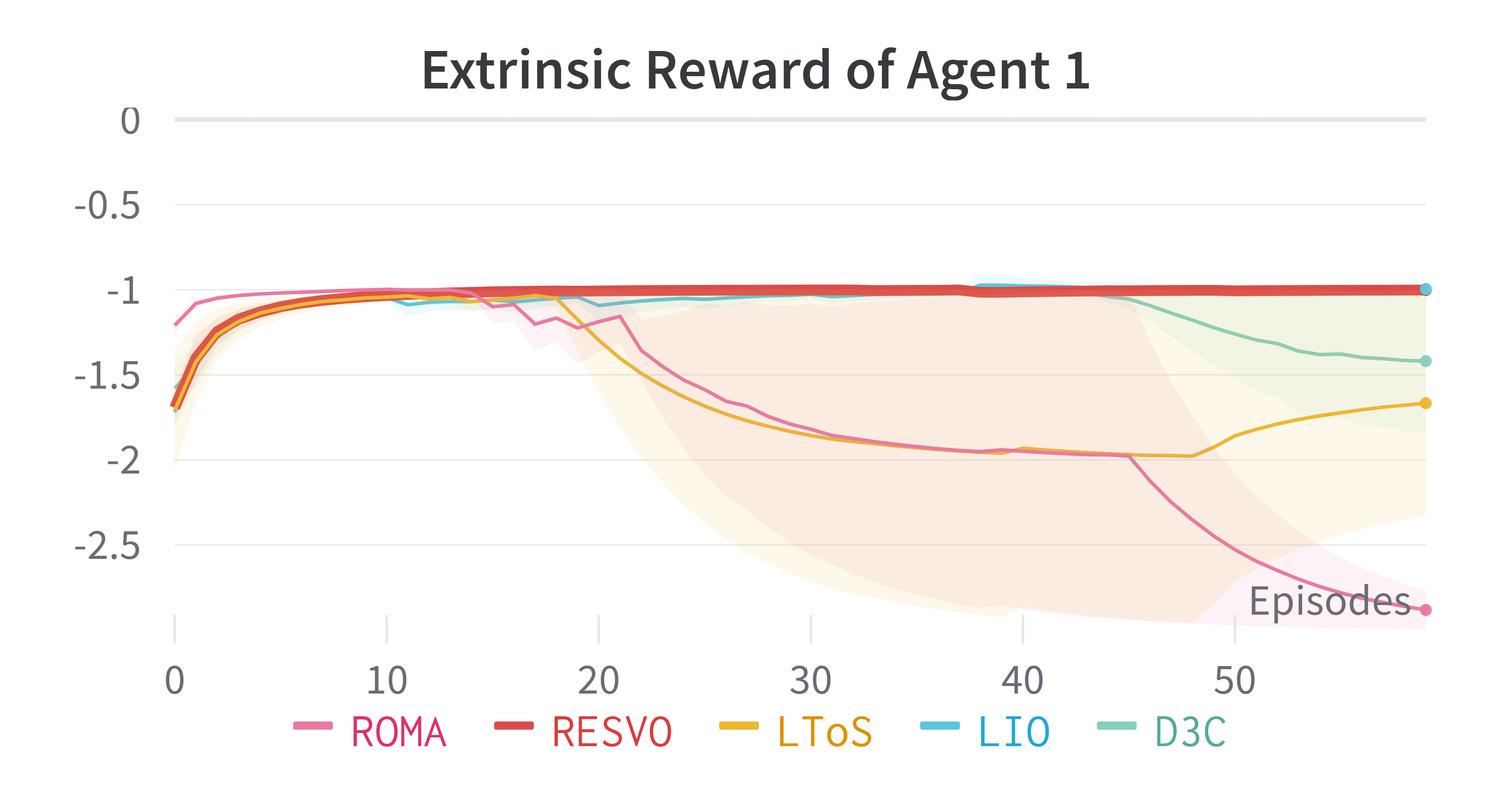}
        \caption{}
    \end{subfigure}
    \begin{subfigure}[b]{0.45\textwidth}
        \includegraphics[width=\textwidth]{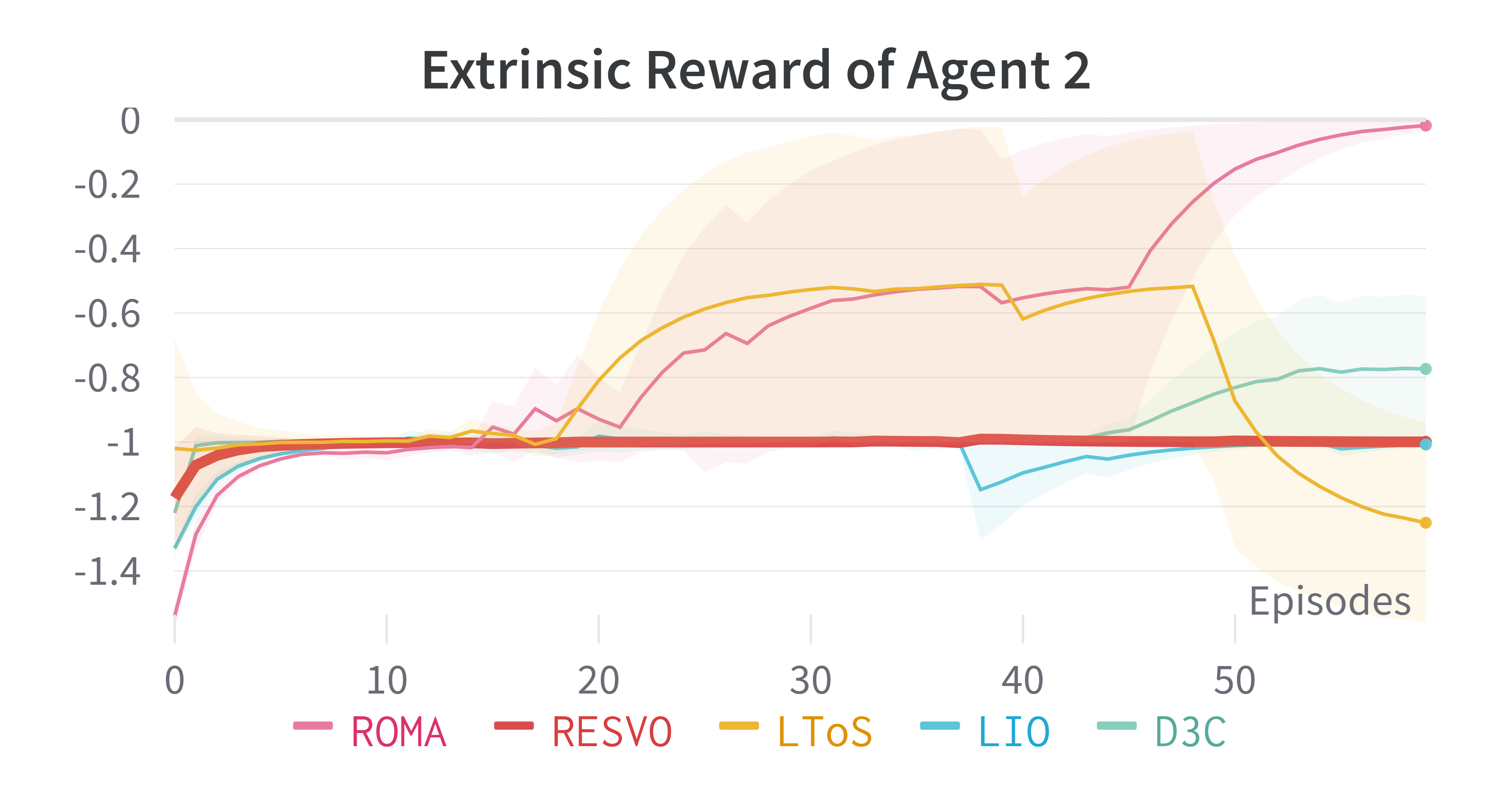}
        \caption{}
    \end{subfigure}
    \begin{subfigure}[b]{0.45\textwidth}
        \includegraphics[width=\textwidth]{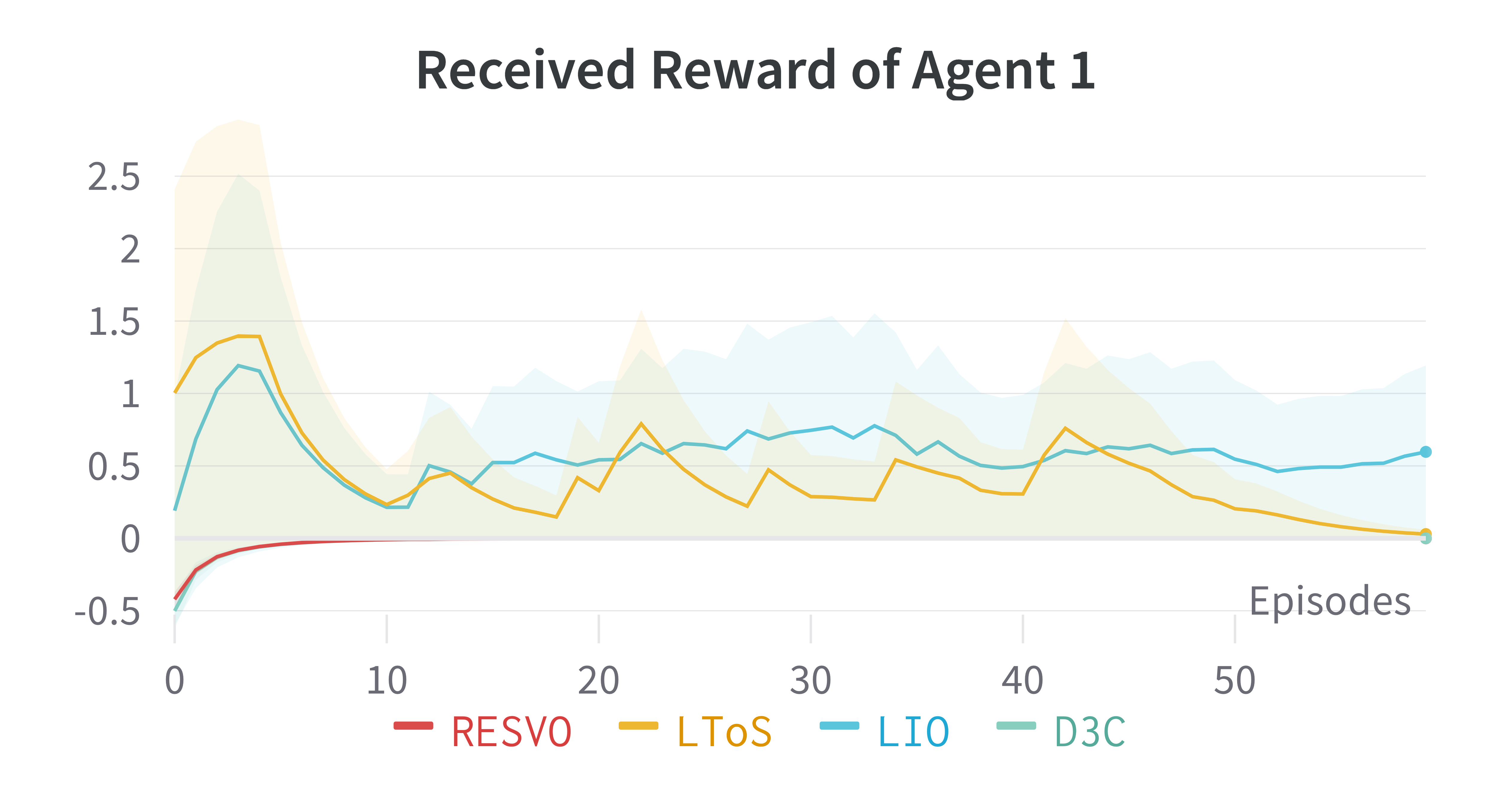}
        \caption{}
    \end{subfigure}
    \begin{subfigure}[b]{0.45\textwidth}
        \includegraphics[width=\textwidth]{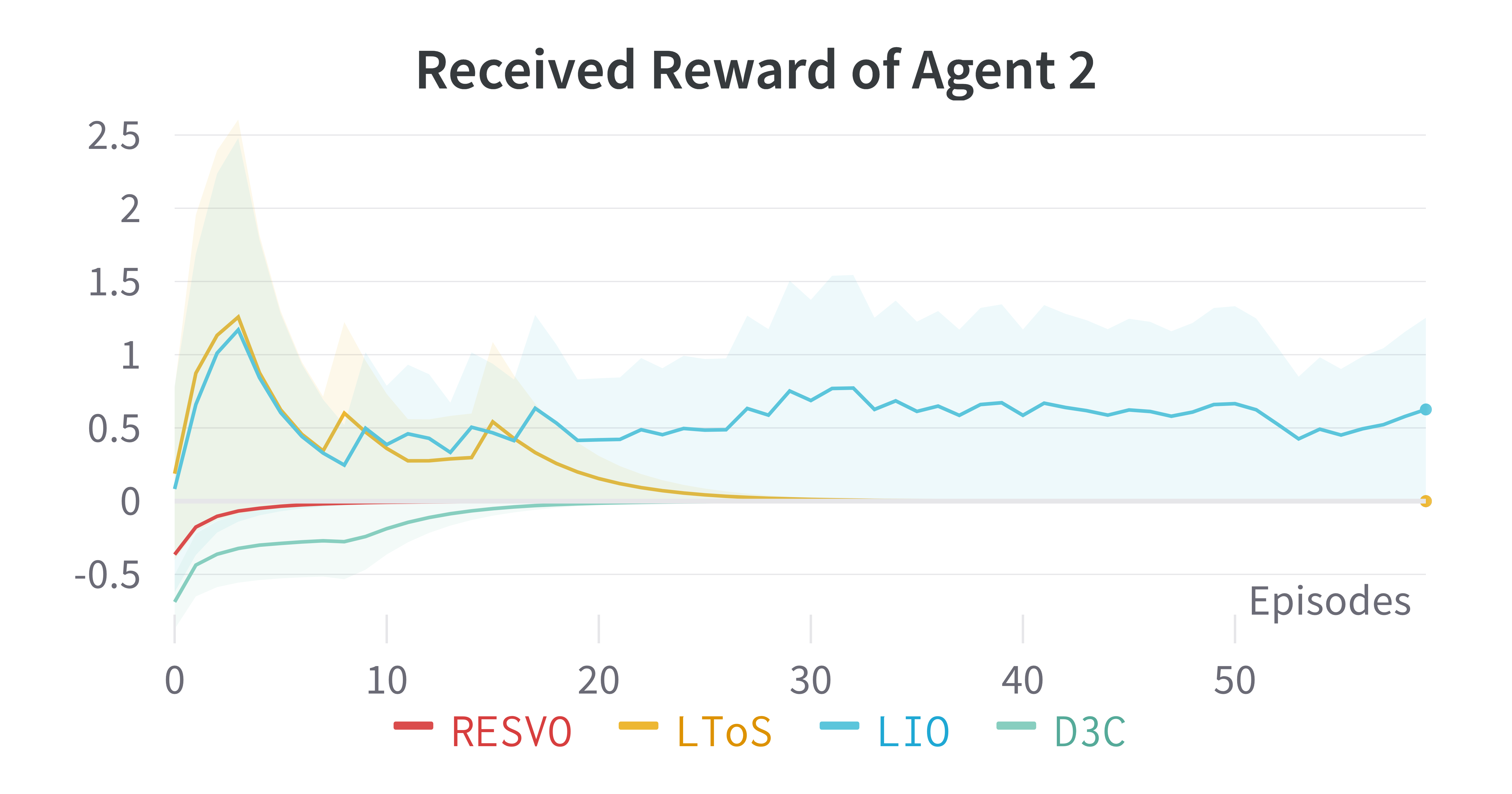}
        \caption{}
    \end{subfigure}
    \caption{Extrinsic (a-b) and received rewards (c-d) of different algorithms in Iterated Prisoner's Dilemma. Since ROMA is not designed based on SVO, rewards will not be transmitted between agents.}
    \label{fig:ipd-main}
\end{figure}
We first analyze the performance of each algorithm on the $2$-players task, IPD.
Figure~\ref{fig:ipd-main} shows the performance and emerged SVOs of RESVO and the performance of four baselines in the IPD environment.
In the IPD, a simplified version of the public good dilemma, we want both agents to be producers (that is, to cooperate with each other) to achieve the highest social welfare. 
Therefore, we set the rank constraint in RESVO to $1$. 
This shows that the optimal joint policy of IPD requires only one role: the producer or the cooperator.
It can be seen from the figure that the ROMA algorithm based only on the division of labor cannot stably converge to mutual cooperation. 
Both players choose the cooperate action and receive a $-1$ environmental reward.
In the early stages of training, ROMA can learn to cooperate.
Nevertheless, once a player chooses the ``defect'' action, ROMA's role learning will fix the player's character, so she will always choose the ``defect'' action and make mutual cooperation unsustainable.

The LToS and D3C algorithms based only on the SVO mechanism can converge to mutual cooperation quickly, but this equilibrium also cannot be maintained for a long time. 
However, the reasons for the inability of these two baselines to maintain mutual cooperation are not the same. 
The LToS algorithm is that because the agent does not form a stable SVO, agent $1$ no longer shares rewards with agent $2$ from the middle of training (Figure~\ref{fig:ipd-main}d), so that agent $2$ begins to explore other strategies.
Although the D3C algorithm forms a stable SVO, the agent's policy and SVO have no conditional dependence, causing the policy to diverge at the end of the training.
The LIO algorithm and the RESVO algorithm show the best results. 
However, LIO still has performance fluctuations after a training period and cannot maintain stable mutual cooperation.
This suggests that the sanction-based approach does not sustain stable cooperation in IPD.

Moreover, an interesting phenomenon can be found in Figure~\ref{fig:ipd-main}(c-d), where RESVO and other SVO-based baselines (i.e., LToS and D3C) learn two completely different SVOs to try to maintain cooperation.
As seen from the figure, after RESVO converges to mutual cooperation, the agent no longer needs to receive rewards, nor does it share rewards.
However, agents trained by other baselines have always received and shared rewards during the training process.
This illustrates that while RESVO maintains cooperation through individualism orientation, other baselines achieve the same result through martyrdom orientation, where the coefficients in the transformation matrices are all negative\footnote{In IPD, the external rewards of the agents are all negative, so only negative coefficients can deliver positive rewards.}.
Our analysis suggests that another reason for the inability of LToS and D3C to maintain stable cooperation may be due to this martyrdom orientation, an SVO that causes rewards to be passed between agents all the time. 
After policies converge to the equilibrium of mutual cooperation, these rewards will become noise and cause instability in the training process.
In contrast, RESVO shifts from a cooperative orientation to an individualism orientation after the policies reach equilibrium, thus making the equilibrium remain stable.
Although LIO is not an SVO-based method, the sanction mechanism also allows reward transmission between agents all the time, thus leading to the same instability of the training process.


\begin{figure}[htb!]
    \centering
    \begin{subfigure}[b]{0.45\textwidth}
        \includegraphics[width=\textwidth]{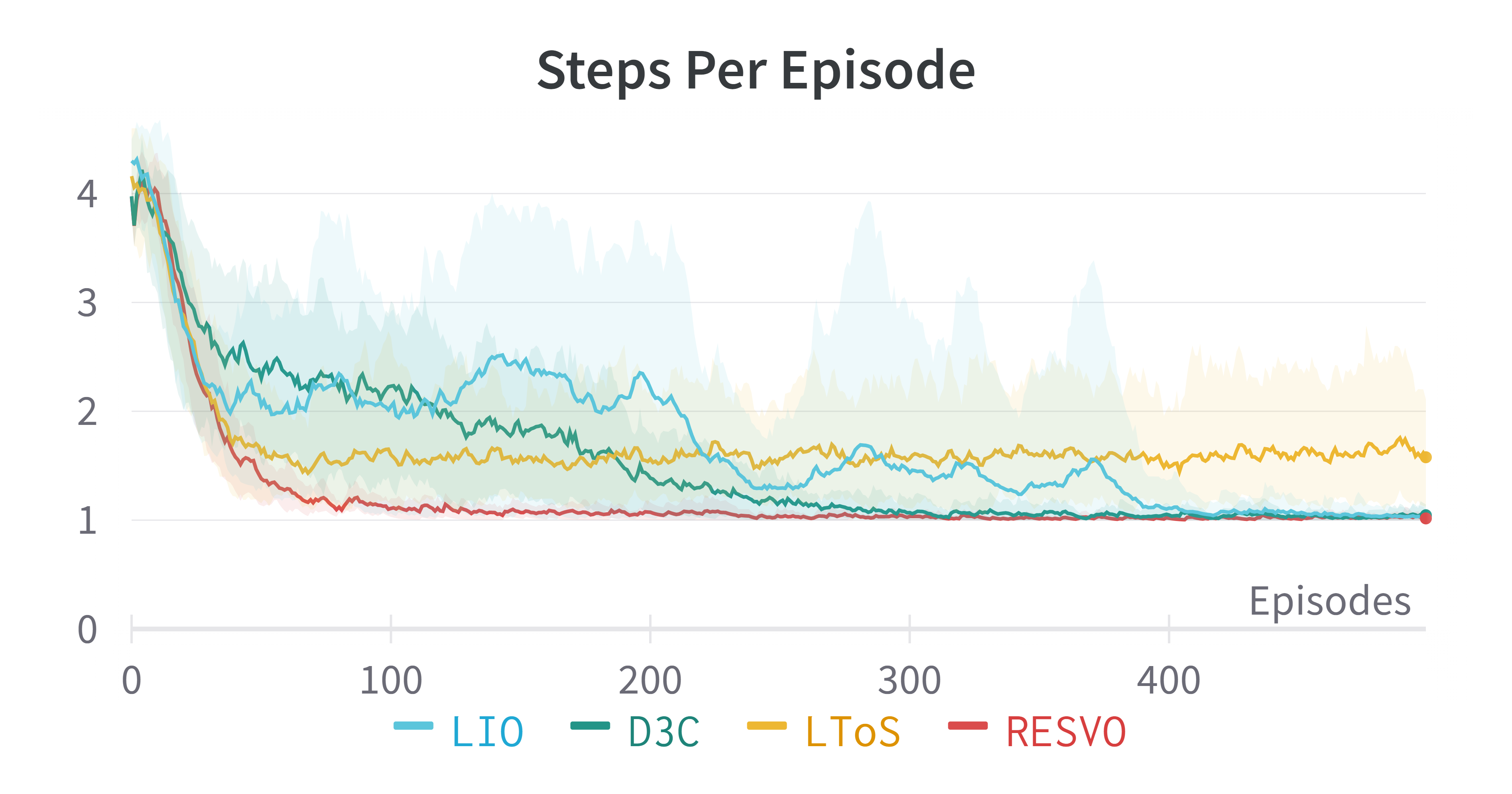}
        \caption{}
    \end{subfigure}
    \begin{subfigure}[b]{0.45\textwidth}
        \includegraphics[width=\textwidth]{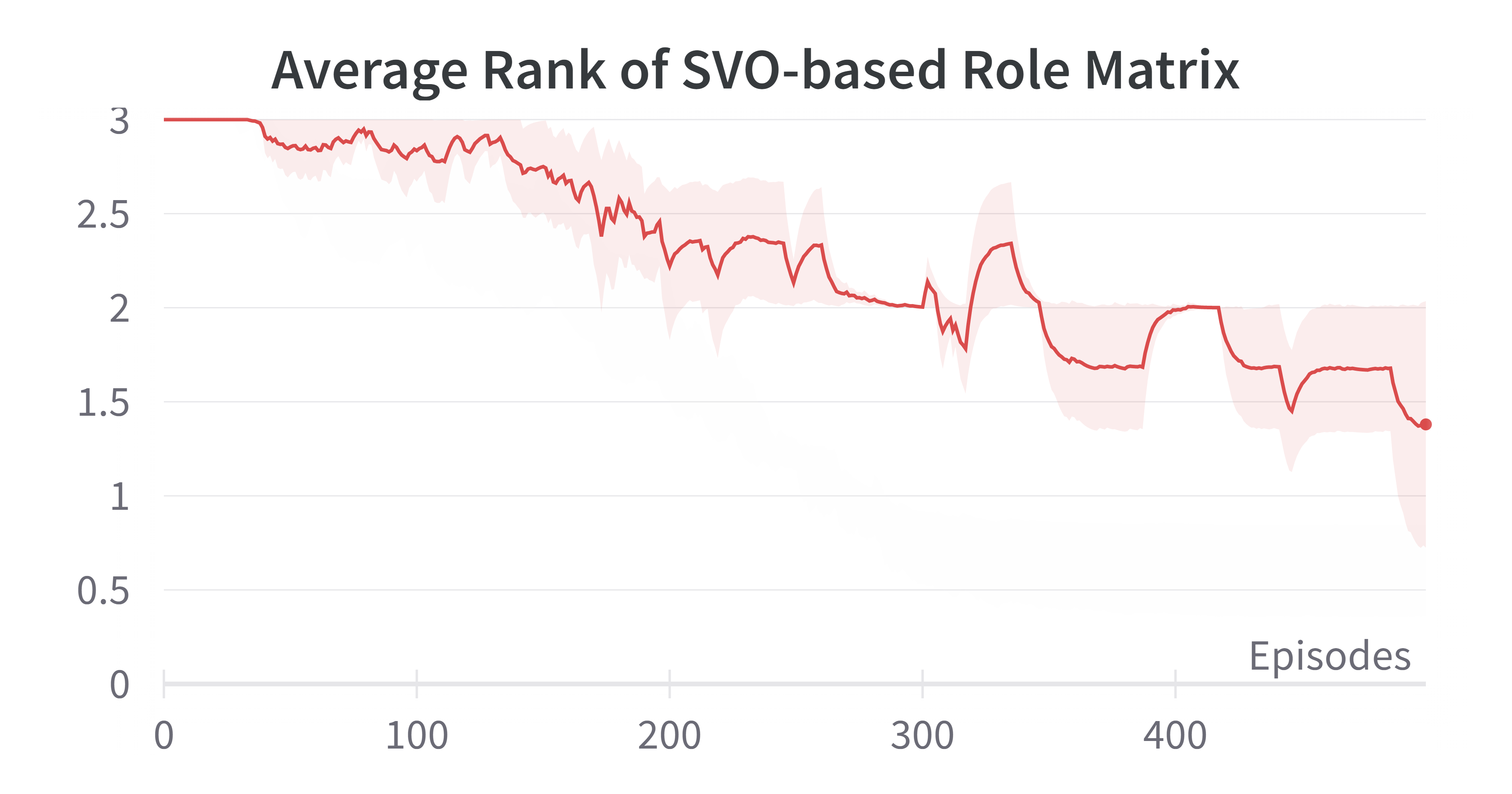}
        \caption{}
    \end{subfigure}
    \begin{subfigure}[b]{0.32\textwidth}
        \includegraphics[width=\textwidth]{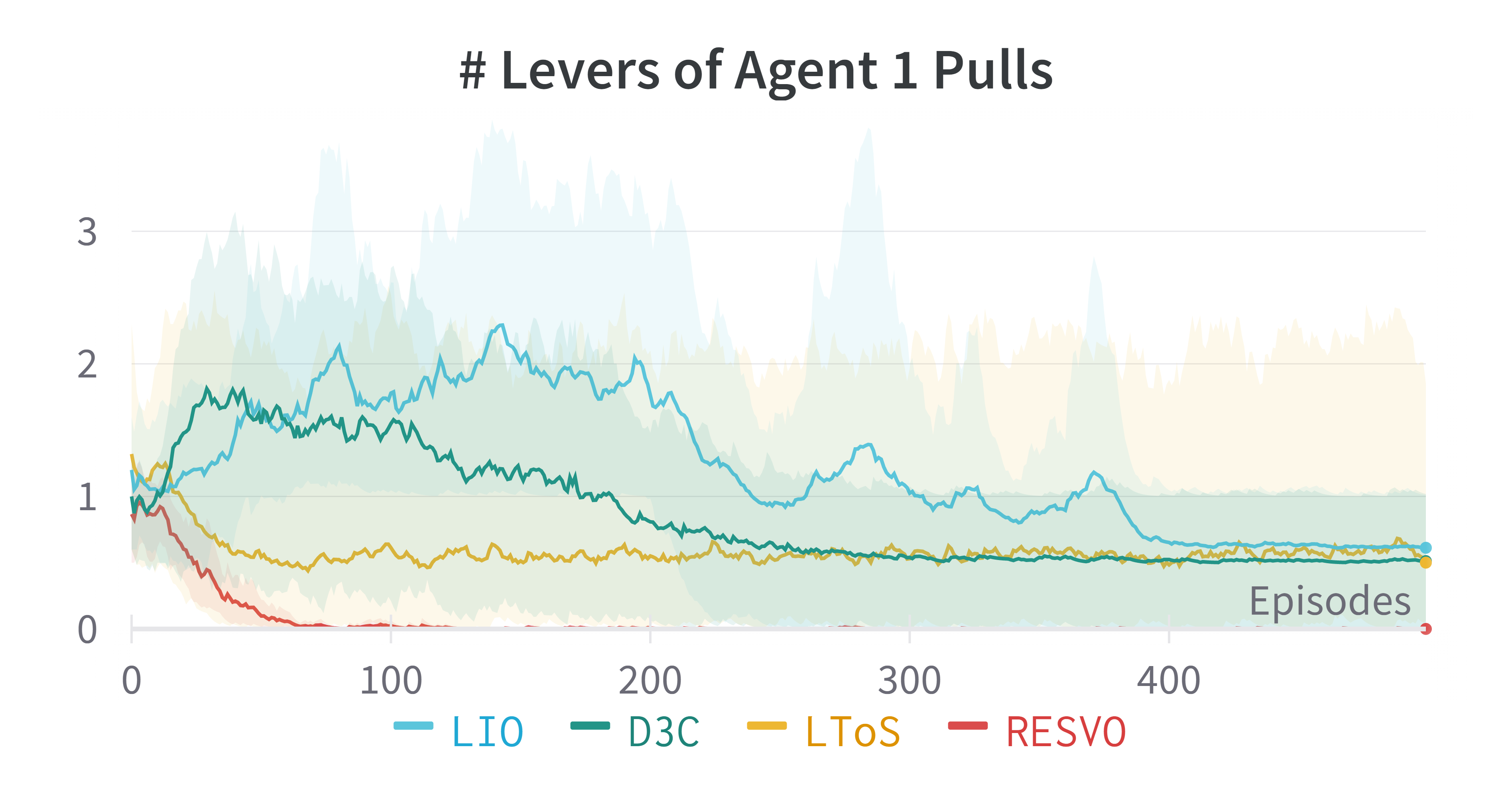}
        \caption{}
    \end{subfigure}
    \begin{subfigure}[b]{0.32\textwidth}
        \includegraphics[width=\textwidth]{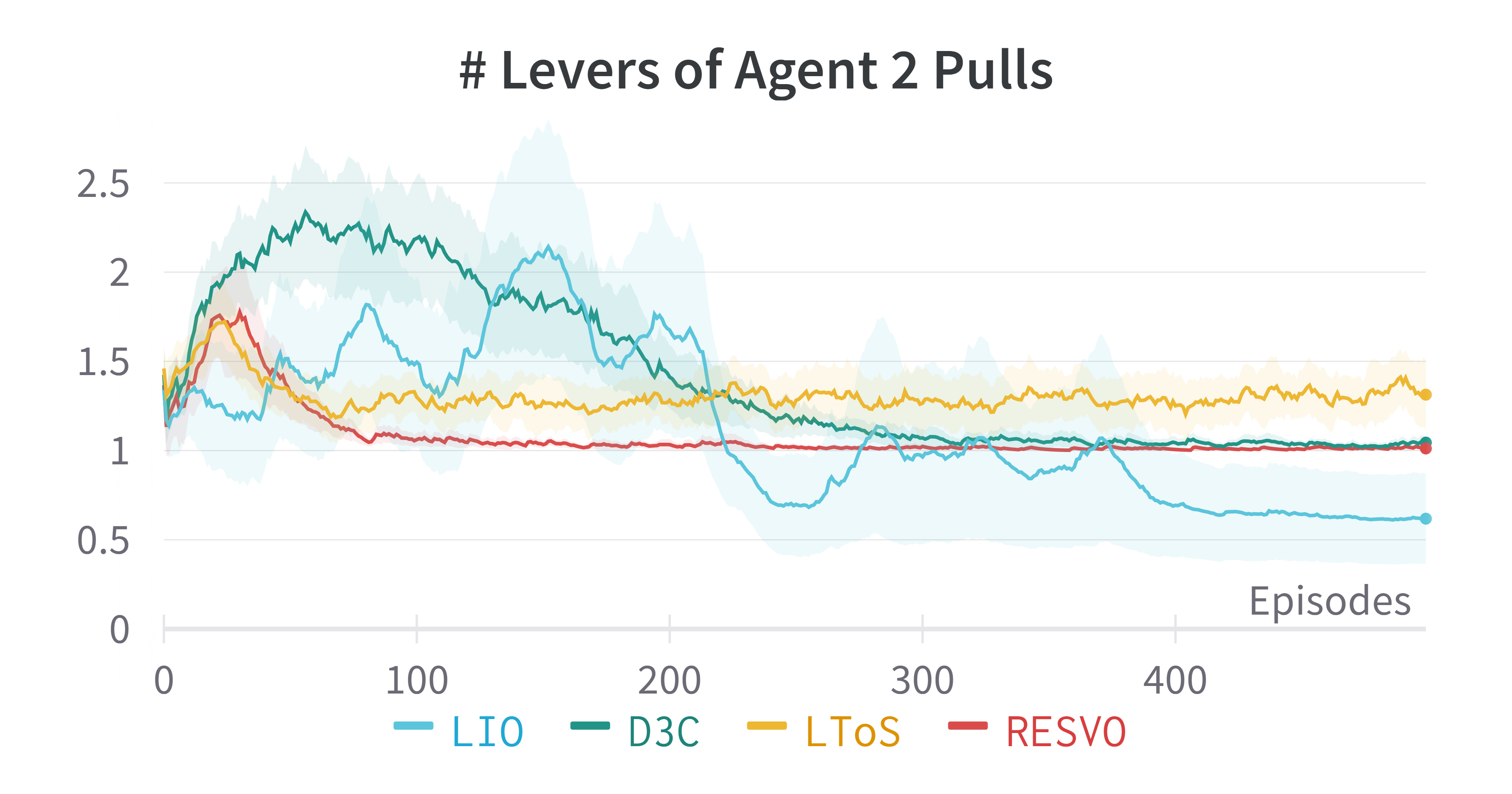}
        \caption{}
    \end{subfigure}
    \begin{subfigure}[b]{0.32\textwidth}
        \includegraphics[width=\textwidth]{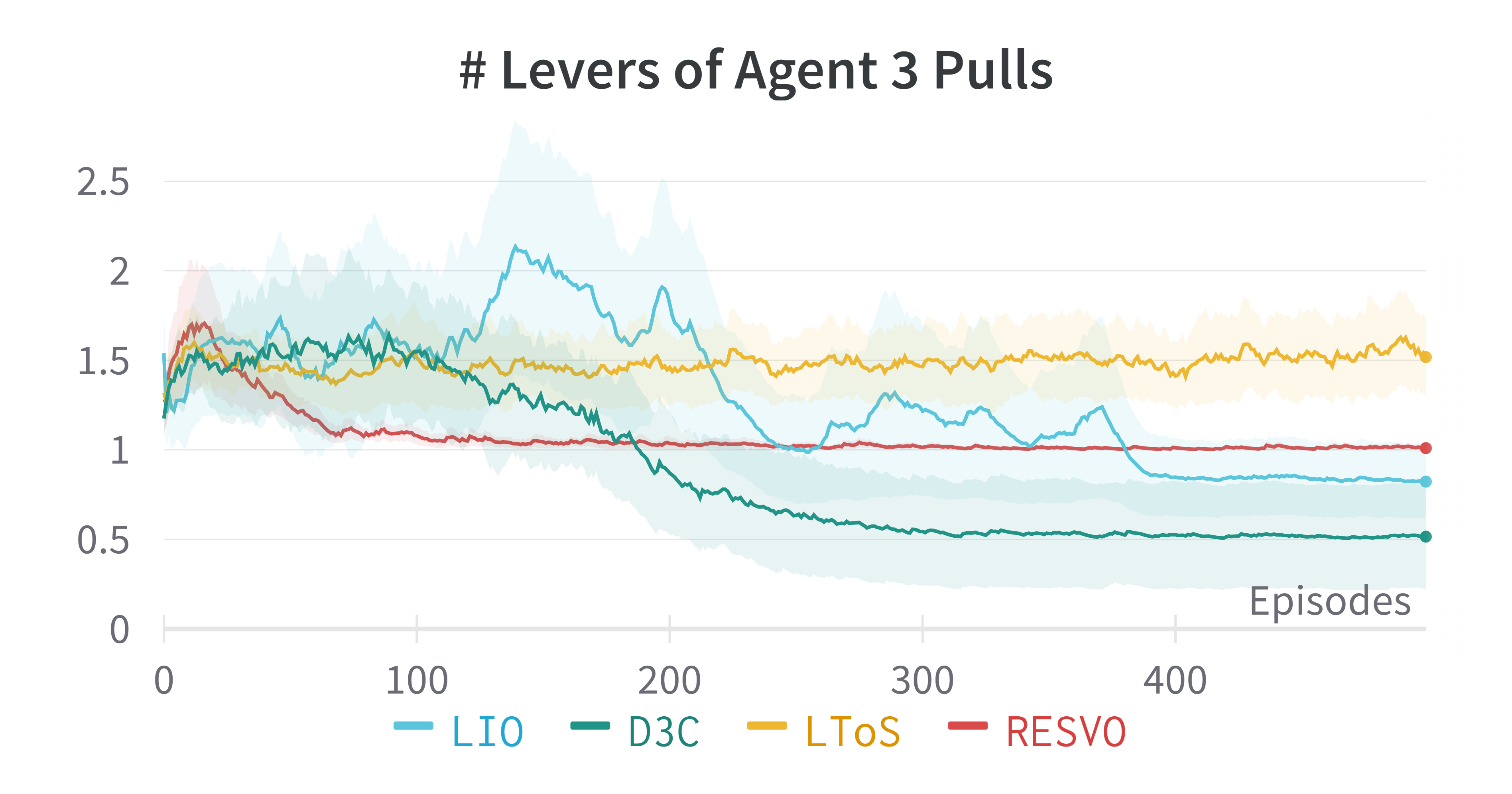}
        \caption{}
    \end{subfigure}
    \caption{(a) Steps per episode. (b) The average rank of the SVO-based role matrix. (c-e) The number of levers each agent pulls of different algorithms in $3$-Player Escape Room.}
    \label{fig:er-main}
\end{figure}

Next, we expand our study scenario into a more complex $3$-player Escape Room.
Figures~\ref{fig:er-main} shows the performance of RESVO and baselines in the $3$-Player Escape Room with $M=2$. 
This means that two agents must pull the lever for the agent to open the door.
Figures~\ref{fig:er-main}(b) shows the satisfaction of the rank constraints of the RESVO algorithm during training. 
In the $3$-player Escape Room, we set the rank constraint of RESVO to $2$.
Specifically, we hope $2$ roles emerge from the three agents through SVO, the lever-puller, and the door-opener.
As can be seen from the figure, as the training progresses, the rank constraint of the RESVO algorithm can be well satisfied.

A simple analysis of this task shows that the optimal policy only requires each agent to perform one-step action. 
This is because it only takes one timestep for any agent to move from the ``initial'' to the ``lever'' or ``gate''.
At the same time, each move will bring a reward of $-1$, and a more than $1$ move will be a social welfare decline.
The Figure~\ref{fig:er-main}(a) records the steps required by different algorithms to complete the task. 
Since ROMA performs poorly on the most straightforward IPD task, we no longer show the performance of ROMA on the more complex Escape Room and Cleanup tasks.
It can be seen that RESVO can converge to the optimal policy fastest.
D3C and LIO take about $4$ times as many samples to converge to equilibrium compared to RESVO. 
LToS falls into a locally optimal solution early and cannot escape from it.

Figure~\ref{fig:er-main}(c-e) shows the division of labor of different agents in the $3$-Player Escape Room with different algorithms.
It can be seen from these $3$ figures that whether the division of labor is formed and whether the division of labor is stable dramatically affects the performance and convergence speed of the algorithm.
RESVO has converged to a stable division of labor: agent $1$ opens the door, and agents $2$ and $3$ pull the lever.
Therefore, RESVO can converge to equilibrium the fastest and maintain it compared with baselines.

For the D3C algorithm, agent $2$ is stably assigned the role of a lever-puller, but there is no stable division of labor between agent $1$ and agent $3$. 
The two oscillate between the roles of lever-puller and door-opener.
None of the three agents have their own fixed roles for the LIO algorithm. 
Moreover, comparing Figure~\ref{fig:er-main}(a) and Figure~\ref{fig:er-main}(c-e), it can be seen that even after the LIO algorithm converges to equilibrium, the three agents are still dynamically allocated between the two roles of lever-puller and door-opener.
The instability of the division of labor between D3C and LIO also affects their convergence speed, and LIO converges more slowly than D3C.
For the LToS, the three agents have been unable to form a correct division of labor, and the average number of lever pulls greater than twice, indicating that the average number of the lever-puller is less than $2$.
This will make an agent need to change from the ``door-opener'' to the ``lever-puller'' to complete the task successfully. 
The role change process mentioned above makes the timestep of the LToS algorithm to complete the task more significant than $1$ on average.

\begin{figure}[htb!]
    \centering
    \begin{subfigure}[b]{0.45\textwidth}
        \includegraphics[width=\textwidth]{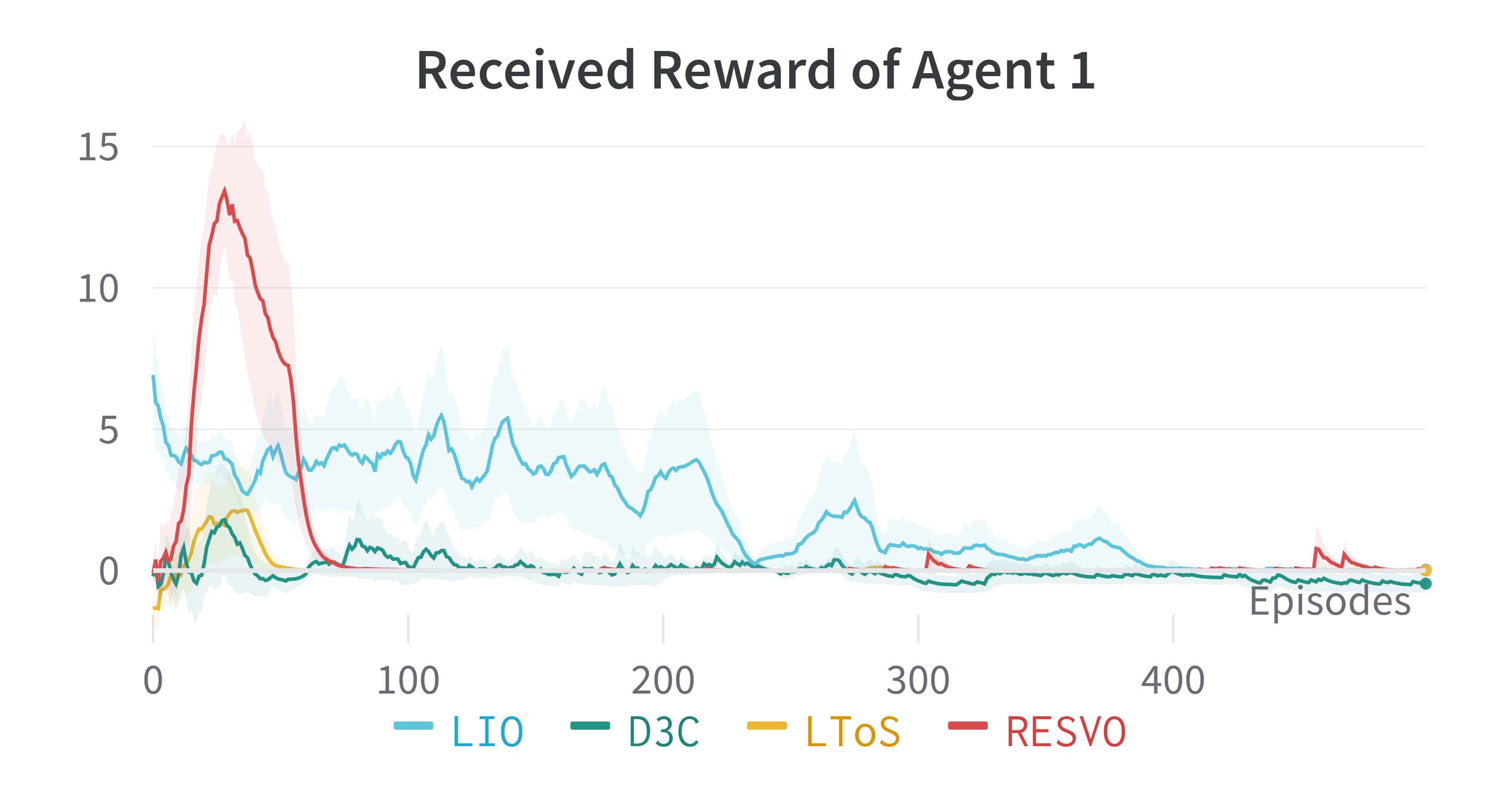}
    \end{subfigure}
    \begin{subfigure}[b]{0.45\textwidth}
        \includegraphics[width=\textwidth]{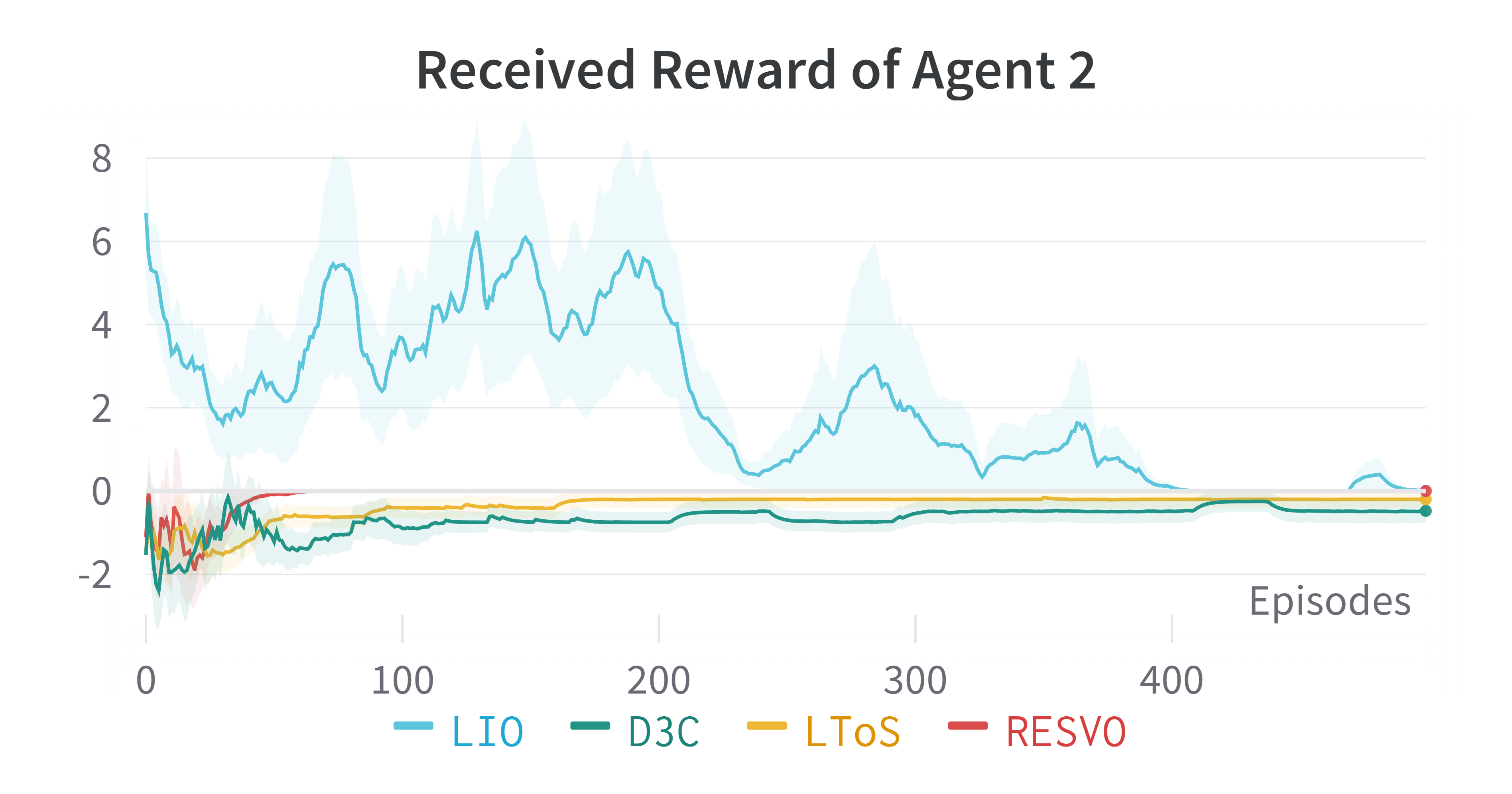}
    \end{subfigure}
    \begin{subfigure}[b]{0.45\textwidth}
        \includegraphics[width=\textwidth]{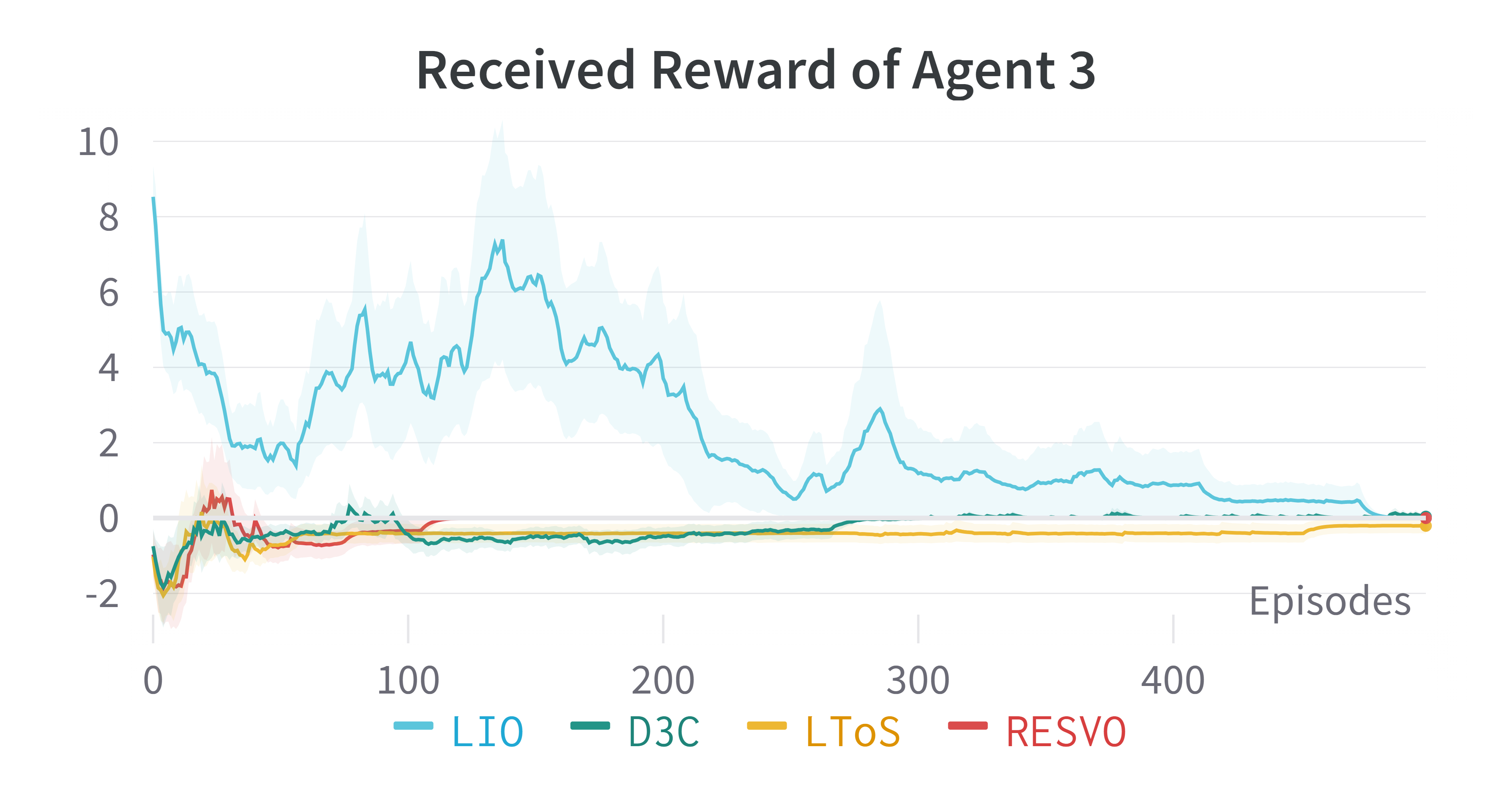}
    \end{subfigure}
    \begin{subfigure}[b]{0.45\textwidth}
        \includegraphics[width=\textwidth]{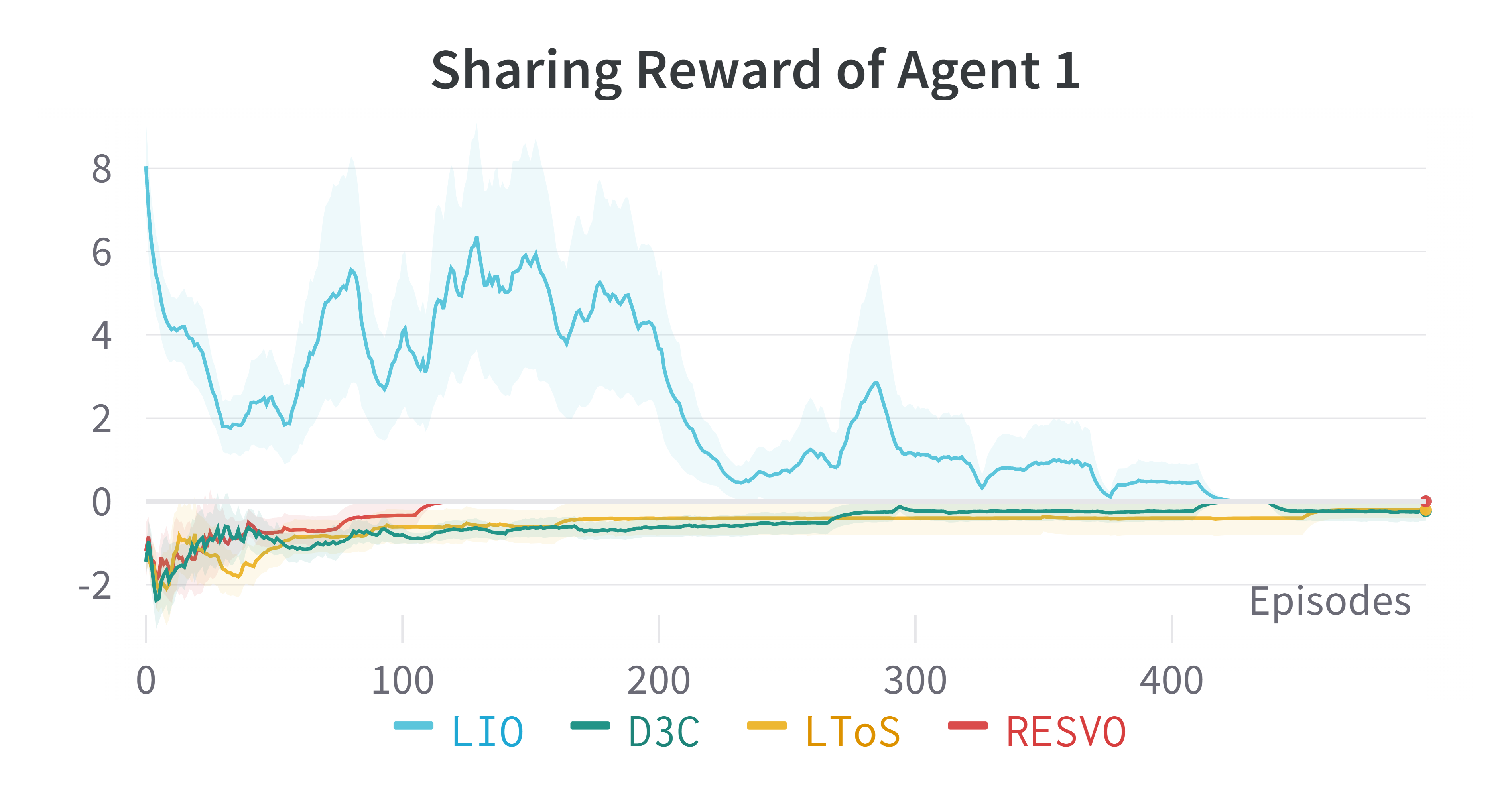}
    \end{subfigure}
    \begin{subfigure}[b]{0.45\textwidth}
        \includegraphics[width=\textwidth]{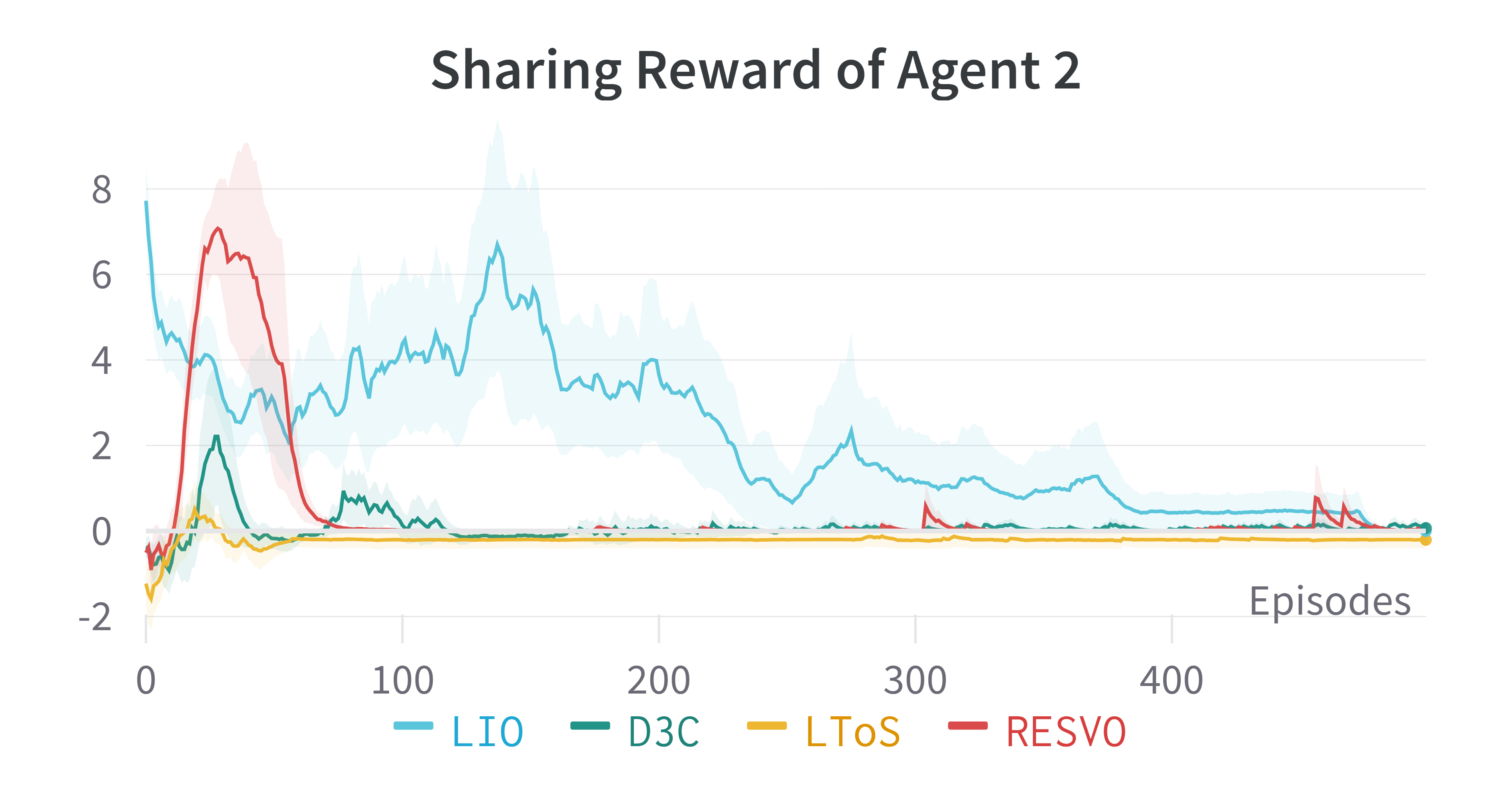}
    \end{subfigure}
    \begin{subfigure}[b]{0.45\textwidth}
        \includegraphics[width=\textwidth]{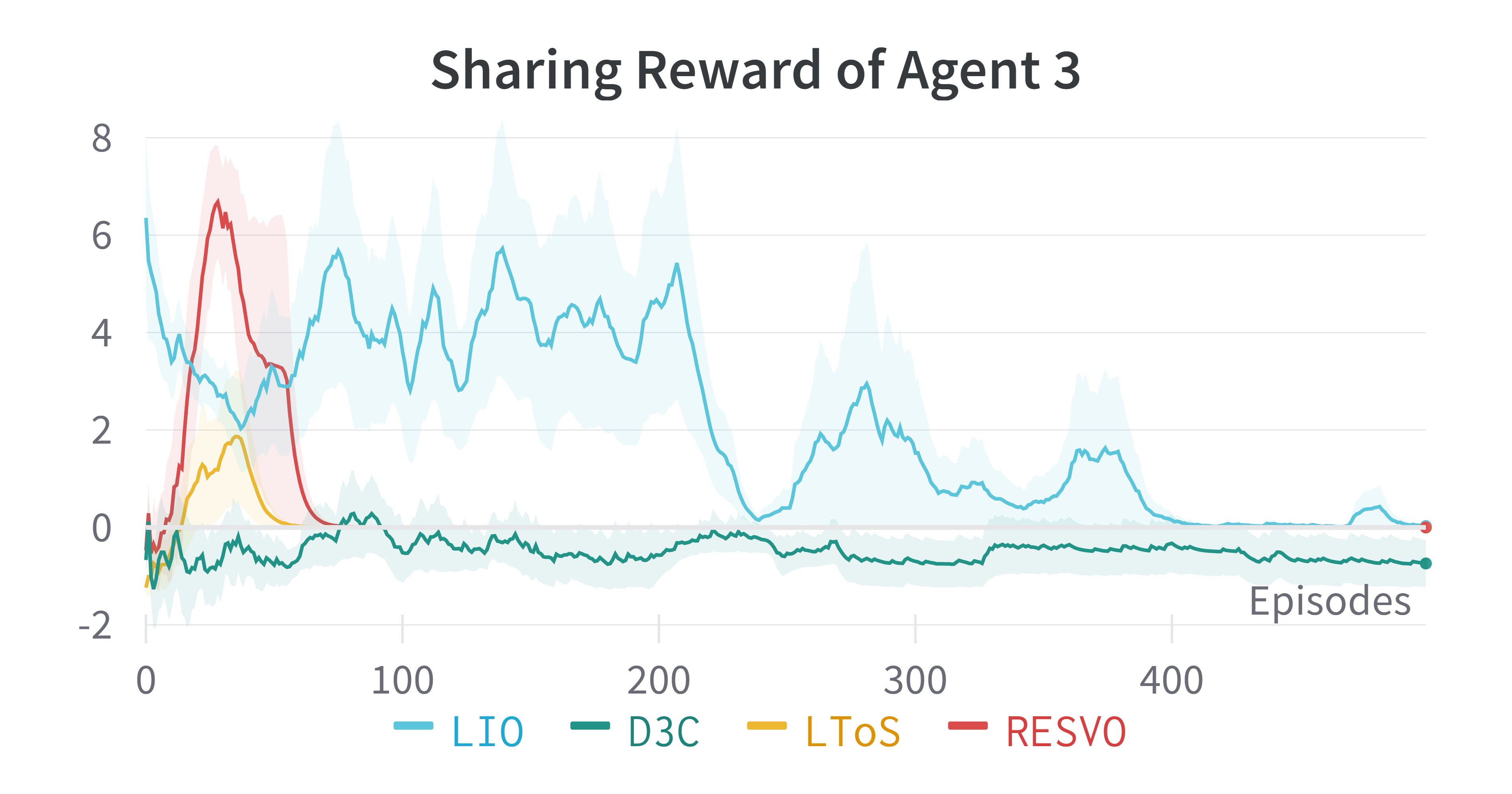}
    \end{subfigure}
    \caption{Shared and received rewards of different algorithms $3$-Player Escape Room.}
    \label{fig:er-append1}
\end{figure}

In order to explore the reason why RESVO can form a stable division of labor, we show in Figure~\ref{fig:er-append1} how different agents share and receive rewards in different algorithms.
That is, the emerged SVOs of three SVO-based methods (including RESVO) and LIO.
It is worth noting that in more complex Escape Room and Cleanup environments, in order to make SVO have powerful representation capabilities, RESVO uses the transition matrix coefficients of multiple consecutive time steps as SVO. 
This differs from the setting in our IPD experiments, making it impossible to classify the agent as a certain kind of SVO. 
Therefore, we indirectly analyze the agent's SVO from the pattern of sharing rewards in the following.

A counter-intuitive phenomenon can be seen in the figure. 
Those who pull the lever have no profit, and those who open the door can obtain more significant benefits. 
Therefore, intuitively, to maintain a stable division of labor, the door opener should share her reward with the lever-puller so both parties can get rewards. 
At the same time, since each agent is self-motivated, this can form a stable role division. 
However, the three SVO-based algorithms, RESVO, LToS, and D3C, share the rewards in turn. 
Those who pull the levers give rewards to those who open the door. 
The difference between these three algorithms is the size of the reward shared.
A plausible explanation is that the SVO-based algorithm learned a more "aggressive" approach. 
The agent that discovers the action of pulling the lever gives the agent that discovers the action of opening the door a positive reward so that the role of the door opener is fixed. 
Then determine the role of the lever-puller. 
This method can instead promote the rapid formation of the division of labor and regional stability for RESVO. 
The rank constraints in RESVO and the policy conditioned on SVO enable another role can also be fixed once a role is fixed. 
Nevertheless, other SVO-based methods do not have this advantage, so the algorithm converges slowly during the training process, and the division of labor cannot be maintained stably.
It can be seen from the figure that at the beginning of training, RESVO agents share a large-scale reward value, thus promoting the algorithm to converge quickly.
After the algorithm converges, similar to the IPD task, the agent almost no longer shares and receives rewards, thus maintaining the stability of the policies and division of labor.

The sanction-based LIO method exhibits a similar pattern to that in the IPD environment. 
During the entire training process, even after the policies converges to equilibrium, the agents continue to transmit large rewards, maintaining the stability of the division of labor through a high ``cost'' way. 
One defect of this method is that the module that generates the sanction will always receive a large gradient during the training process, which makes the optimization process unstable. 
The algorithm converges slowly and cannot maintain a stable division of labor for a long time.

\subsection{Role Emergence in the Cleanup}

\begin{figure}[htb!]
    \centering
    \begin{subfigure}[b]{0.45\textwidth}
        \includegraphics[width=\textwidth]{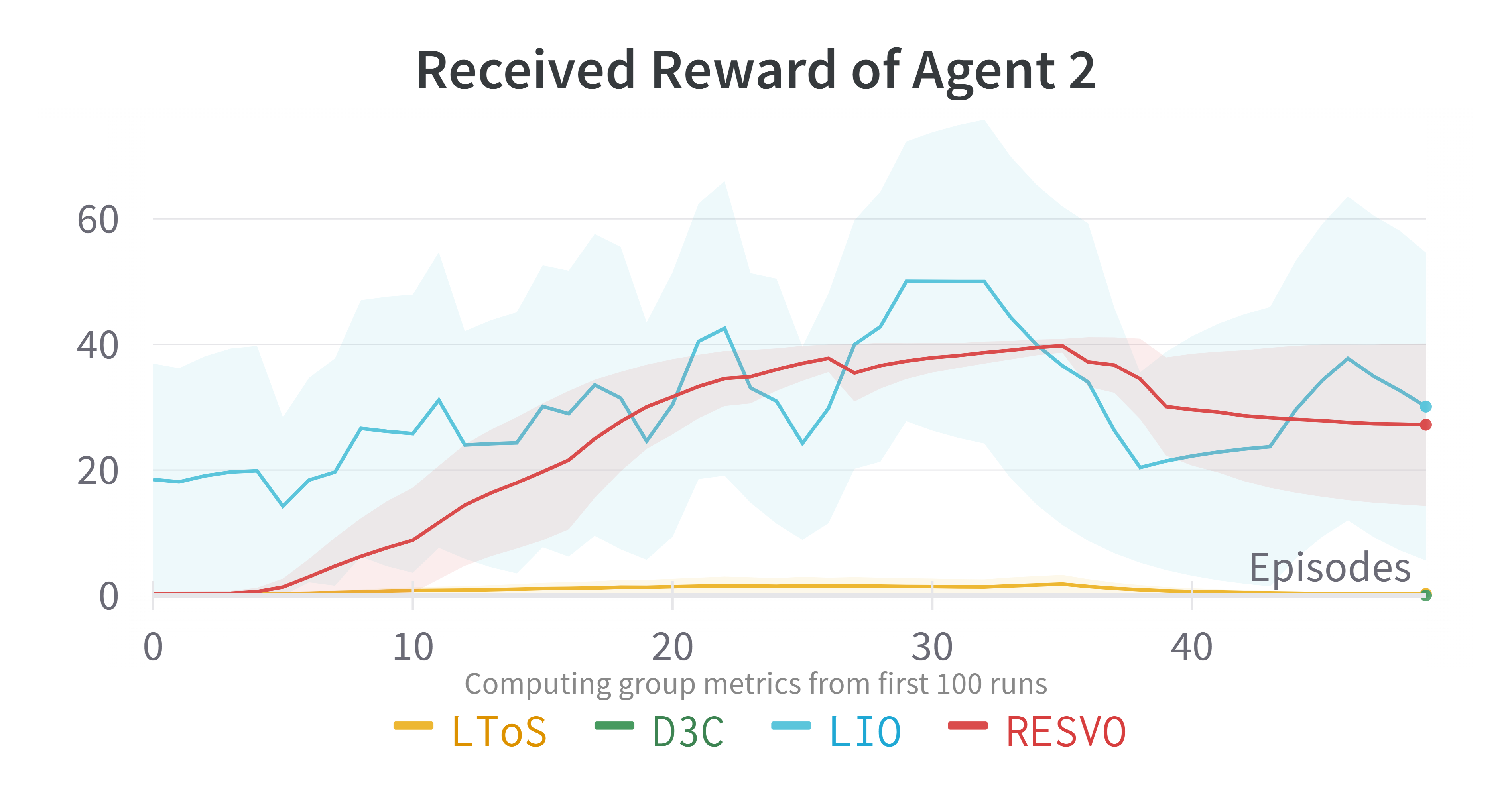}
        \caption{}
    \end{subfigure}
    \begin{subfigure}[b]{0.45\textwidth}
        \includegraphics[width=\textwidth]{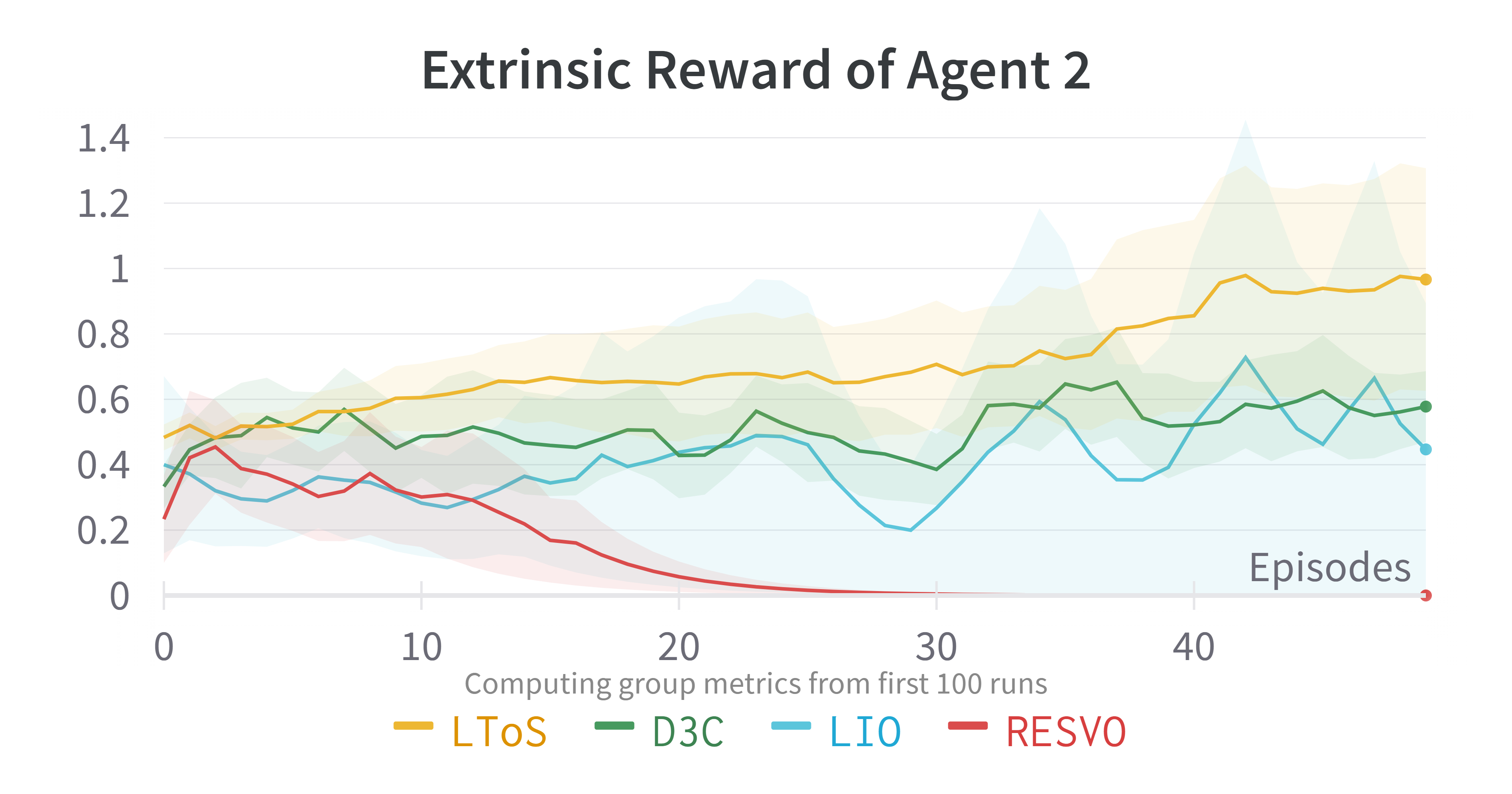}
        \caption{}
    \end{subfigure}
    \begin{subfigure}[b]{0.45\textwidth}
        \includegraphics[width=\textwidth]{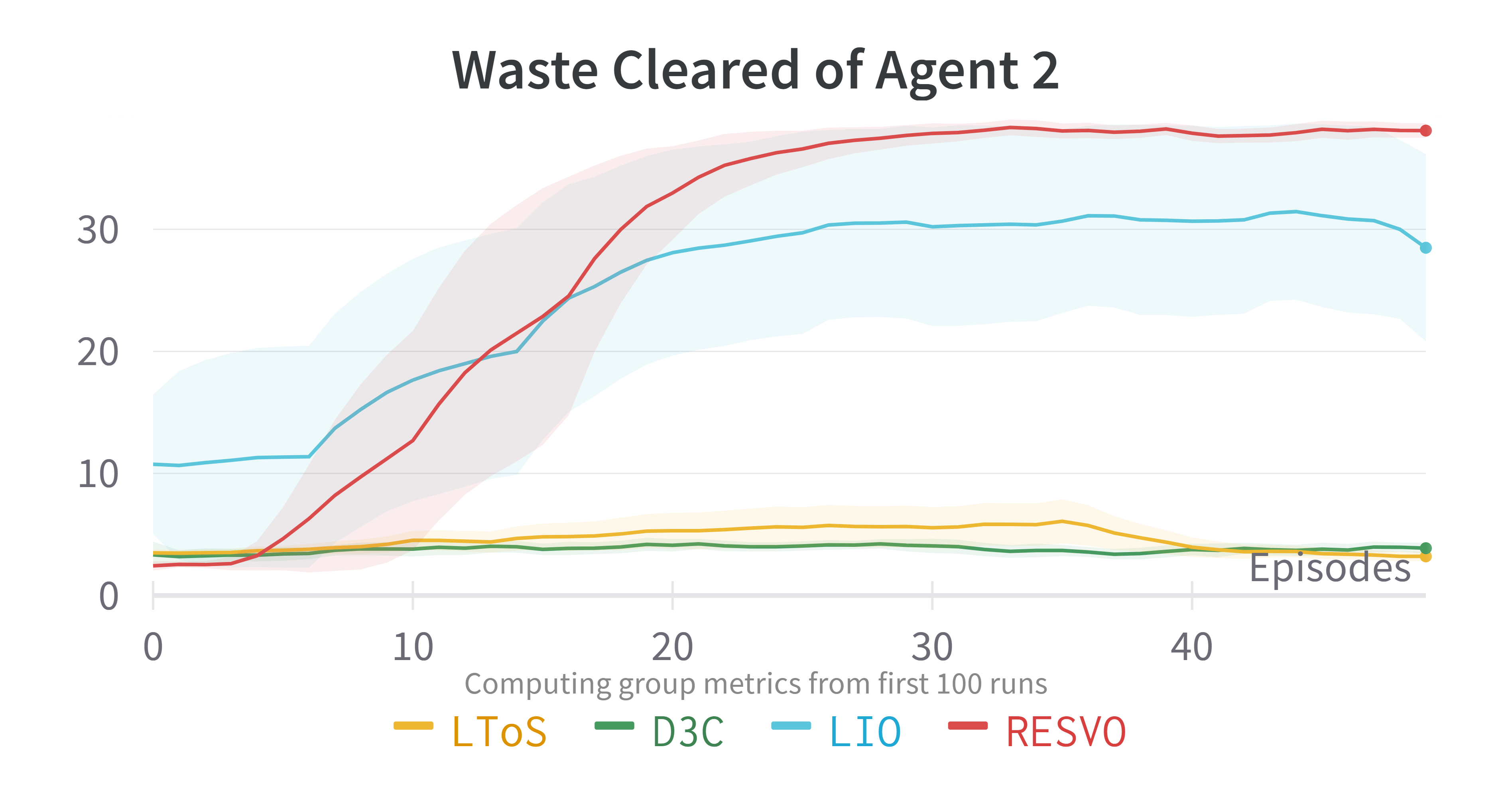}
        \caption{}
    \end{subfigure}
    \begin{subfigure}[b]{0.45\textwidth}
        \includegraphics[width=\textwidth]{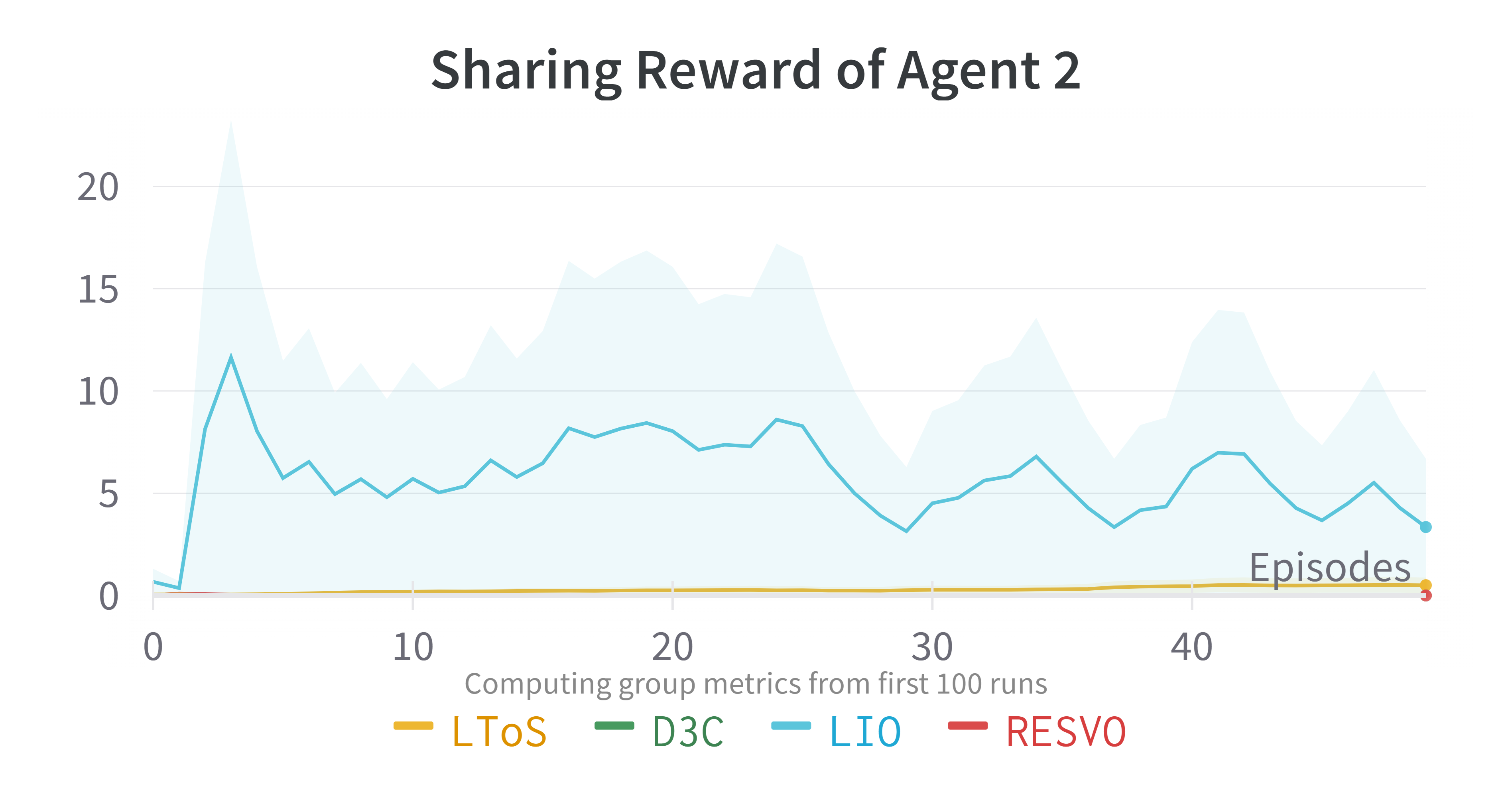}
        \caption{}
    \end{subfigure}
    \begin{subfigure}[b]{0.45\textwidth}
        \includegraphics[width=\textwidth]{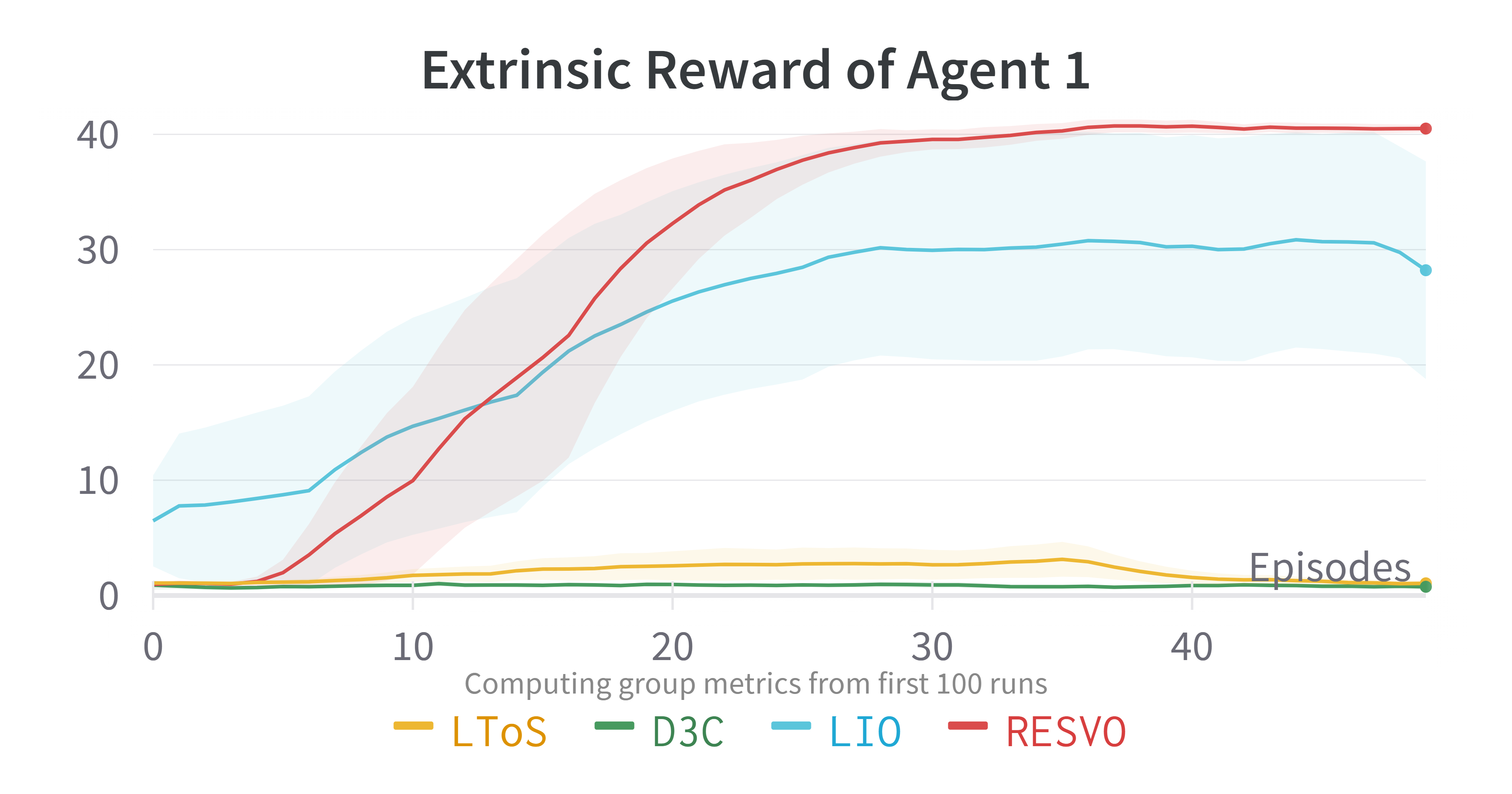}
        \caption{}
    \end{subfigure}
    \begin{subfigure}[b]{0.45\textwidth}
        \includegraphics[width=\textwidth]{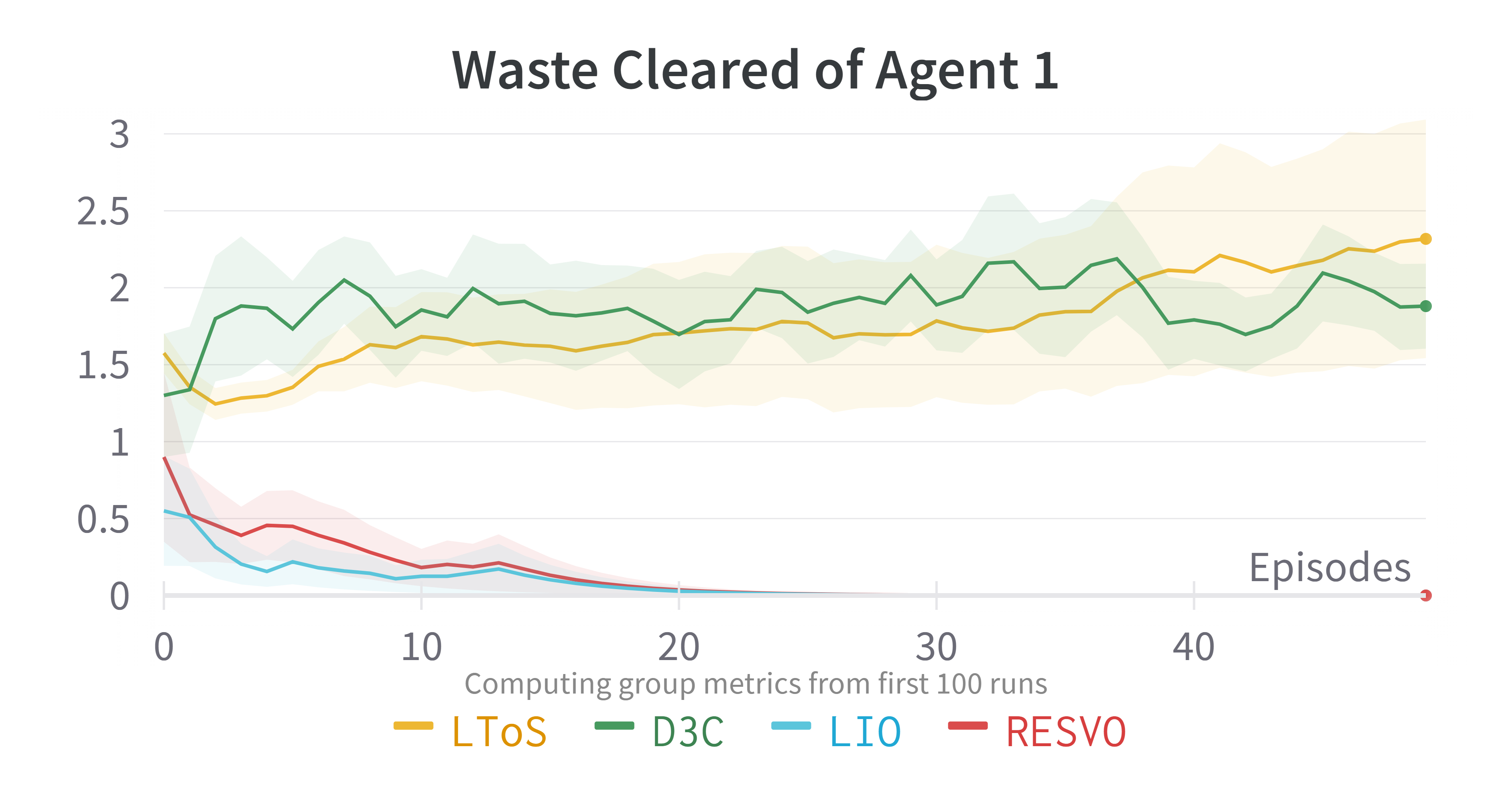}
        \caption{}
    \end{subfigure}
    \caption{Shared and received rewards, extrinsic rewards, and waste cleared of different algorithms in $10\times 10$ map size, $2$-Player Cleanup.}
    \label{fig:clean-main}
\end{figure}

We finally test our method in Cleanup and start from a map size of $10$ by $10$ and the number of agents $2$.
Although there are only $2$ agents, it is still much more complicated than $N$-Players Escape Room because Cleanup has far more timesteps per episode than the former.
In this task, similar to the $3$-Player Escape Room, there is also an apparent division of labor between the two agents under the optimal cooperative policy: one agent needs to clean up wastes (producer), and the other agent needs to collect apples (consumer).
We can judge the division of labor or roles of the two agents from the amount of waste they clean up.

As seen from Figure~\ref{fig:clean-main}, different algorithms also show remarkable differences in the $2$-agents Cleanup task.
On the one hand, LToS and D3C have not learned a good division of labor, and both agents hardly clean up waste (Figure~\ref{fig:clean-main}(c) and (f)), so no apples grow. 
This makes the extrinsic reward for both agents small (Figure~\ref{fig:clean-main}(b) and (e)), and the joint policy suffers from a public good dilemma.
As can be seen from Figure~\ref{fig:clean-main}(a) and (d), the two agents hardly share rewards, and their SVOs are almost the same, which is why LToS and D3C cannot learn a good division of labor or roles.
On the other hand, RESVO and LIO have successfully formed a reasonable division of labor.
Agent $1$ is responsible for collecting apples (consumer), and agent $2$ is responsible for cleaning up wastes (producer, see Figure~\ref{fig:clean-main}(c) and (f)).
At the same time, to maintain a stable division of labor or roles, agent $1$ as a consumer will continue to share rewards with agent $2$ as a producer (Figure~\ref{fig:clean-main}(a) and (d)), to achieve a larger average extrinsic reward, or the social welfare.
That is, agent $2$ maintains a stable division of labor by showing a cooperative orientation to agent $1$.

However, similar to the previous tasks, the sanction-based LIO method is very different from RESVO in the way that roles emerge. 
The two algorithms exhibit different robustness in maintaining the division of labor.
As can be seen from Figure~\ref{fig:clean-main}(d), on the one hand, for LIO, agent $2$, which is the producer, also needs to share the reward with agent $1$. 
RESVO, on the other hand, uses a more "energy-efficient" approach to promoting role emergence: 
Since producers receive no rewards, there is no need to share rewards with consumers who can receive large extrinsic rewards. 
This sparsity of reward sharing between agents also enables RESVO to maintain a more stable division of labor while receiving greater social welfare.
It can be seen from Figure~\ref{fig:clean-main}(c) that the agent $2$ trained by the LIO algorithm does not always clean up the waste, and its behavior shows a significant variance. 
This also makes the external reward of agent $1$ unable to maintain a high level all the time and also has a significant variance (Figure~\ref{fig:clean-main}(e)).

\begin{figure}[htb!]
    \centering
    \begin{subfigure}[b]{0.64\textwidth}
        \includegraphics[width=\textwidth]{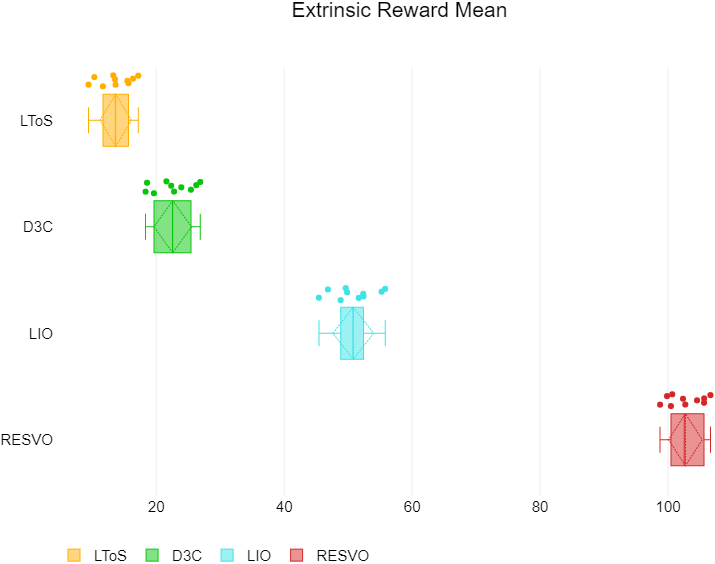}
        \caption{}
    \end{subfigure}\\
    \begin{subfigure}[b]{0.45\textwidth}
        \includegraphics[width=\textwidth]{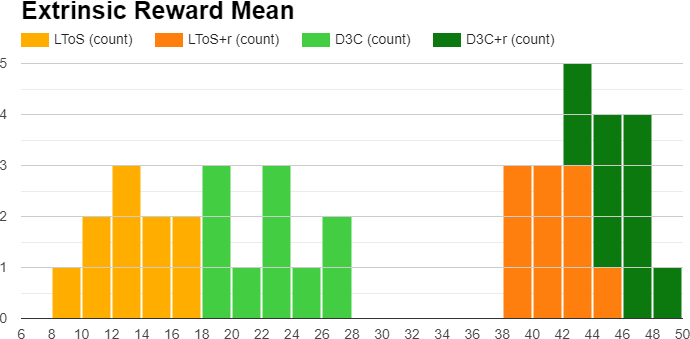}
        \caption{}
    \end{subfigure}\hfill
    \begin{subfigure}[b]{0.45\textwidth}
        \includegraphics[width=\textwidth]{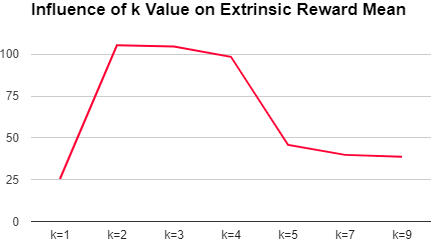}
        \caption{}
    \end{subfigure}
    \caption{The (a) boxplot, (b) histogram of performance, and the (c) influence of $k$ value for extrinsic rewards under $10$ runs of different algorithms in $48\times 18$ map size, $10$-Player Cleanup.}
    \label{fig:clean10-main}
\end{figure}

\subsection{Static versus Dynamic Division of Labor}

In the three tasks in the previous two sections, we find that the sanction-based LIO can also effectively learn a reasonable division of labor or roles. 
However, due to the difference between the sanction mechanism and SVO, LIO needs to transmit rewards densely between agents, which makes it impossible to maintain a stable division of labor. 
In other words, RESVO realizes a \textit{static} division of labor through the emergence of SVO, but LIO realizes a \textit{dynamic} division of labor. 
In the static division of labor, the role of each agent is fixed while completing the task; 
on the contrary, the agent's role will change in the dynamic division of labor.

In addition, from the experiments in the previous two sections, we find preliminary evidence that static division of labor can lead to better social welfare.
Nevertheless, the above tasks only contain $2-3$ agents, and the impact of dynamic division of labor cannot be fully reflected. 
For example, for the IPD or Cleanup task that only consists of $2$ agents, the dynamic division of labor is the role reversal of the two agents.
To this end, we compare the performance of the algorithms in a larger Cleanup environment with a map size of $48$ by $18$ and many agents of $10$.
There are only two roles in the Cleanup: the waste cleaner and the apple picker. 
When the number of agents exceeds $2$, multiple agents will have the same role. 
Cooperation among roles and agents with the same role is required to get rid of the public good dilemma and achieve greater social welfare. 
At this time, for the cooperation of the same role, the static role or division of labor has more advantages. 
Because dynamic roles involve role changes, that is, the composition of group members with the same role changes. 
This will make the cooperation of the same role unstable, affecting the algorithm's performance.
The larger the number of agents, the more severe the problem of unstable cooperation will become, which will also cause more significant performance degradation.
To test the above hypothesis, we conducted multiple random experiments in $10$-player Cleanup to ensure the reliability of the results.
As seen from Figure~\ref{fig:clean10-main}, RESVO shows a clear performance advantage (about two times) compared to LIO in the more complex Cleanup environment.

\begin{figure}[htb!]
    \centering
    \begin{subfigure}[b]{0.3\textwidth}
        \includegraphics[width=\textwidth]{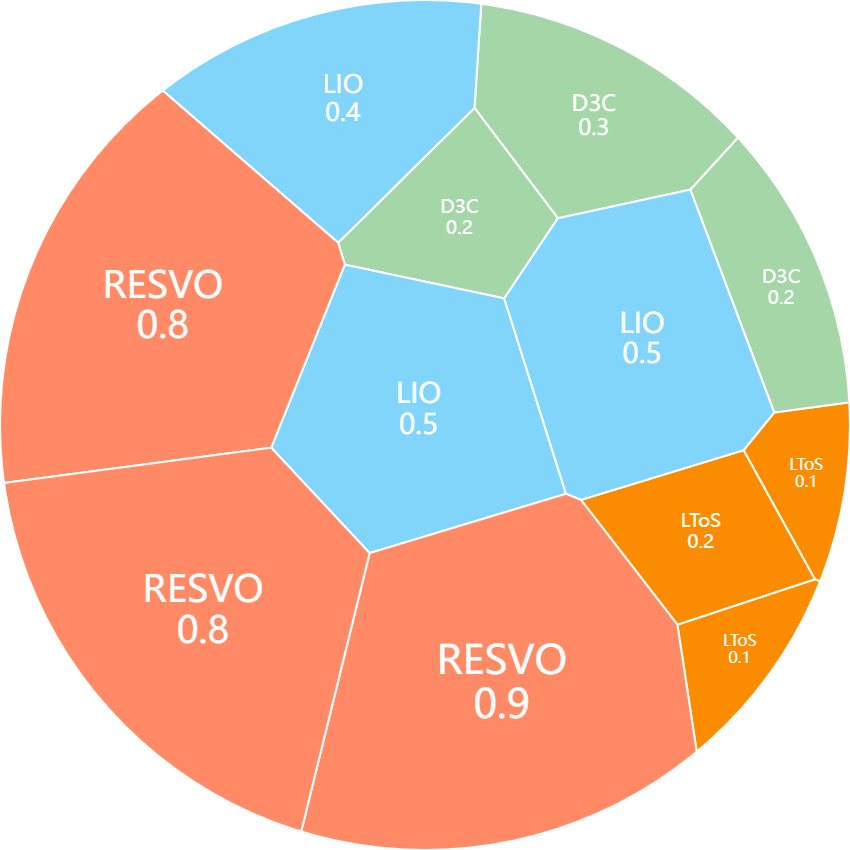}
    \end{subfigure}\hfill
    \begin{subfigure}[b]{0.3\textwidth}
        \includegraphics[width=\textwidth]{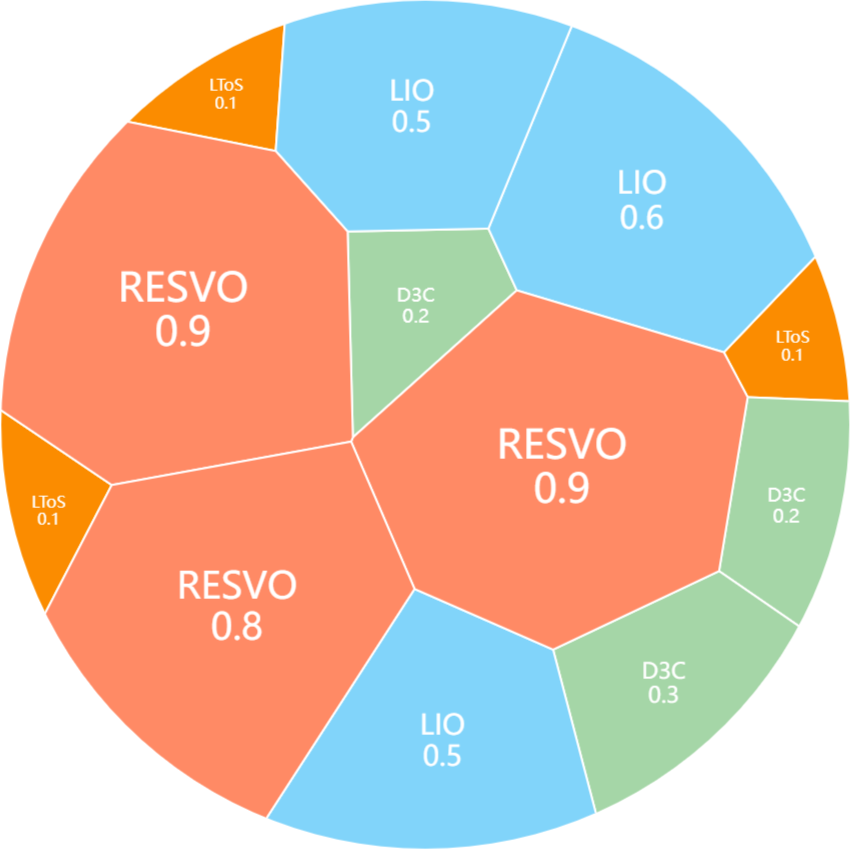}
    \end{subfigure}\hfill
    \begin{subfigure}[b]{0.3\textwidth}
        \includegraphics[width=\textwidth]{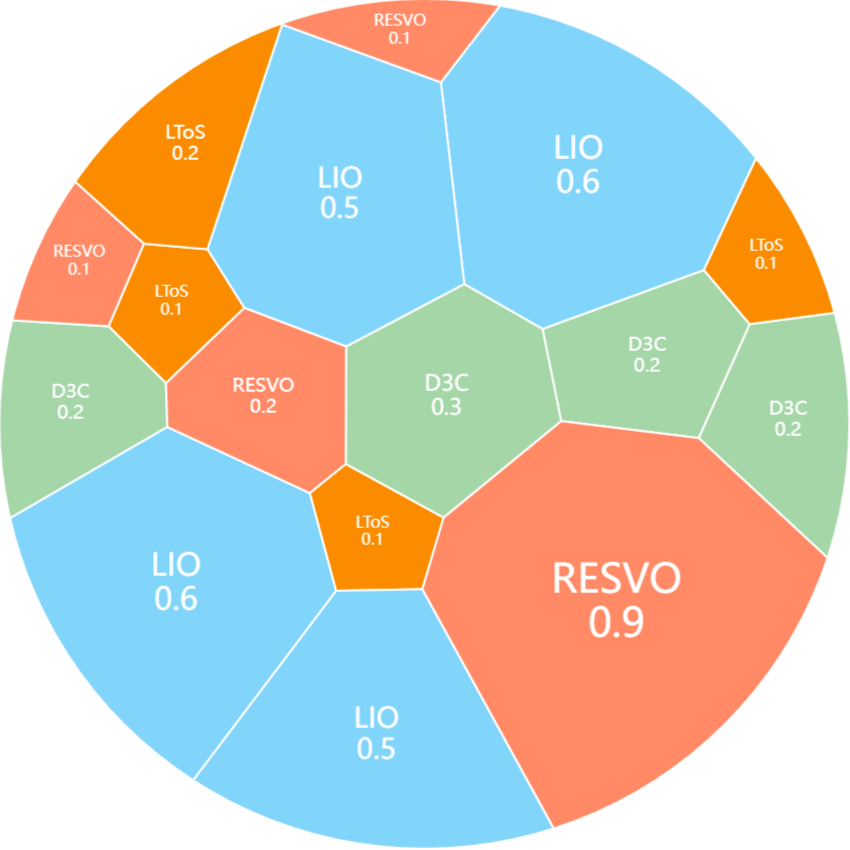}
    \end{subfigure}
    \caption{The role composition of different algorithms in $48\times 18$ map size, $10$-Player Cleanup. The numbers in the color blocks represent the number of times in $10$ experiments assigned to cleaning up wastes. The size of the color block is proportional to the number.}
    \label{fig:clean10-append1}
\end{figure}

Figure~\ref{fig:clean10-main} verifies the impact of the dynamic division of labor in the task completion process of different algorithms on performance. 
We also find that the dynamics of the division of labor are not only reflected in the completion of one task but also in the completion of different tasks by recording the dynamics of labor division of different algorithms under different random seeds.
Specifically, for each random experiment, we count the number of waste cleaned by each agent, similar to Figure~\ref{fig:clean-main}(c) and (f). 
If the agent is cleaning a low amount of waste (closer to $0$), the counter for the agent's role as a cleaner is incremented by one. Otherwise, the picker counter is incremented by one. 
Figure~\ref{fig:clean10-append1} shows the average dynamic where $10$ agents are assigned the role of cleaning wastes by different algorithms under $10$ random seeds.
As seen from the figure, in different random experiments, the SVO-based methods, LToS, D3C, and RESVO, all show better static division of labor than the sanction-based LIO.
However, LToS and D3C converge to a suboptimal static division of labor.
Most of the agents are assigned the role of collecting apples.
Both LIO and RESVO converge to a better division of labor. 
More agents choose to clean up waste, but the former maintains its dynamics.
Combined with the results of Figure~\ref{fig:clean10-main}, it can be seen that in tasks involving more complex optimal division of labor patterns, the static division of labor learned by RESVO can be more efficient than the dynamic division of labor in LIO.

\begin{figure}[htb!]
    \centering
    \includegraphics[width=\textwidth]{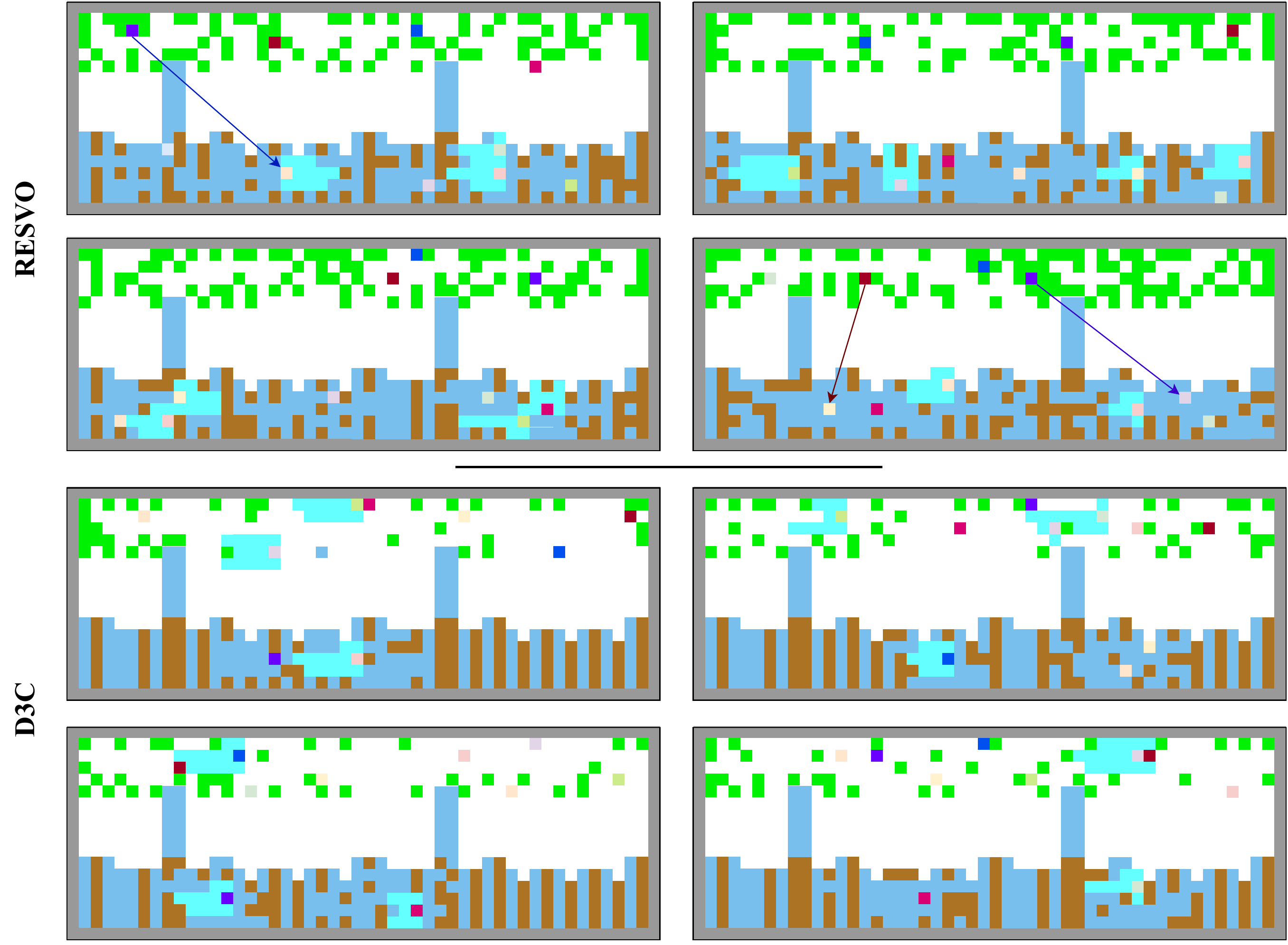}
    \caption{Selected keyframes from the rendered results of RESVO and D3C in the $10$-player Cleanup. Arrows indicate reward sharing.}
    \label{fig:cleanup10-vis1}
\end{figure}

\begin{figure}[htb!]
    \centering
    \includegraphics[width=\textwidth]{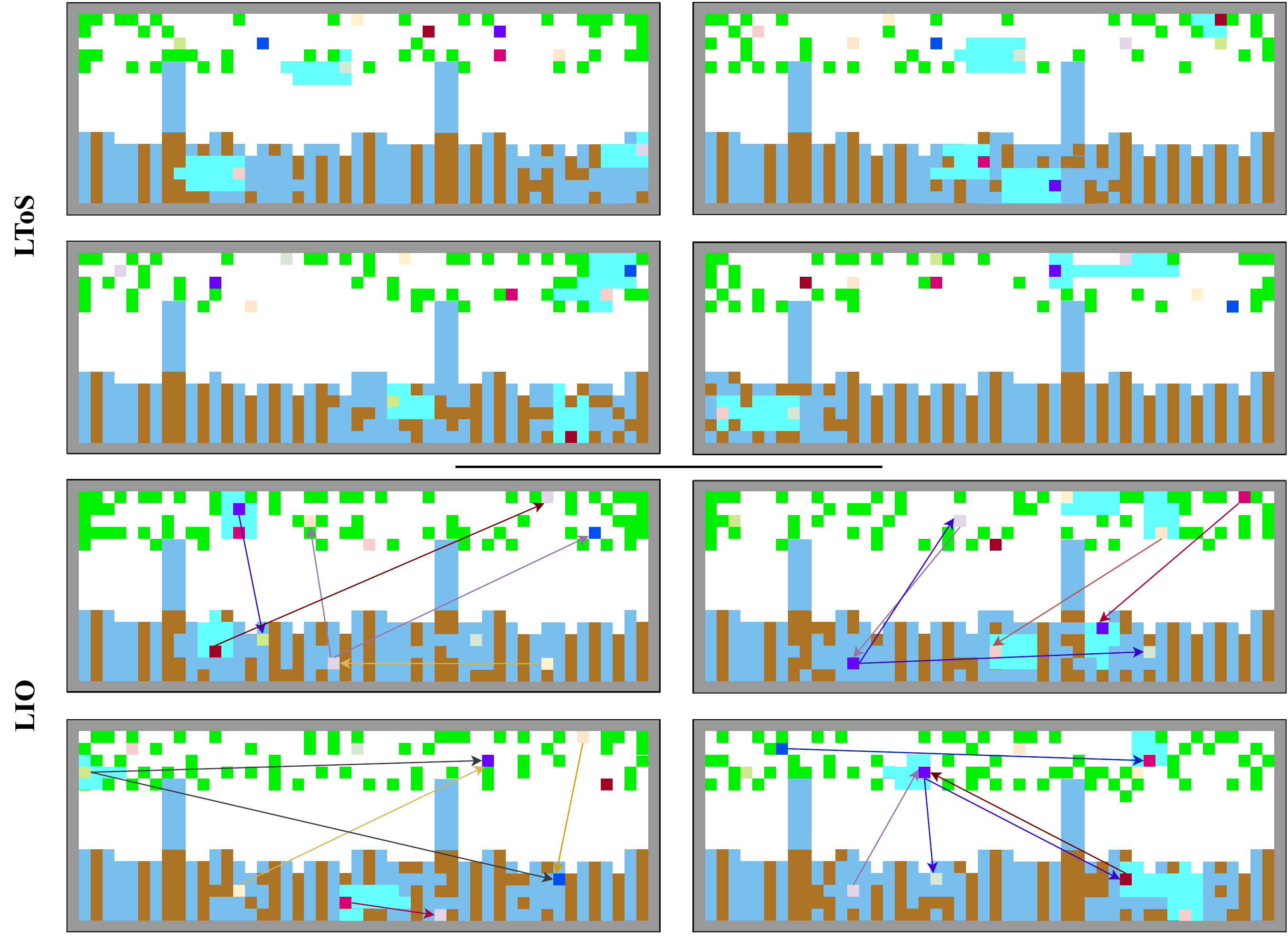}
    \caption{Selected keyframes from the rendered results of LToS and LIO in the $10$-player Cleanup. Arrows indicate reward sharing.}
    \label{fig:cleanup10-vis2}
\end{figure}

To more intuitively demonstrate the policies learned by different algorithms in the $10$-player Cleanup, we selected keyframes from the rendered results, as shown in Figure~\ref{fig:cleanup10-vis1} and~\ref{fig:cleanup10-vis2}.
The arrow in the figure indicates the reward sharing.
The policies shown in the figure match the performance results and related analysis presented earlier.

\subsection{SVO Emergence as a Plug-and-Play Module}

As introduced in Introduction and Method, RESVO is based on the independence theory, which transforms the learning of social value orientation, i.e., the learning of transformation matrix, into a symmetric learning to share problem. 
In this way, the role representation of the agent is transformed from the transformation matrix to the sharing coefficient matrix. 
To constrain the number of emergent roles, RESVO introduces a novel rank constraint. 
At the same time, to make the agent's behavior subject to role constraints, RESVO introduces a conditional policy based on role representation and a mutual information constraint to accelerate the learning of this augmented policy.
The above core idea of RESVO shows that the rank constraint and the conditional policy to satisfy the mutual information constraint are the two main modules of RESVO. 
Both modules are common to other SVO-based methods. 
Therefore, we can naturally add them to the SVO-based baselines, namely LToS and D3C, i.e., The SVO-based role emergence mechanism in RESVO can be introduced into other SVO-based methods as a plug-and-play module.

The numerical experimental results of LToS and D3C with the addition of the rank constraint and the conditional policy based on the mutual information constraint are shown in Figure~\ref{fig:clean10-main}(b).
We conduct different randomized experiments in the 10-player Cleanup environment under $10$ different random seeds and count the average extrinsic reward of the agents after the algorithm converges for each experiment.
As can be seen from the figure, both the LToS+r algorithm and the D3C+r algorithm show significant and stable performance improvements.

\subsection{The Impact of Rank Constraints}

Finally, we perform an ablation analysis of rank constraints in RESVO.
Intuitively, the number of roles, or the pattern of division of labor, will primarily affect the completion of tasks.
The rank constraint $k$ represents a priori knowledge of the optimal number of roles in the task.
In IPD, $3$-player Escape Room, and Cleanup of different complexity, we set $k$, or the number of roles, to $1$, $2$, and $2$, respectively.
In this section, we want to verify the sensitivity of RESVO to the hyperparameter $k$. 
Similarly, we conduct $10$ randomized experiments in the $10$-player Cleanup environment for different $k$ and count the average external reward of the agents, and the results are shown in Figure~\ref{fig:clean10-main}(c).

It can be seen from the results in the figure that the RESVO algorithm is sensitive to the size of $k$. 
When the selection of $k$ is too large or too small, the performance will decrease significantly. 
The optimal number of roles in the $10$-player Cleanup environment should be $2$. 
However, it can be seen from the figure that when $k$ is set to $3$ or $4$, the algorithm can also show promising results, indicating that the RESVO algorithm can show good robustness \textit{near} the optimal $k$.
We then present a more in-depth visual analysis of the SVO-based embeddings of emergent roles, which more intuitively shows the robustness of the RESVO algorithm around the optimal value of $k$.

\begin{figure}[htb!]
    \centering
    \includegraphics[width=0.8\textwidth]{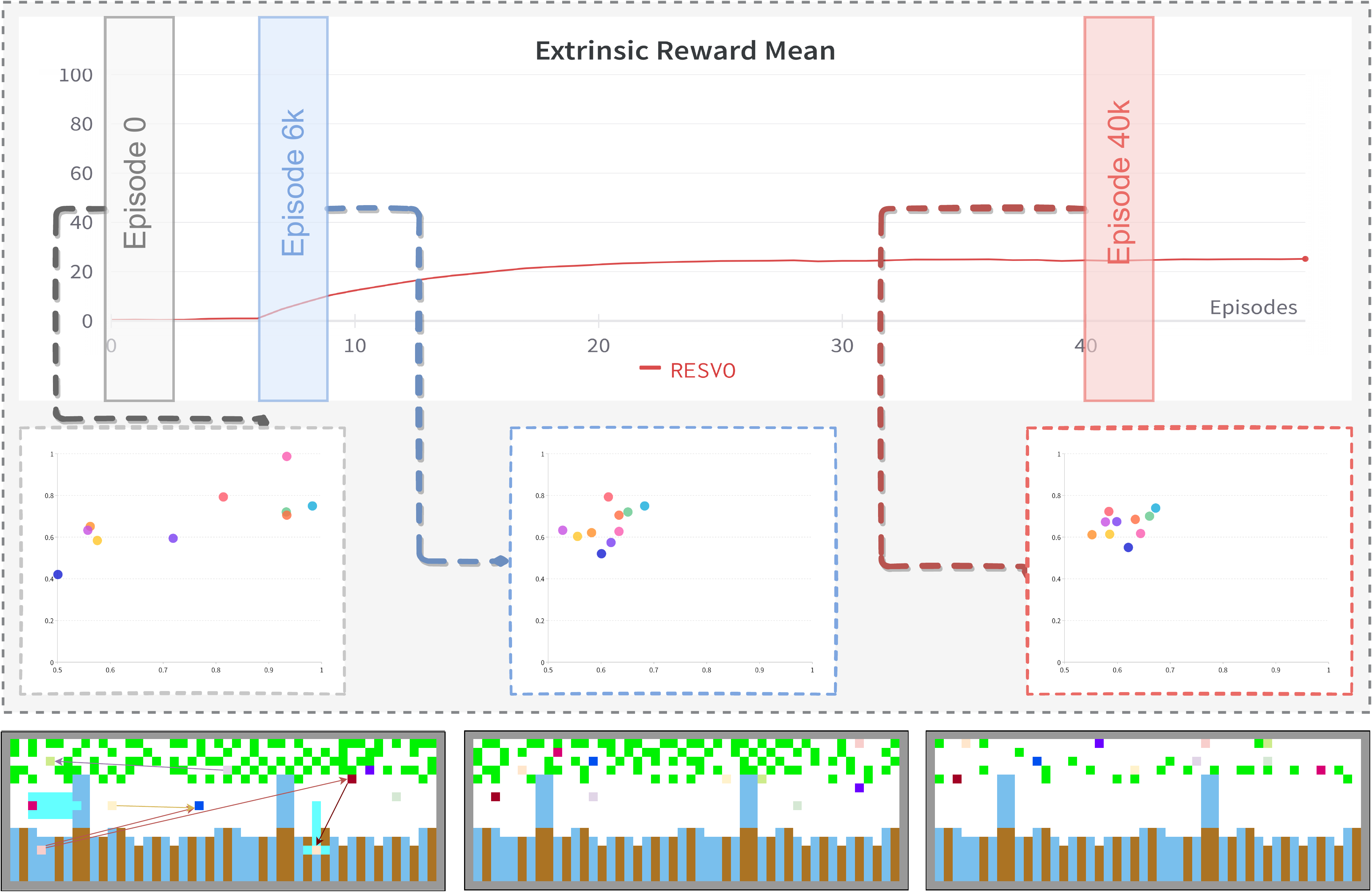}
    \caption{SVO-based role emergence with the rank $k=1$ constrain.}
    \label{fig:rank1-progress}
\end{figure}

\begin{figure}[htb!]
    \centering
    \includegraphics[width=0.8\textwidth]{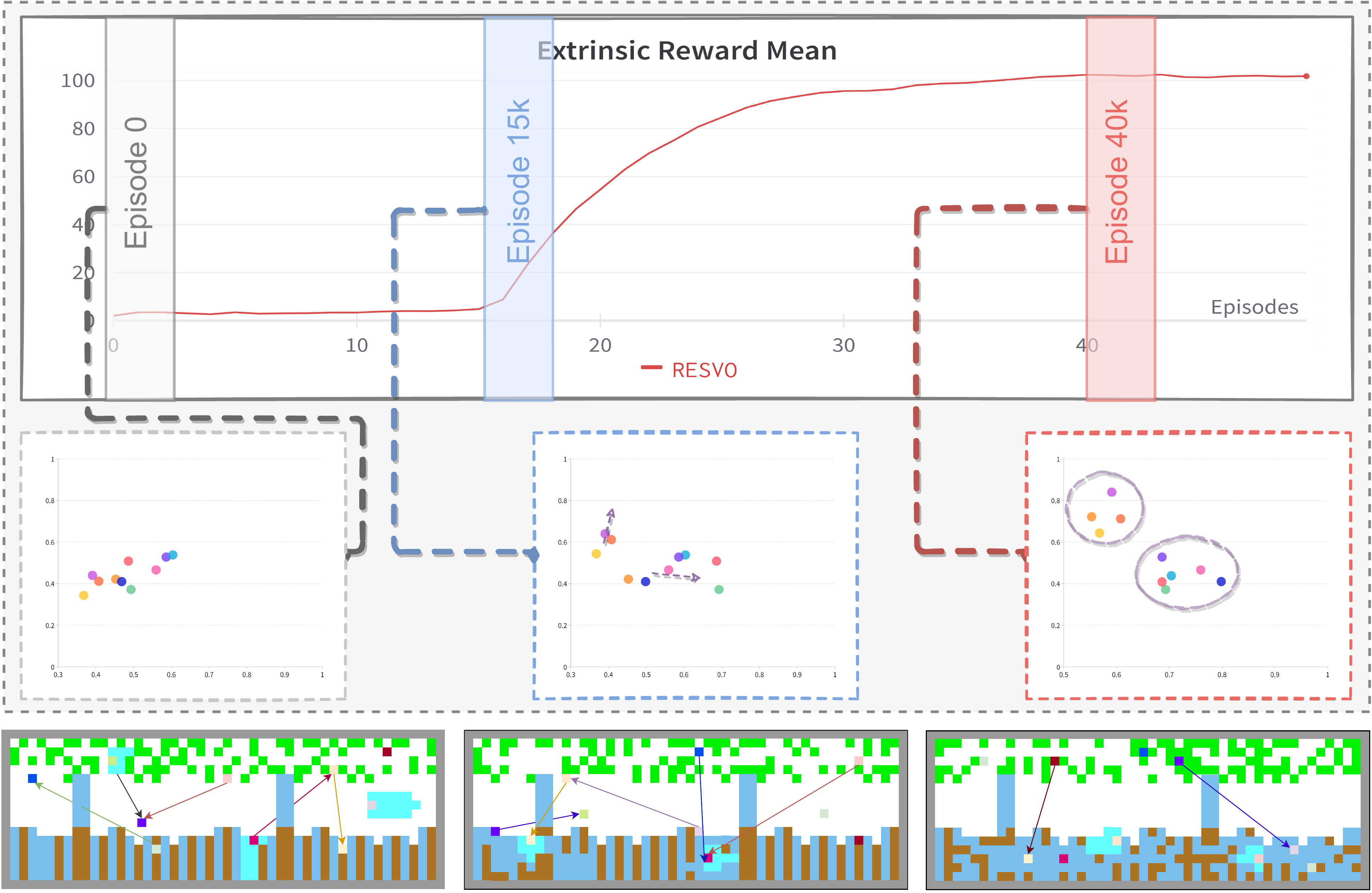}
    \caption{SVO-based role emergence with the rank $k=2$ constrain.}
    \label{fig:rank2-progress}
\end{figure}

\begin{figure}[htb!]
    \centering
    \includegraphics[width=0.8\textwidth]{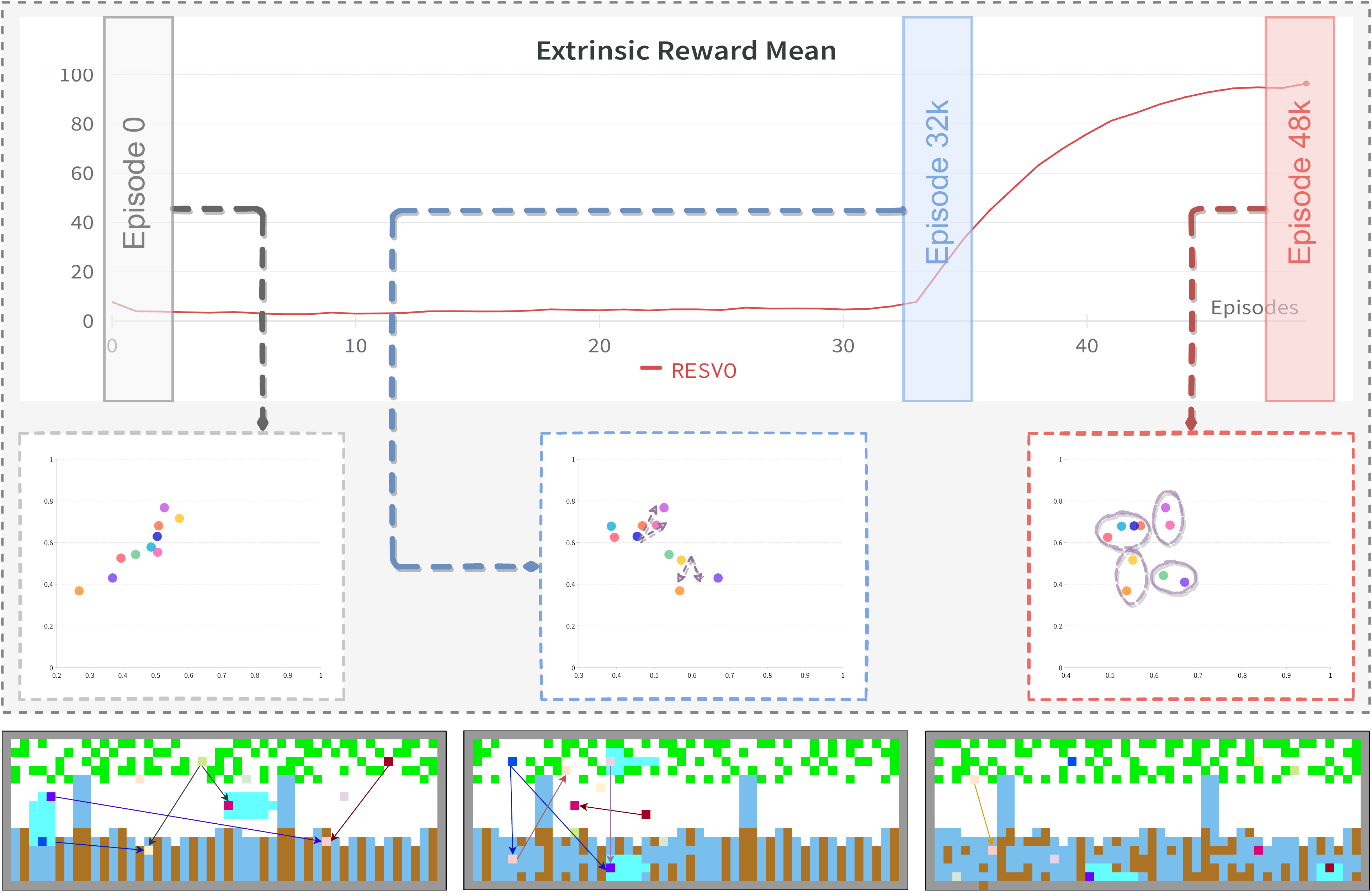}
    \caption{SVO-based role emergence with the rank $k=4$ constrain.}
    \label{fig:rank4-progress}
\end{figure}

\begin{figure}[htb!]
    \centering
    \includegraphics[width=0.8\textwidth]{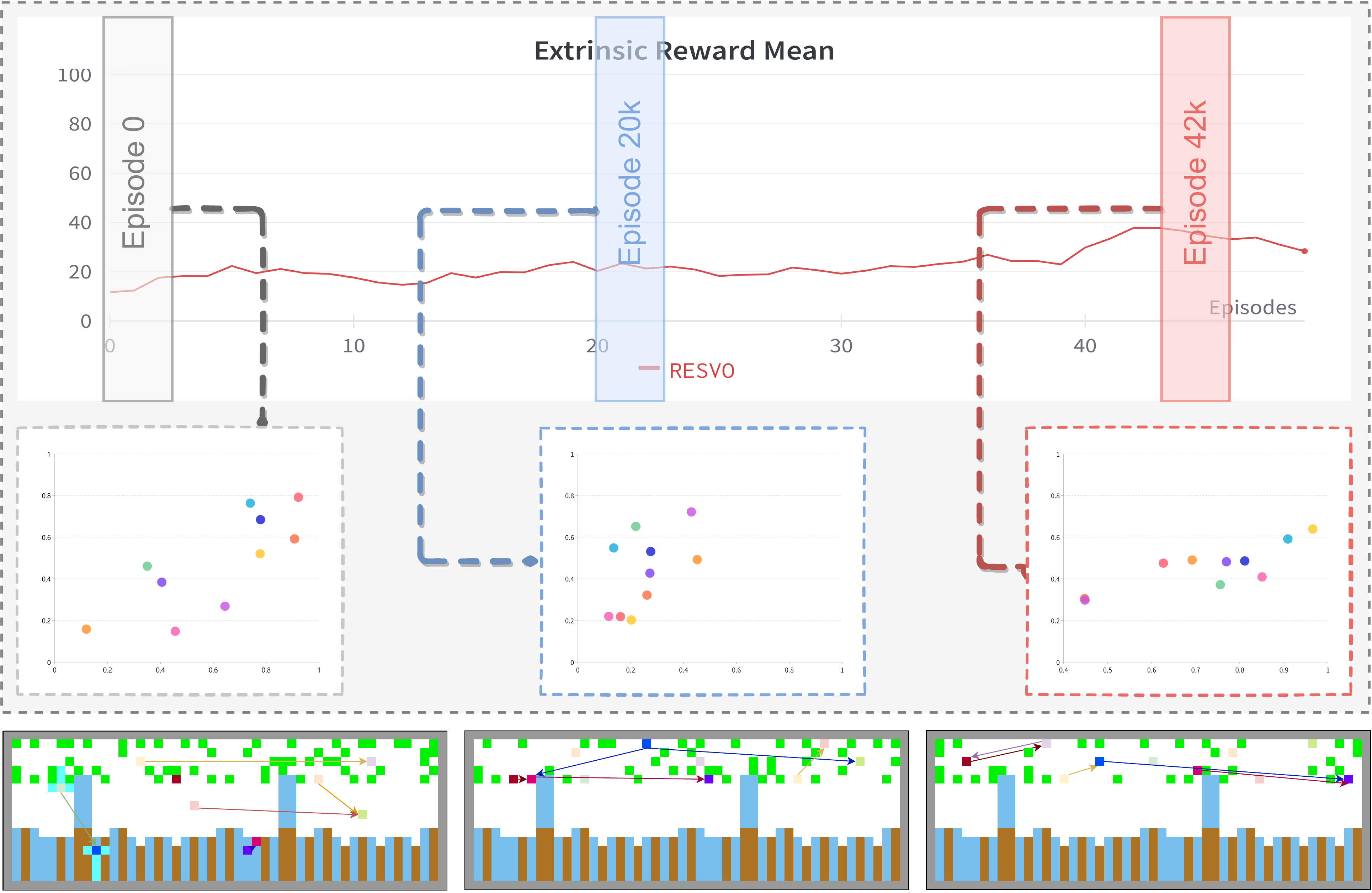}
    \caption{SVO-based role emergence with the rank $k=9$ constrain.}
    \label{fig:rank9-progress}
\end{figure}

Figure~\ref{fig:rank1-progress}-\ref{fig:rank9-progress} show the SVO emergence in the training procedure of RESVO under different rank constraints in Cleanup with a map size of $48\times 18$ and $10$ agents.
We visualize the SVO of each agent at the first timestep of a particular episode. 
We randomly map the SVO embedding of each agent to a $2D$ space using a fixed random neural network. 
This dimensionality reduction method can ensure that similar SVOs are also close together in $2D$ space.
As seen from Figure~\ref{fig:rank1-progress}, when the rank constraint is very low, all agents learn similar SVOs, that is, similar roles. 
This either means that all agents are free-riders or all agents have a composite role that needs to both collect apples and clean up waste. 
In either case, the overall performance will be poor.

When the size of the rank constraint is within a reasonable range, the algorithm can converge to the best performance, as shown in Figures~\ref{fig:rank2-progress} and~\ref{fig:rank4-progress}. 
In previous results, the algorithm can converge to a better result when the rank constraint is near the optimal value ($k=2$). 
Through the visualization results in Figures~\ref{fig:rank2-progress} and~\ref{fig:rank4-progress}, we can propose an explanation experimentally. 
It can be seen from the two figures that although the agent is divided into more roles when $k=4$, the similarity of these roles is different. 
Some roles are more similar, and others are less similar. 
Therefore, in the Cleanup, when $k=4$, some two roles may show a similar division of labor, so the performance will not be affected when there are more roles.

However, when the rank constraint increases ($k=9$), the situation worsens. 
As shown in Figure~\ref{fig:rank9-progress}, when the agent has too many roles, the SVO of the agent presents strong randomness, which makes the policy of the agent constrained by the SVO also present greater randomness. 
In the Cleanup task, the number of agents that collect apples or clean up wastes will be small, and some agents will follow random policies and do useless work, which will not improve the algorithm's performance.

\section{Closing Remarks}\label{sec:discuss}

In this paper, we introduce a typical mechanism of human society, i.e., division of labor, to solve intertemporal social dilemmas (ISDs) in multi-agent reinforcement learning.
A novel learning framework (RESVO) transforms role learning into a social value orientation emergence problem. 
RESVO solves it symmetrically by endowing agents with altruism to learn to share rewards with different weights with other agents.
Numerical experiments on three tasks with different complexity in the presence of ISD show that RESVO can emerge stable roles and efficiently solve the ISD through the division of labor.
Meanwhile, the SVO-based role emergence mechanism in RESVO can be introduced into other SVO-based methods as a plug-and-play module and bring a significant performance boost.

For the classic example, iterated prisoner's dilemma, we provide formal analyses and numerical results for the effect of social value orientation and division of labor in Appendix~\ref{sec:ana-ipd}. 
However, for more complex ISDs, i.e., $N$-Player Escape Room and Cleanup, we only provide empirical results in Section.~\ref{sec:exps}.
We aim to formally analyze the evolutionary dynamics of environmental behaviors, social value orientations, and roles of ISDs.
However, the environmental policies, roles, and sharing policies are interdependent, and it is not easy to accurately account for the effect of the division of labor when considering environmental policy updates.
These challenges have perplexed the researchers for a long time~\citep{hirsch2012differential,gould2016differentiating,dong2021birds}, and we believe that the solution to these questions is an essential and promising direction for future work.

In experiments, we find that the sanction-based LIO, the SVO-based LToS, D3C, and our RESVO take a completely different approach to maintain the division of labor. 
The former achieves a dynamic division of labor by continuously passing rewards among the agents, while the latter achieves a static division of labor by more sparse reward passing. 
In the IPD, $3$-player Escape Room, and Cleanup tasks of varying complexity involved in the experiments, we find that static division of labor exhibits better performance compared to dynamic division of labor in tasks where the same role corresponds to multiple agents, such as the $3$-player Escape Room, and the $10$-player Cleanup, converging faster to better social welfare.

However, in this paper, we obtain the above conclusions by the social welfare of the algorithm only in a limited task. 
We believe the dynamic division of labor will be more advantageous than the static division of labor in specific tasks and certain evaluation metrics. 
For example, dynamic division of labor may be more robust in tasks that involve roles that change dynamically; 
furthermore, static division of labor may pose fairness issues because some agents receive lower extrinsic rewards than others.
We leave the above questions for future exploration.


\bibliographystyle{ims}
\bibliography{sample}
\newpage

\appendix

\section{Implementation Details}

Our method is built on the open-sourced codebase LIO~\citep{yang2020learning}, ROMA~\citep{Wang2020ROMAMR} and Sequential Social Dilemma Games (SSDG)~\citep{vinitsky2019open}.
We test our method on two Cleanup maps with different numbers of agents (i.e., $2$ and $10$), and the details of each map are shown in Table~\ref{tab:cleanup-setttings}.

\begin{table}[htb!]
\centering
\caption{Environmental settings for Cleanup with different numbers of agents.}
\label{tab:cleanup-setttings}
\resizebox{0.5\textwidth}{!}{%
\begin{tabular}{lrr}
\hline
Parameter & n=$2$ & n=$10$ \\ \hline
Map Size & $10\times 10$ & $48\times 18$ \\
View Size & $7$ & $15$ \\
Max Steps & $50$ & $400$ \\
Apple Respawn Probability & $0.3$ & $0.15$ \\
Depletion Threshold & $0.4$ & $0.99$ \\
Restoration Threshold & $0.0$ & $0.0$ \\
Waste Spawn Probability & $0.5$ & $0.15$ \\ \hline
\end{tabular}%
}
\end{table}

We use fully-connected neural and convolutional networks for function approximation in the [IPD, ER] and Cleanup, respectively.
All algorithms use the same neural architecture without sharing parameters within each intertemporal social dilemma. 
The output of each agent's orientation function is scaled element-wise by a multiplier $w_{\max}$, which is bounded in $\left[-w_{\max}, w_{\max }\right]$.
The multiplier $w_{\max }$ is set to $[3,2,2]$ for [IPD, ER, Cleanup], respectively. 
We use a policy gradient algorithm to train the policy in IPD, ER, and actor-critic for Cleanup.
An exploration lower bound $\epsilon$ is used to perform sufficient exploration, with $\epsilon$ decaying linearly from $\epsilon_{\text {start }}$ to $\epsilon_{\text {end }}$ by $\epsilon_{\text {div }}$ episodes.
The values of these hyperparameters in different tasks are consistent with LIO~\citep{yang2020learning}.
We use discount factor $\gamma=0.99$, the gradient descent for policy optimization, and the Adam~\citep{kingma2015adam} optimizer for training value and orientation functions.

For the trajectory encoder, 
we build the causal transformer~\citep{chen2021decision} implementation for Cleanup off of minGPT\footnote{\url{https://github.com/karpathy/minGPT}.}, a publicly available reimplementation of GPT.
We use most of the hyperparameters from Decision Transformer~\citet{chen2021decision}\footnote{\url{https://github.com/kzl/decision-transformer}, MIT License.}.
Moreover, for Iterated Prisoner's Dilemma and $N$-Player Escape Room, our code is based on the Huggingface Transformers library~\citep{wolf2020transformers}.
Our hyperparameters on these tasks are also the same with~\citet{chen2021decision}.

We compare our method against various baselines. 
For ROMA~\citep{Wang2020ROMAMR}, LIO~\citep{yang2020learning}, LToS~\citep{yi2021learning}, and D3C~\citep{gemp2020d3c}, we use the codes provided by the authors and the hyper-parameters that have been fine-tuned on the Sequential Social Dilemma Games (SSDG)~\citep{vinitsky2019open}.
The ablations LToS+r and D3C+r use the exact implementation and hyperparameter settings as LToS and D3C, respectively, but with the only difference that the role emergence mechanism is introduced, which is the same as in our method.
Learning curves are smoothed by averaging over a window of $11$ episodes.
The hardware used in the experiment is a server with $128$G memory and $4$ NVIDIA $1080$Ti graphics cards with $11$G video memory.
The code and license of baselines are shown in the following list:
\begin{itemize}
    \item ROMA~\citep{Wang2020ROMAMR}: \url{https://github.com/TonghanWang/ROMA}, Apache-2.0 License;
    \item LIO~\citep{yang2020learning}: \url{https://github.com/011235813/lio}, MIT License;
    \item SSDG~\citep{vinitsky2019open}: \url{https://github.com/eugenevinitsky/sequential_social_dilemma_games}, MIT License.
\end{itemize}

\section{Derivations}\label{sec:derivation}

\subsection{Role Emergence}

Here we deduce the (\ref{eq:svo-loss}) in detail. 
For logical integrity, we have merged and reorganized some of the contents of Section~\ref{sec:svo-based}  and Section~\ref{sec:role-policy}.
At each time step $t$, each agent $j$ receives a total reward
\begin{equation}
    r_j(\boldsymbol{\eta}, \boldsymbol{r}):=w^{j}_{\eta_j}[j] \cdot r_{j} + \sum_{i \neq j} w^{i}_{\eta_i}[j] \cdot r_{i},
\end{equation}
Each agent $j$ learns a SVO-based role conditioned policy $\pi_j(\cdot\mid o_j,e_j(\boldsymbol{\eta}))$ parameterized by $\theta_j$, where $e_j(\cdot)$ is the SVO-based role embedding, to maximize the objective
\begin{equation}\label{eq:policy-update-a}
    \max_{\theta_j} J^{\mathrm{policy}}(\theta_j,\{e_j\}):=\mathbb{E}_{\boldsymbol{\pi}(\cdot\mid\cdot,\{e_j\})}\left[\sum_{t=0}^{T}\gamma^t r_j^t(\boldsymbol{\eta},\boldsymbol{r})\right].
\end{equation}
Upon experiencing trajectories $\boldsymbol{\tau}:=\{\left(s^{0}, \mathbf{a}^{0}, \mathbf{r}^{0}, \ldots, s^{T}\right)\}$, the recipient carries out an update
\begin{equation}\label{eq:pol-params-update-a}
    \hat{\theta}_{j} \leftarrow \theta_{j}+\beta f\left(\boldsymbol{\tau}, \theta_{j}, \eta\right).
\end{equation}
that adjusts its policy parameters with learning rate $\beta$.
Assuming policy optimization learners in this work and choosing policy gradient for exposition, the update function is
\begin{equation}\label{eq:pg-update-func}
    f\left(\boldsymbol{\tau}, \theta_{j}, \eta\right)=\sum_{t=0}^{T} \nabla_{\theta_{j}} \log \pi_{j}\left(a^{t}_{j} \mid o^{t}_{j}, e^{t}_j(\boldsymbol{\eta})\right) G^{t}_{j}\left(\boldsymbol{\tau} ; \boldsymbol{\eta}\right),
\end{equation}
where the return $G^{t}_{j}\left(\boldsymbol{\tau}, \boldsymbol{\eta}\right)=\sum_{l=t}^{T} \gamma^{l-t} r_{j}\left(\boldsymbol{\eta}, \mathbf{r}_l\right)$ depends on orientation parameters $\boldsymbol{\eta}$.
After each agent has updated its policy to $\hat{\pi}_{j}$, parameterized by new $\hat{\theta}_{j}$, it generates a new trajectory $\hat{\tau}_{j}$. 
Using these trajectories, each sharing agent $i$ updates shared orientation function parameters $\eta_i$ to maximize the following individual objective:
\begin{equation}\label{eq:svo-rank-u-a}
    \max_{\eta_i} J_{i}^{\mathrm{svo}}\left(\hat{\tau}_{i}, \hat{\boldsymbol{\theta}}, \boldsymbol{\eta}, \eta_i\right):=\mathbb{E}_{\hat{\boldsymbol{\pi}}}\left[\sum_{t=0}^{T}\gamma^t \left(\hat{r}_i^t-\alpha\|W^{i,t}_{\eta_i}-W_k^{i,t}\|^{2}_{2}\right)\right],
\end{equation}
where the superscription $i$ denotes the $i$-th column of the matrix $W^{i,t}_{\eta_i}$.
The penalty term introduced by the rank constraint can be regarded as an intrinsic reward $-\alpha\|W^{i,t}_{\eta}-W_k^{i,t}\|^{2}_{2}$.
And the gradient of $J_{i}^{\mathrm{svo}}\left(\hat{\tau}_{i}, \hat{\boldsymbol{\theta}}, \boldsymbol{\eta}, \eta_i\right)$ w.r.t. $\eta_i$ is:
\begin{equation}\label{eq:three-parts}
    \begin{aligned}
        &\nabla_{\eta_i}J_{i}^{\mathrm{svo}}\left(\hat{\tau}_{i}, \hat{\boldsymbol{\theta}}, \boldsymbol{\eta}, \eta_i\right) \\
        =& \nabla_{\eta_i}\hat{\boldsymbol{\theta}}\nabla_{\hat{\boldsymbol{\theta}}}J_{i}^{\mathrm{svo}}\left(\hat{\tau}^{i}, \hat{\boldsymbol{\theta}}, \boldsymbol{\eta}, \eta_i\right) + \nabla_{\eta_i}\boldsymbol{\eta}\nabla_{\boldsymbol{\eta}}J_{i}^{\mathrm{svo}}\left(\hat{\tau}_{i}, \hat{\boldsymbol{\theta}}, \boldsymbol{\eta}, \eta_i\right) +  \nabla_{\eta_i}J_{i}^{\mathrm{svo}}\left(\hat{\tau}_{i}, \hat{\boldsymbol{\theta}}, \boldsymbol{\eta}, \eta_i\right) \\
        =& \underbrace{\textcolor{blue}{\sum_{j=1}^{N} (\nabla_{\eta_i} \hat{\theta}_j)^{T} \nabla_{\hat{\theta}_j} J_{i}^{\mathrm{svo}}\left(\hat{\tau}_{i}, \hat{\boldsymbol{\theta}}, \boldsymbol{\eta}, \eta_i\right)}}_{\textcolor{blue}{\text{blue part}}} + \underbrace{\textcolor{orange}{(\nabla_{\eta_i}\boldsymbol{\eta})^{T}\nabla_{\boldsymbol{\eta}}J_{i}^{\mathrm{svo}}\left(\hat{\tau}_{i}, \hat{\boldsymbol{\theta}}, \boldsymbol{\eta}, \eta_i\right)}}_{\textcolor{orange}{\text{orange part}}} + \\
        & \underbrace{\textcolor{purple}{\nabla_{\eta_i}J_{i}^{\mathrm{svo}}\left(\hat{\tau}_{i}, \hat{\boldsymbol{\theta}}, \boldsymbol{\eta}, \eta_i\right)}}_{\textcolor{purple}{\text{purple part}}}.
    \end{aligned}
\end{equation}

The first factor of each term in the summation of the \textcolor{blue}{blue} part follows directly from (\ref{eq:pol-params-update-a}) and (\ref{eq:pg-update-func}):
\begin{equation}
    \begin{aligned}
        \nabla_{\eta_i} \hat{\theta}_j=\beta\sum_{t=0}^{T} 
        &\nabla_{\eta_i}\nabla_{\theta_{j}} \log \pi_{j}\left(a^{t}_{j} \mid o^{t}_{j}, e_j(\boldsymbol{\eta})\right) {G}^{t}_{j}\left(\boldsymbol{\tau} ; \boldsymbol{\eta}\right) + \\
        &\nabla_{\theta_{j}} \log \pi_{j}\left(a^{t}_{j} \mid o^{t}_{j}, e_j(\boldsymbol{\eta})\right) \left(\nabla_{\eta_i} {G}^{t}_{j}\left(\boldsymbol{\tau} ; \boldsymbol{\eta}\right)\right)^{T}.
    \end{aligned}
\end{equation}
Note that (\ref{eq:pol-params-update-a}) does not contain recursive dependence of $\theta_{j}$ on $\eta$ since $\theta_{j}$ is a function of incentives in \textit{previous} episodes and orientation parameters, not those in trajectory $\tau_{i}$ and current $\eta$. 
The second factor of the \textcolor{blue}{blue} part can be derived similarly as standard policy gradients~\citep{Sutton1999PolicyGM}:
\begin{equation}
    \begin{aligned}
        &\nabla_{\hat{\theta}_{j}} J_{i}^{\mathrm{svo}}\left(\hat{\tau}_{i}, \hat{\boldsymbol{\theta}}, \boldsymbol{\eta}, \eta_i\right)=\nabla_{\hat{\theta}^{j}} \tilde{V}_{i}^{\boldsymbol{\hat{\pi}}}\left(\hat{s}_{0},\eta_i\right)=\nabla_{\hat{\theta}^{j}} \sum_{\mathbf{\hat{a}}} \boldsymbol{\hat{\pi}}\left(\mathbf{\hat{a}} \mid \hat{s}_{0}, \boldsymbol{\hat{e}}(\boldsymbol\eta)\right) \tilde{Q}_{i}^{\boldsymbol{\hat{\pi}}}\left(\hat{s}_{0}, \mathbf{\hat{a}}, \eta_i\right)\\
        =& \sum_{\mathbf{\hat{a}}} \hat{\pi}^{-j}_{\hat{\theta}_{-j}} \left(\left(\nabla_{\hat{\theta}_{j}} \hat{\pi}^{j}_{\hat{\theta}_{j}}\right) \tilde{Q}_{i}^{\boldsymbol{\hat{\pi}}}\left(\hat{s}_{0}, \mathbf{\hat{a}}, \eta_i\right)+\hat{\pi}^{j}_{\hat{\theta}_{j}} \nabla_{\hat{\theta}_{j}} \tilde{Q}_{i}^{\boldsymbol{\hat{\pi}}}\left(\hat{s}_{0}, \mathbf{\hat{a}}, \eta_i\right)\right) \\
        =& \sum_{\mathbf{\hat{a}}} \hat{\pi}^{-j}_{\hat{\theta}_{-j}} \left(\left(\nabla_{\hat{\theta}_{j}} \hat{\pi}^{j}_{\hat{\theta}_{j}}\right) \tilde{Q}_{i}^{\boldsymbol{\hat{\pi}}}\left(\hat{s}_{0}, \mathbf{\hat{a}}, \eta_i\right) + \hat{\pi}^{j}_{\hat{\theta}_{j}} \nabla_{\hat{\theta}_{j}}\left(\hat{r}_{i}+\gamma \sum_{\hat{s}^{\prime}} P\left(\hat{s}^{\prime} \mid \hat{s}_{0}, \mathbf{\hat{a}}\right) \tilde{V}_{i}^{\boldsymbol{\pi}}\left(\hat{s}^{\prime},\eta_i\right)\right)\right) \\
        =& \sum_{\mathbf{\hat{a}}} \hat{\pi}^{-j}_{\hat{\theta}_{-j}} \left(\left(\nabla_{\hat{\theta}_{j}} \hat{\pi}^{j}_{\hat{\theta}_{j}}\right) \tilde{Q}_{i}^{\boldsymbol{\hat{\pi}}}\left(\hat{s}_{0}, \mathbf{\hat{a}}, \eta_i\right) + \gamma \hat{\pi}^{j}_{\hat{\theta}_{j}} \sum_{\hat{s}^{\prime}} P\left(\hat{s}^{\prime} \mid \hat{s}_{0}, \mathbf{\hat{a}}\right) \nabla_{\hat{\theta}_{j}}\tilde{V}_{i}^{\boldsymbol{\pi}}\left(\hat{s}^{\prime},\eta_i\right)\right) \\
        =& \sum_{x} \sum_{k=0}^{T} P\left(\hat{s}_{0} \rightarrow x, k, \boldsymbol{\hat{\pi}}\right) \gamma^{k} \sum_{\mathbf{\hat{a}}} \hat{\pi}^{-j}_{\hat{\theta}_{-j}} \nabla_{\hat{\theta}_{j}} \hat{\pi}^{j}_{\hat{\theta}_{j}} \tilde{Q}_{i}^{\boldsymbol{\hat{\pi}}}(x, \mathbf{\hat{a}}, \eta_{i}) \\
        =& \sum_{\hat{s}} d^{\boldsymbol{\hat{\pi}}}(\hat{s})\sum_{\mathbf{\hat{a}}} \hat{\pi}^{-j}_{\hat{\theta}_{-j}} \nabla_{\hat{\theta}_{j}} \hat{\pi}^{j}_{\hat{\theta}_{j}} \tilde{Q}_{i}^{\boldsymbol{\hat{\pi}}}(\hat{s}, \mathbf{\hat{a}}, \eta_{i}) \\
        =& \sum_{\hat{s}} d^{\boldsymbol{\hat{\pi}}}(\hat{s})\sum_{\mathbf{\hat{a}}} \hat{\pi}^{-j}_{\hat{\theta}_{-j}} \hat{\pi}^{j}_{\hat{\theta}_{j}}\nabla_{\hat{\theta}_{j}} \log\hat{\pi}^{j}_{\hat{\theta}_{j}} \tilde{Q}_{i}^{\boldsymbol{\hat{\pi}}}(\hat{s}, \mathbf{\hat{a}}, \eta_{i}) \\
        =&\mathbb{E}_{\hat{\boldsymbol{\pi}}}\left[\nabla_{\hat{\theta}^{j}} \log \hat{\pi}^{j}_{\hat{\theta}_{j}}\left(\hat{a}_{j} \mid \hat{o}_{j}, \hat{e}_{j}(
        \boldsymbol{\eta})\right) \tilde{Q}^{i, \hat{\boldsymbol{\pi}}}(\hat{s}, \hat{\mathbf{a}}, \eta_i)\right],
    \end{aligned}
\end{equation}
where $\tilde{Q}_{i}^{\hat{\boldsymbol{\pi}}}(\hat{s}, \hat{\mathbf{a}}, \boldsymbol{\eta})$ represents the $Q$-function of the augmented reward which is augmented by the intrinsic reward $-\alpha\|W^{i,t}_{\eta_i}-W_k^{i,t}\|^{2}_{2}$, $\hat{\pi}^{-j}_{\hat{\theta}_{-j}}:=\hat{\pi}^{-j}_{\hat{\theta}_{-j}}\left(\hat{a}_{-j} \mid \hat{s}_{0}, \hat{e}_{-j}(\boldsymbol\eta)\right)$, $\hat{\pi}^{j}_{\hat{\theta}_{j}}:=\hat{\pi}^{j}_{\hat{\theta}_{j}}\left(\hat{a}_{j} \mid \hat{s}_{0}, \hat{e}_{j}(\boldsymbol\eta)\right)$.
Similar to the derivation of the second factor of the \textcolor{blue}{blue} part, the second factor of the \textcolor{orange}{orange} part can be derived as
\begin{equation}
    \begin{aligned}
        &\nabla_{\boldsymbol{\eta}} J_{i}^{\mathrm{svo}}\left(\hat{\tau}_{i}, \hat{\boldsymbol{\theta}}, \boldsymbol{\eta}, \eta_i\right) = \nabla_{\boldsymbol{\eta}} \tilde{V}_{i}^{\boldsymbol{\hat{\pi}}}\left(\hat{s}_{0},\eta_i\right) = \nabla_{\boldsymbol{\eta}} \sum_{\mathbf{\hat{a}}} \boldsymbol{\hat{\pi}}\left(\mathbf{\hat{a}} \mid \hat{s}_{0}, \boldsymbol{\hat{e}}(\boldsymbol\eta)\right) \tilde{Q}_{i}^{\boldsymbol{\hat{\pi}}}\left(\hat{s}_{0}, \mathbf{\hat{a}}, \eta_i\right)\\
        =& \sum_{\mathbf{\hat{a}}} \left(\left(\nabla_{\boldsymbol{\eta}} \boldsymbol{\hat{\pi}}_{\boldsymbol{\hat{\theta}}}\right) \tilde{Q}_{i}^{\boldsymbol{\hat{\pi}}}\left(\hat{s}_{0}, \mathbf{\hat{a}}, \eta_i\right) + \boldsymbol{\hat{\pi}}_{\boldsymbol{\hat{\theta}}} \nabla_{\boldsymbol{\eta}} \tilde{Q}_{i}^{\boldsymbol{\hat{\pi}}}\left(\hat{s}_{0}, \mathbf{\hat{a}}, \eta_i\right)\right) \\
        =& \sum_{\mathbf{\hat{a}}} \left(\left(\nabla_{\boldsymbol{\eta}} \boldsymbol{\hat{\pi}}_{\boldsymbol{\hat{\theta}}}\right) \tilde{Q}_{i}^{\boldsymbol{\hat{\pi}}}\left(\hat{s}_{0}, \mathbf{\hat{a}}, \eta_i\right) + \boldsymbol{\hat{\pi}}_{\boldsymbol{\hat{\theta}}} \nabla_{\boldsymbol{\eta}} \left(\hat{r}_{i}+\gamma \sum_{\hat{s}^{\prime}} P\left(\hat{s}^{\prime} \mid \hat{s}_{0}, \mathbf{\hat{a}}\right) \tilde{V}_{i}^{\boldsymbol{\pi}}\left(\hat{s}^{\prime},\eta_i\right)\right)\right) \\
        =& \sum_{\mathbf{\hat{a}}} \left(\left(\nabla_{\boldsymbol{\eta}} \boldsymbol{\hat{\pi}}_{\boldsymbol{\hat{\theta}}}\right) \tilde{Q}_{i}^{\boldsymbol{\hat{\pi}}}\left(\hat{s}_{0}, \mathbf{\hat{a}}, \eta_i\right) + \gamma\boldsymbol{\hat{\pi}}_{\boldsymbol{\hat{\theta}}} \sum_{\hat{s}^{\prime}} P\left(\hat{s}^{\prime} \mid \hat{s}_{0}, \mathbf{\hat{a}}\right) \nabla_{\boldsymbol{\eta}}\tilde{V}_{i}^{\boldsymbol{\pi}}\left(\hat{s}^{\prime},\eta_i\right)\right) \\
        =& \sum_{x} \sum_{k=0}^{T} P\left(\hat{s}_{0} \rightarrow x, k, \boldsymbol{\hat{\pi}}\right) \gamma^{k} \sum_{\mathbf{\hat{a}}} \nabla_{\boldsymbol{\eta}} \boldsymbol{\hat{\pi}}_{\boldsymbol{\hat{\theta}}} \tilde{Q}_{i}^{\boldsymbol{\hat{\pi}}}(x, \mathbf{\hat{a}}, \eta_{i}) \\
        =& \sum_{\hat{s}} d^{\boldsymbol{\hat{\pi}}}(\hat{s})\sum_{\mathbf{\hat{a}}} \nabla_{\boldsymbol{\eta}} \boldsymbol{\hat{\pi}}_{\boldsymbol{\hat{\theta}}} \tilde{Q}_{i}^{\boldsymbol{\hat{\pi}}}(\hat{s}, \mathbf{\hat{a}}, \eta_{i}) \\
        =& \sum_{\hat{s}} d^{\boldsymbol{\hat{\pi}}}(\hat{s})\sum_{\mathbf{\hat{a}}} \boldsymbol{\hat{\pi}}_{\boldsymbol{\hat{\theta}}} \nabla_{\boldsymbol{\eta}} \log \boldsymbol{\hat{\pi}}_{\boldsymbol{\hat{\theta}}} \tilde{Q}_{i}^{\boldsymbol{\hat{\pi}}}(\hat{s}, \mathbf{\hat{a}}, \eta_{i}) \\
        =&\mathbb{E}_{\hat{\boldsymbol{\pi}}}\left[\left(\sum_{j=1}^{N}\nabla_{\boldsymbol{\eta}} \log \hat{\pi}^{j}_{\hat{\theta}_{j}}\left(\hat{a}_{j} \mid \hat{o}_{j}, \hat{e}_{j}(\boldsymbol{\eta})\right)\right) \tilde{Q}^{i, \hat{\boldsymbol{\pi}}}(\hat{s}, \hat{\mathbf{a}}, \eta_i)\right].
    \end{aligned}
\end{equation}
Finally, the \textcolor{purple}{purple} part can be directly derived as
\begin{equation}
    \nabla_{\eta_i} J_{i}^{\mathrm{svo}}\left(\hat{\tau}_{i}, \hat{\boldsymbol{\theta}}, \boldsymbol{\eta}, \eta_i\right)=\mathbb{E}_{\hat{\boldsymbol{\pi}}}\left[\sum_{t=0}^{T}-2\alpha\gamma^t\nabla_{\eta_i}W^{i,t}_{\eta_i}\left(W^{i,t}_{\eta_i}-W_k^{i,t}\right)\right].
\end{equation}

\subsection{Policy Optimization}
Introducing SVO-based role embedding and conditioning individual policies on this embedding explicitly establish the connection between the role and the individual policies to encourage the division of labor through the diversity of roles. 
Nevertheless, this does not enable the role representation to exhibit sufficient responsibilities. 
That is, the role can constrain the long-term behavior of the agent.
Drawing inspiration from \citet{eysenbach2018diversity, Wang2020ROMAMR}, we propose to learn SVO-based roles that are identifiable by agents' long-term behaviors, which can be achieved by maximizing $I(\tau_i;e_i\mid\boldsymbol{o},\boldsymbol{a})$, the conditional mutual information between the individual trajectory and the role given the current \textit{joint} observation and \textit{joint} action.

Based on variational inference and the derivation process in~\citet{Wang2020ROMAMR}, a variational posterior estimator can be proposed to derive a tractable lower bound for the mutual information objective
\begin{equation}
    \begin{aligned}
    &I\left(e_{i}^{t} ; \tau_{i}^{t-1} \mid \boldsymbol{o}^{t},\boldsymbol{a}^{t}\right) \\
    &= \mathbb{E}_{e_{i}^{t}, \tau_{i}^{t-1}, \boldsymbol{o}^{t},\boldsymbol{a}^{t}}\left[\log \frac{p\left(\rho_{i}^{t} \mid \tau_{i}^{t-1}, \boldsymbol{o}^{t},\boldsymbol{a}^{t}\right)}{p\left(e_{i}^{t} \mid \boldsymbol{o}^{t},\boldsymbol{a}^{t}\right)}\right] \\
    &=\mathbb{E}_{e_{i}^{t}, \tau_{i}^{t-1},\boldsymbol{o}^{t},\boldsymbol{a}^{t}}\left[\log \frac{q_{\phi}\left(e_{i}^{t} \mid \tau_{i}^{t-1}, \boldsymbol{o}^{t},\boldsymbol{a}^{t}\right)}{p\left(e_{i}^{t} \mid \boldsymbol{o}^{t},\boldsymbol{a}^{t}\right)}\right] \\
    &\quad+\mathbb{E}_{\tau_{i}^{t-1}, \boldsymbol{o}^{t},\boldsymbol{a}^{t}}\left[D_{\mathrm{KL}}\left(p\left(e_{i}^{t} \mid \tau_{i}^{t-1}, \boldsymbol{o}^{t},\boldsymbol{a}^{t}\right) \| q_{\phi}\left(e_{i}^{t} \mid \tau_{i}^{t-1}, \boldsymbol{o}^{t},\boldsymbol{a}^{t}\right)\right)\right] \\
    &\geq \mathbb{E}_{e_{i}^{t}, \tau_{i}^{t-1}, \boldsymbol{o}^{t},\boldsymbol{a}^{t}}\left[\log \frac{q_{\phi}\left(e_{i}^{t} \mid \tau_{i}^{t-1}, \boldsymbol{o}^{t},\boldsymbol{a}^{t}\right)}{p\left(e_{i}^{t} \mid \boldsymbol{o}^{t},\boldsymbol{a}^{t}\right)}\right],
    \end{aligned}
\end{equation}
where the last inequality holds via non-negativity of the KL divergence. 
Then it follows that:
\begin{equation}
    \begin{aligned}
        & \mathbb{E}_{e_{i}^{t}, \tau_{i}^{t-1}, \boldsymbol{o}^{t},\boldsymbol{a}^{t}}\left[\log \frac{q_{\phi}\left(e_{i}^{t} \mid \tau_{i}^{t-1}, \boldsymbol{o}^{t},\boldsymbol{a}^{t}\right)}{p\left(e_{i}^{t} \mid \boldsymbol{o}^{t},\boldsymbol{a}^{t}\right)}\right] \\
        =& \mathbb{E}_{e_{i}^{t}, \tau_{i}^{t-1}, \boldsymbol{o}^{t},\boldsymbol{a}^{t}}\left[\log q_{\phi}\left(e_{i}^{t} \mid \tau_{i}^{t-1}, \boldsymbol{o}^{t},\boldsymbol{a}^{t}\right)\right]-\mathbb{E}_{e_{i}^{t}, \boldsymbol{o}^{t},\boldsymbol{a}^{t}}\left[\log p\left(e_{i}^{t} \mid \boldsymbol{o}^{t},\boldsymbol{a}^{t}\right)\right] \\
        =& \mathbb{E}_{e_{i}^{t}, \tau_{i}^{t-1}, \boldsymbol{o}^{t},\boldsymbol{a}^{t}}\left[\log q_{\phi}\left(e_{i}^{t} \mid \tau_{i}^{t-1}, \boldsymbol{o}^{t},\boldsymbol{a}^{t}\right)\right]+\mathbb{E}_{\boldsymbol{o}^{t},\boldsymbol{a}^{t}}\left[H\left(e_{i}^{t} \mid \boldsymbol{o}^{t},\boldsymbol{a}^{t}\right)\right] \\
        =& \mathbb{E}_{\tau_{i}^{t-1}, \boldsymbol{o}^{t},\boldsymbol{a}^{t}}\left[\int p\left(e_{i}^{t} \mid \tau_{i}^{t-1}, \boldsymbol{o}^{t},\boldsymbol{a}^{t}\right) \log q_{\phi}\left(e_{i}^{t} \mid \tau_{i}^{t-1}, \boldsymbol{o}^{t},\boldsymbol{a}^{t}\right) d e_{i}^{t}\right]\\
        &\qquad\qquad+\mathbb{E}_{\boldsymbol{o}^{t},\boldsymbol{a}^{t}}\left[H\left(e_{i}^{t} \mid \boldsymbol{o}^{t},\boldsymbol{a}^{t}\right)\right].
    \end{aligned}
\end{equation}
The role encoder is conditioned on the joint observations and actions, so given the observations and actions, the distributions of roles, $p\left(e_{i}^{t}\right)$, are independent from the local histories. 
Thus, we have
\begin{equation}
    \begin{aligned}
        I\left(e_{i}^{t} ; \tau_{i}^{t-1} \mid \boldsymbol{o}^{t},\boldsymbol{a}^{t}\right) &\geq -\mathbb{E}_{\tau_{i}^{t-1}, \boldsymbol{o}^{t},\boldsymbol{a}^{t}}\left[\mathcal{C E}\left[p\left(e_{i}^{t} \mid \boldsymbol{o}^{t},\boldsymbol{a}^{t}\right) \| q_{\phi}\left(e_{i}^{t} \mid \tau_{i}^{t-1}, \boldsymbol{o}^{t},\boldsymbol{a}^{t}\right)\right]\right] \\
        &\qquad+ \mathbb{E}_{\boldsymbol{o}^{t},\boldsymbol{a}^{t}}\left[H\left(e_{i}^{t} \mid \boldsymbol{o}^{t},\boldsymbol{a}^{t}\right)\right].
    \end{aligned}
\end{equation}
In SVO-based role emergence, we use orientation functions to act as the role encoder. 
Therefore, the above formula can be transformed into
\begin{equation}
    \begin{aligned}
        I\left(e_{i}^{t} ; \tau_{i}^{t-1} \mid \boldsymbol{o}^{t},\boldsymbol{a}^{t}\right) &\geq -\mathbb{E}_{\tau_{i}^{t-1}, \boldsymbol{o}^{t},\boldsymbol{a}^{t}}\left[\mathcal{C E}\left[W_{\boldsymbol{\eta}}\left(e_{i}^{t} \mid \boldsymbol{o}^{t},\boldsymbol{a}^{t}\right) \| q_{\phi}\left(e_{i}^{t} \mid \tau_{i}^{t-1}, \boldsymbol{o}^{t},\boldsymbol{a}^{t}\right)\right]\right] \\
        &\qquad+ \mathbb{E}_{\boldsymbol{o}^{t},\boldsymbol{a}^{t}}\left[H_{\boldsymbol{\eta}}\left(e_{i}^{t} \mid \boldsymbol{o}^{t},\boldsymbol{a}^{t}\right)\right].
    \end{aligned}
\end{equation}
In practice, we use a replay buffer $\mathcal{D}$ and minimize the following summation of all agents
\begin{equation}
    \begin{aligned}
        &\operatorname{L}^{\mathrm{mi}}\left(\boldsymbol{\tau};\boldsymbol{\eta},\phi\right) 
        = \frac{1}{n}\sum_{i=1}^{N}\operatorname{L}_{i}^{\mathrm{mi}}\left(\tau_i;\boldsymbol{\eta},\phi\right)\\
        :=&\mathbb{E}_{\left(\tau_{i}^{t-1}, \boldsymbol{o}^{t},\boldsymbol{a}^{t}\right) \sim \mathcal{D}}\left[\mathcal{C} \mathcal{E}\left[W_{\boldsymbol{\eta}}\left(e_{i}^{t} \mid \boldsymbol{o}^{t},\boldsymbol{a}^{t}\right) \| q_{\phi}\left(e_{i}^{t} \mid \tau_{i}^{t-1}, \boldsymbol{o}^{t},\boldsymbol{a}^{t}\right)\right]-H_{\boldsymbol{\eta}}\left(e_{i}^{t} \mid \boldsymbol{o}^{t},\boldsymbol{a}^{t}\right)\right] \\
        =& \mathbb{E}\left[D_{\mathrm{KL}}\left[W_{\boldsymbol{\eta}}\left(e_{i}^{t}\mid \boldsymbol{o}^{t},\boldsymbol{a}^{t}\right) \| q_{\phi}\left(e_{i}^{t} \mid \tau_{i}^{t-1}, \boldsymbol{o}^{t},\boldsymbol{a}^{t}\right)\right]\right],
    \end{aligned}
\end{equation}
In addition, in the implementation, the orientation functions model $p$ as a normal distribution and output the mean and variance of the SVO-based role embedding, which can be sampled from the corresponding normal distribution. 
However, in the training process of role emergence, we only use the mean of the output of orientation functions.

\subsection{Practice Implementation}

From the derivation of (\ref{eq:three-parts}), it can be seen that the gradient calculation involves the calculation of multi-part partial derivatives and complex matrix-vector multiplication.
Alternatively, one may rely on automatic differentiation in modern machine learning frameworks~\citep{tensorflow} to compute the chain rule (\ref{eq:three-parts}) via direct minimization of a loss function.
This loss function is derived as follows:

\begin{equation}
    \begin{aligned}
        &\nabla_{\eta_i} J_{i}^{\mathrm{svo}}\left(\hat{\tau}_{i}, \hat{\boldsymbol{\theta}}, \boldsymbol{\eta}, \eta_i\right) = \nabla_{\eta_i} \tilde{V}_{i}^{\boldsymbol{\hat{\pi}}}\left(\hat{s}_{0},\eta_i\right) = \nabla_{\eta_i} \sum_{\mathbf{\hat{a}}} \boldsymbol{\hat{\pi}}\left(\mathbf{\hat{a}} \mid \hat{s}_{0}, \boldsymbol{\hat{e}}(\boldsymbol\eta)\right) \tilde{Q}_{i}^{\boldsymbol{\hat{\pi}}}\left(\hat{s}_{0}, \mathbf{\hat{a}}, \eta_i\right)\\
        =& \sum_{\mathbf{\hat{a}}} \left(\left(\nabla_{\eta_i} \boldsymbol{\hat{\pi}}_{\boldsymbol{\hat{\theta}}}\right) \tilde{Q}_{i}^{\boldsymbol{\hat{\pi}}}\left(\hat{s}_{0}, \mathbf{\hat{a}}, \eta_i\right) + \boldsymbol{\hat{\pi}}_{\boldsymbol{\hat{\theta}}} \nabla_{\eta_i} \tilde{Q}_{i}^{\boldsymbol{\hat{\pi}}}\left(\hat{s}_{0}, \mathbf{\hat{a}}, \eta_i\right)\right) \\
        =& \sum_{\mathbf{\hat{a}}} \left(\left(\nabla_{\eta_i} \boldsymbol{\hat{\pi}}_{\boldsymbol{\hat{\theta}}}\right) \tilde{Q}_{i}^{\boldsymbol{\hat{\pi}}}\left(\hat{s}_{0}, \mathbf{\hat{a}}, \eta_i\right) + \boldsymbol{\hat{\pi}}_{\boldsymbol{\hat{\theta}}} \nabla_{\eta_i} \left(\hat{r}_{i}-\right.\right.\\
        &\quad\quad\quad\left.\left.\alpha\|W^{i}_{\eta_i}-W_k^{i}\|^{2}_{2}+\gamma \sum_{\hat{s}^{\prime}} P\left(\hat{s}^{\prime} \mid \hat{s}_{0}, \mathbf{\hat{a}}\right) \tilde{V}_{i}^{\boldsymbol{\pi}}\left(\hat{s}^{\prime},\eta_i\right)\right)\right) \\
        =& \sum_{\mathbf{\hat{a}}} \left(\left(\nabla_{\eta_i} \boldsymbol{\hat{\pi}}_{\boldsymbol{\hat{\theta}}}\right) \tilde{Q}_{i}^{\boldsymbol{\hat{\pi}}}\left(\hat{s}_{0}, \mathbf{\hat{a}}, \eta_i\right) - 2\boldsymbol{\hat{\pi}}_{\boldsymbol{\hat{\theta}}}\alpha\nabla_{\eta_i}W^{i}_{\eta_i}\left(W^{i}_{\eta_i}-W_k^{i}\right) + \right.\\
        &\quad\quad\quad\left.\gamma\boldsymbol{\hat{\pi}}_{\boldsymbol{\hat{\theta}}} \sum_{\hat{s}^{\prime}} P\left(\hat{s}^{\prime} \mid \hat{s}_{0}, \mathbf{\hat{a}}\right) \nabla_{\eta_i}\tilde{V}_{i}^{\boldsymbol{\pi}}\left(\hat{s}^{\prime},\eta_i\right)\right) \\
        =& \sum_{x} \sum_{k=0}^{T} P\left(\hat{s}_{0} \rightarrow x, k, \boldsymbol{\hat{\pi}}\right) \gamma^{k} \sum_{\mathbf{\hat{a}}} \left(\nabla_{\eta_i} \boldsymbol{\hat{\pi}}_{\boldsymbol{\hat{\theta}}} \tilde{Q}_{i}^{\boldsymbol{\hat{\pi}}}(x, \mathbf{\hat{a}}, \eta_{i})-2\boldsymbol{\hat{\pi}}_{\boldsymbol{\hat{\theta}}}\alpha\nabla_{\eta_i}W^{i}_{\eta_i}\left(W^{i}_{\eta_i}-W_k^{i}\right)\right) \\
        =& \sum_{\hat{s}} d^{\boldsymbol{\hat{\pi}}}(\hat{s})\sum_{\mathbf{\hat{a}}} \left(\nabla_{\eta_i} \boldsymbol{\hat{\pi}}_{\boldsymbol{\hat{\theta}}} \tilde{Q}_{i}^{\boldsymbol{\hat{\pi}}}(\hat{s}, \mathbf{\hat{a}}, \eta_{i})-2\boldsymbol{\hat{\pi}}_{\boldsymbol{\hat{\theta}}}\alpha\nabla_{\eta_i}W^{i}_{\eta_i}\left(W^{i}_{\eta_i}-W_k^{i}\right)\right)\\
        =& \sum_{\hat{s}} d^{\boldsymbol{\hat{\pi}}}(\hat{s})\sum_{\mathbf{\hat{a}}} \boldsymbol{\hat{\pi}}_{\boldsymbol{\hat{\theta}}} \left(\nabla_{\eta_i} \log \boldsymbol{\hat{\pi}}_{\boldsymbol{\hat{\theta}}} \tilde{Q}_{i}^{\boldsymbol{\hat{\pi}}}(\hat{s}, \mathbf{\hat{a}}, \eta_{i})-2\alpha\nabla_{\eta_i}W^{i}_{\eta_i}\left(W^{i}_{\eta_i}-W_k^{i}\right)\right) \\
        =&\mathbb{E}_{\hat{\boldsymbol{\pi}}}\left[\left(\sum_{j=1}^{N}\nabla_{\boldsymbol{\eta}} \log \hat{\pi}^{j}_{\hat{\theta}_{j}}\left(\hat{a}_{j} \mid \hat{o}_{j}, \hat{e}_{j}(\boldsymbol{\eta})\right)\right) \tilde{Q}^{i, \hat{\boldsymbol{\pi}}}(\hat{s}, \hat{\mathbf{a}}, \eta_i)-2\alpha\nabla_{\eta_i}W^{i}_{\eta_i}\left(W^{i}_{\eta_i}-W_k^{i}\right)\right].
    \end{aligned}
\end{equation}
Hence descending a stochastic estimate of this gradient is equivalent to minimizing the following loss:
\begin{equation}
-\sum_{t=0}^{T}\sum_{j=1}^{N} \log \pi^{j}_{\hat{\theta}^{j}}\left(\hat{a}^{t}_{j} \mid \hat{o}^{t}_{j}, \hat{e}^{t}_{j}(\boldsymbol{\eta})\right) \sum_{\ell=t}^{T} \gamma^{\ell-t} \left(\hat{r}_i^{\ell}-\alpha\|\Delta^{i,\ell}(W,k)\|^{2}_{2}\right)-2\alpha\nabla_{\eta_i}W^{i,t}_{\eta_i}\Delta^{i,t}(W,k),
\end{equation}
where $\Delta^{i,\ell}(W,k):=W^{i,\ell}_{\eta_i}-W_k^{i,\ell}$ and $\Delta^{i,t}(W,k):=W^{i,t}_{\eta_i}-W_k^{i,t}$.

\section{Analysis in Iterated Prisoner’s Dilemma}\label{sec:ana-ipd}

\begin{table}[htb!]
\centering
\caption{Prisoner’s Dilemma.}
\label{tab:pd}
\resizebox{0.4\textwidth}{!}{%
\begin{tabular}{|c|c|c|}
\hline
A1/A2 & C & D \\ \hline
C & $(-1, -1)$ & $(-3, 0)$ \\ \hline
D & $(0, -3)$ & $(-2, -2)$ \\ \hline
\end{tabular}%
}
\end{table}

\begin{figure}[htb!]
    \centering
    \begin{subfigure}[b]{0.32\textwidth}
        \includegraphics[width=\textwidth]{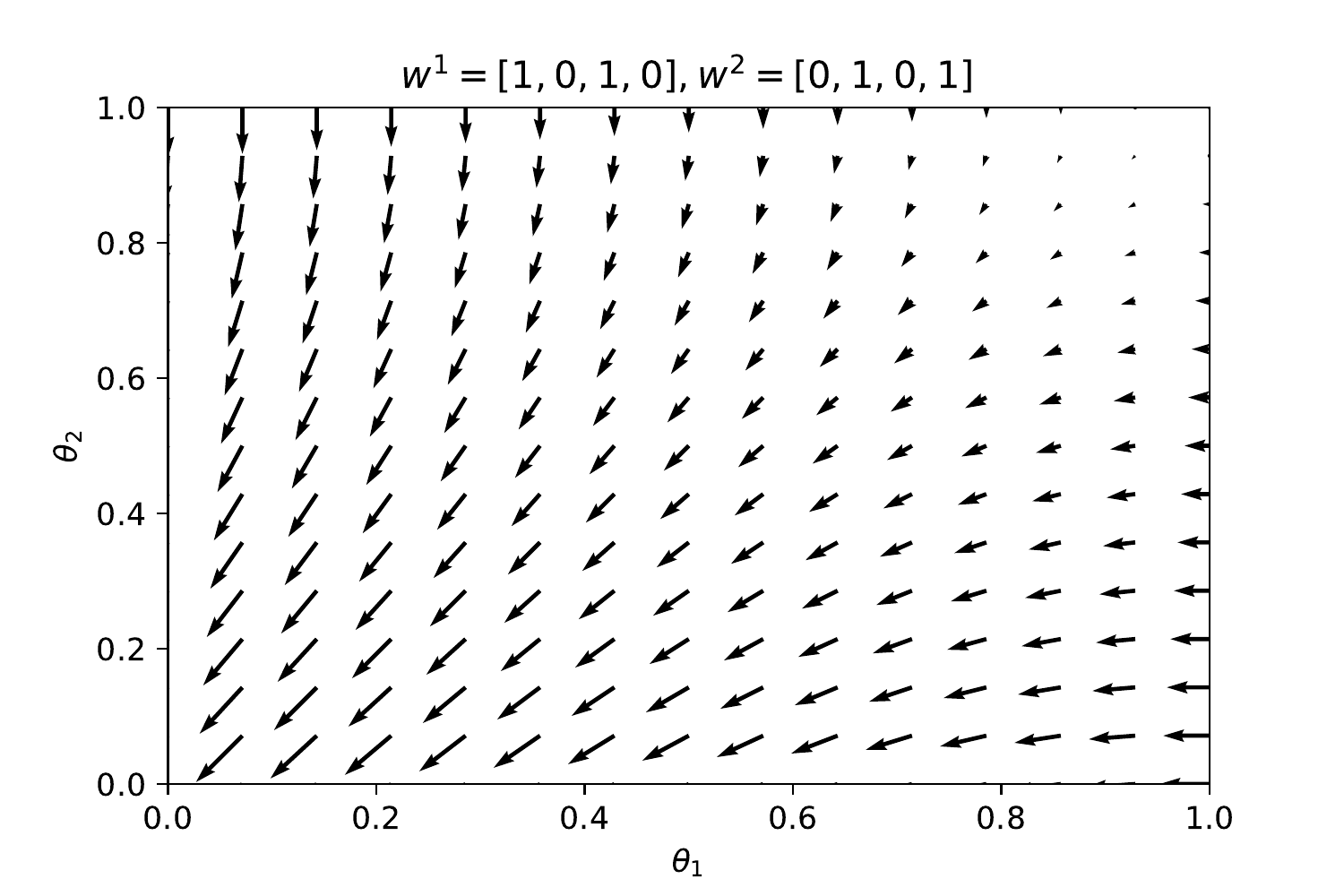}
        \caption{$1$-th round update.}
        \label{fig:p0-m}
    \end{subfigure}
    \begin{subfigure}[b]{0.32\textwidth}
        \includegraphics[width=\textwidth]{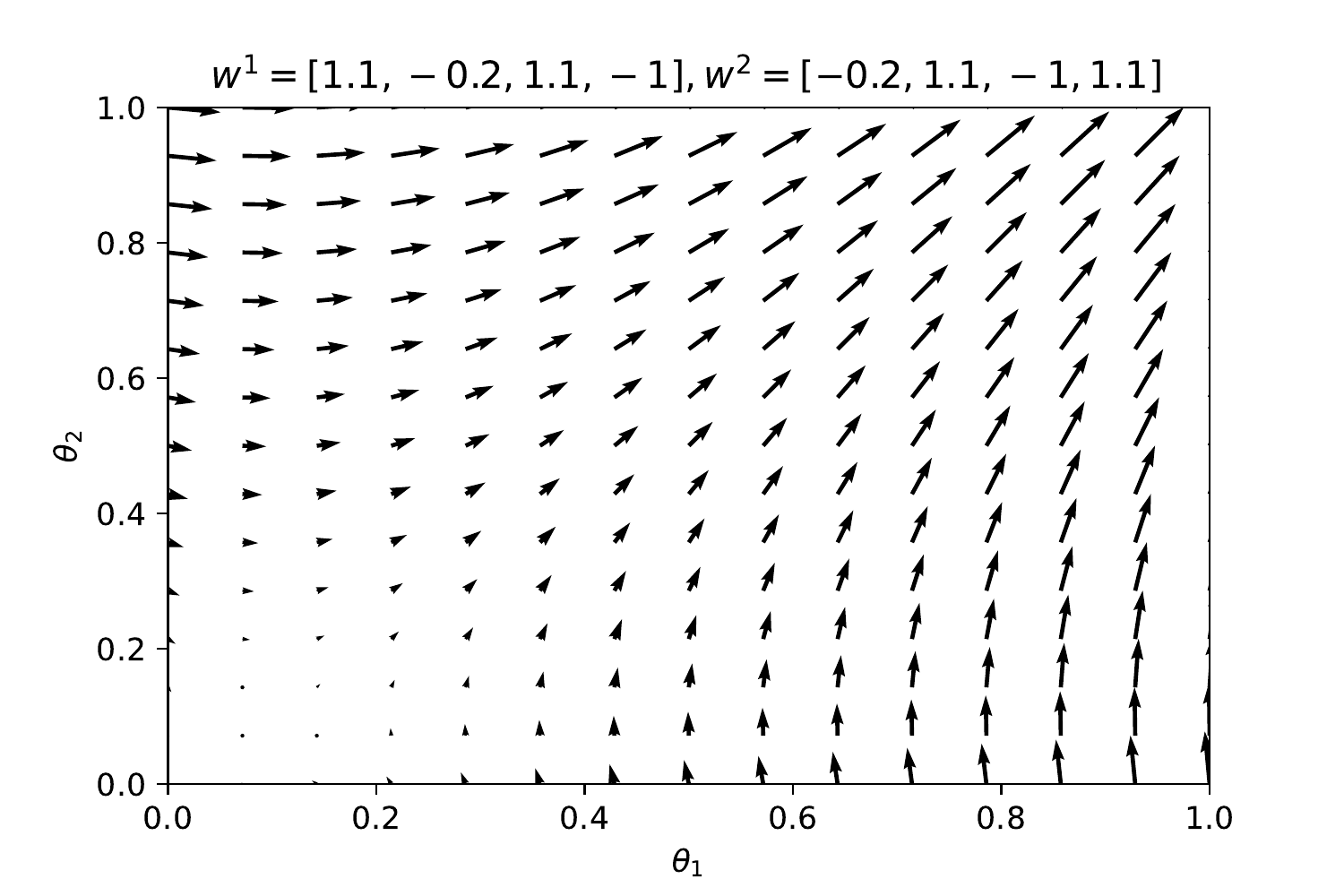}
        \caption{$2$-th round update.}
        \label{fig:p2-m}
    \end{subfigure}
    \begin{subfigure}[b]{0.32\textwidth}
        \includegraphics[width=\textwidth]{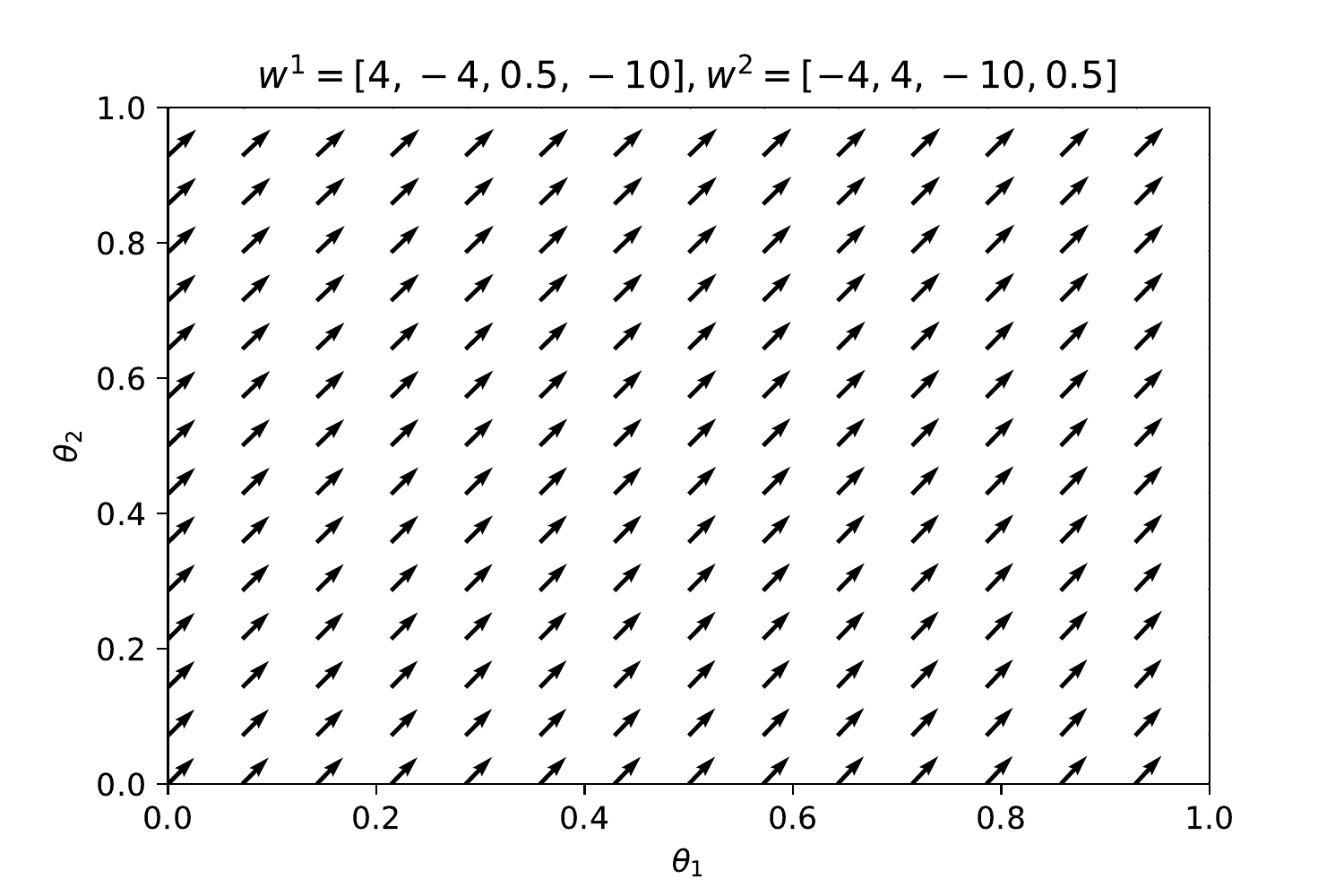}
        \caption{$k$-th round update.}
        \label{fig:p3-m}
    \end{subfigure}
    \caption{Vector fields of the cooperation probability of each agent at the different rounds.}
    \label{fig:p-change-m}
\end{figure}

Similar to~\citet{yang2020learning}, we conduct a complete analysis (Appendix~\ref{sec:ana-ipd}) using closed-form gradient descent without policy gradient approximation in repeated matrix games.
In the stateless Iterated Prisoner’s Dilemma (IPD), for example, with payoff matrix in Table~\ref{tab:pd}.
Figure~\ref{fig:p-change-m} shows the vector field of the two agents' cooperation probabilities after setting the probability of the two agents taking a cooperative action at 0.5 and updating the policy using RESVO. 
As can be seen from the figure, RESVO can converge to mutual cooperation.

\begin{proposition}
RESVO converges to mutual cooperation in the Iterated Prisoner’s Dilemma.
\end{proposition}

\begin{proof}

We prove this by deriving closed-form expressions for the updates to parameters of policies and incentive functions.
Let $\theta_{i}$ for $i \in\{1,2\}$ denote each agent's probability of taking the cooperative action. 
Let $w^{1}:=\left[w^{11}, w^{12}\right]$, $w^{11}:=\left[w_{C}^{11}, w_{D}^{11}\right] \in \mathbb{R}^{2}$ and $w^{12}:=\left[w_{C}^{12}, w_{D}^{12}\right] \in \mathbb{R}^{2}$ denote agent $1$'s orientation function, where the ratios are given to itself and agent $2$ respectively when agent $2$ takes action $a_{2}=C$ or $a_{2}=D$. 
Similarly, let $w^{1}$, $w^{21}$ and $w^{22}$ denote agent $2$'s orientation function. 
The value function for each agent is defined by
\begin{equation}
    \begin{aligned}
        V_{i}\left(\theta_1, \theta_2\right) &= \sum_{t=0}^{T} \gamma^{t} p^{T} r_{i}\approx\frac{1}{1-\gamma} p^{T} r_{i}, \\
        \text{where } p &= \left[\theta_1 \theta_2, \theta_1\left(1-\theta_2\right),\left(1-\theta_1\right) \theta_2,\left(1-\theta_1\right)\left(1-\theta_2\right)\right].
    \end{aligned}
\end{equation}
The total reward received by each agent is
\begin{equation}
    \begin{aligned}
        r_{1} &=\left[-w_{C}^{11}-w_{C}^{21},-3w_{D}^{11}+0, 0-3w_{D}^{21},-2w_{D}^{11}-2w_{D}^{21}\right], \\
        r_{2} &=\left[-w_{C}^{22}-w_{C}^{12}, 0-3w_{D}^{12},-3w_{D}^{22}+0,-2w_{D}^{22}-2w_{D}^{12}\right].
    \end{aligned}
\end{equation}
Agent $2$ updates its policy via the update
\begin{equation}\label{eq:p-update1-a}
    \begin{aligned}
        \hat{\theta}_{2}=& \theta_{2}+\zeta \nabla_{\theta_{2}} V_{2}\left(\theta^{1}, \theta^{2}\right) \\
        =& \theta_{2}+\frac{\zeta}{1-\gamma} \nabla_{\theta_{2}}\left(\theta_{1} \theta_{2}\left(-w_{C}^{22}-w_{C}^{12}\right)-\theta_{1}\left(1-\theta_{2}\right) 3w_{D}^{12}\right.\\
        &\left.-\left(1-\theta_{1}\right) \theta_{2}3w_{D}^{22}+\left(1-\theta_{1}\right)\left(1-\theta_{2}\right)\left(-2w_{D}^{22}-2w_{D}^{12}\right)\right) \\
        =& \theta_{2}+\frac{\zeta}{1-\gamma} \nabla_{\theta_{2}}\left[\left(w_{D}^{12}+w_{D}^{22}-w_{C}^{12}-w_{C}^{22}\right)\theta_1\theta_2 +\left(2w_{D}^{22}-w_{D}^{12}\right)\theta_1 + \right. \\ 
        &\left.\qquad\left(2w_{D}^{12}-w_{D}^{22}\right)\theta_2 - 2\left(w_{D}^{12}+w_{D}^{22}\right)\right] \\
        =& \theta_{2}+\frac{\zeta}{1-\gamma} \left[\left(3w_{D}^{22}-w_{C}^{12}-w_{C}^{22}\right)\theta_1 - 2\left(w_{D}^{12}+w_{D}^{22}\right)\right].
    \end{aligned}
\end{equation}
And likewise, for agent $1$:
\begin{equation}\label{eq:p-update2-a}
    \begin{aligned}
        \hat{\theta}_{1}=& \theta_{1}+\zeta \nabla_{\theta_{1}} V_{1}\left(\theta^{1}, \theta^{2}\right) \\
        =& \theta_{1}+\frac{\zeta}{1-\gamma} \nabla_{\theta_{1}}\left(\theta_{1} \theta_{2}\left(-w_{C}^{11}-w_{C}^{21}\right)-\theta_{1}\left(1-\theta_{2}\right) 3w_{D}^{11}\right.\\
        &\left.-\left(1-\theta_{1}\right) \theta_{2}3w_{D}^{21}+\left(1-\theta_{1}\right)\left(1-\theta_{2}\right)\left(-2w_{D}^{11}-2w_{D}^{21}\right)\right) \\
        =& \theta_{1}+\frac{\zeta}{1-\gamma} \nabla_{\theta_{1}}\left[\left(w_{D}^{11}+w_{D}^{21}-w_{C}^{11}-w_{C}^{21}\right)\theta_1\theta_2 + \left(2w_{D}^{21}-w_{D}^{11}\right)\theta_1 + \right.\\ 
        &\left.\qquad\left(2w_{D}^{11}-w_{D}^{21}\right)\theta_2 - 2\left(w_{D}^{11}+w_{D}^{21}\right)\right]\\
        =& \theta_{1}+\frac{\zeta}{1-\gamma} \left[\left(3w_{D}^{11}-w_{C}^{11}-w_{C}^{21}\right)\theta_2 - 2\left(w_{D}^{11}+w_{D}^{21}\right)\right].
    \end{aligned}
\end{equation}
Let $\hat{p}$ denote the joint action probability under updated policies $\hat{\theta}_{1}$ and $\hat{\theta}_{2}$, let $\Delta_{2}:=\left[\left(3w_{D}^{22}-w_{C}^{12}-w_{C}^{22}\right)\theta_1 - 2\left(w_{D}^{12}+w_{D}^{22}\right)\right]\zeta /(1-\gamma)$ denote agent $2$'s policy update and let $\Delta_{1}:=\left[\left(3w_{D}^{11}-w_{C}^{11}-w_{C}^{21}\right)\theta_2 - 2\left(w_{D}^{11}+w_{D}^{21}\right)\right]\zeta /(1-\gamma)$ denote agent $1$'s policy update.
Agent $1$ updates its orientation function parameters via
\begin{equation}\label{eq:svo-update1-a}
    \begin{aligned}
        w^{1} & \leftarrow w^{1}+\beta \nabla_{w^{1}} \frac{1}{1-\gamma} \hat{p}^{T} r^{\text{env}}_1 \\
        =& w^{1}+\frac{\beta}{1-\gamma} \nabla_{w^{1}}\left[-\left(\theta_{1}+\Delta_{1}\right)\left(\theta_{2}+\Delta_{2}\right)-3\left(\theta_{1}+\Delta_{1}\right)\left(1-\theta_{2}-\Delta_{2}\right)\right.\\
        &\left.-2\left(1-\theta_{1}-\Delta_{1}\right)\left(1-\theta_{2}-\Delta_{2}\right)\right] \\
        =& w^{1}+\frac{\beta}{1-\gamma} \nabla_{w^{1}}\left[5\theta_1-\Delta_1-2+2\theta_2+2\Delta_2\right] \\
        =& w^{1}+\frac{\beta}{1-\gamma} \left[
        \begin{array}{l}
            \nabla_{w_{C}^{11}}\left(-\Delta_1\right)\\
            \nabla_{w_{C}^{12}}\left(2\Delta_2\right)\\
            \nabla_{w_{D}^{11}}\left(-\Delta_1\right)\\
            \nabla_{w_{D}^{12}}\left(2\Delta_2\right)
        \end{array}
        \right]
        = w^{1}+\frac{\beta}{1-\gamma} \left[
        \begin{array}{l}
            \frac{\zeta}{1-\gamma}\theta_2\\
            -\frac{2\zeta}{1-\gamma}\theta_1\\
            -\frac{\zeta}{1-\gamma}\left(3\theta_2-2\right)\\
            -\frac{4\zeta}{1-\gamma}\\
        \end{array}
        \right] \\
        =& w^{1}+\frac{\zeta \beta}{(1-\gamma)^{2}} \left[
        \begin{array}{l}
            \theta_2\\
            -2\theta_1\\
            -(3\theta_2-2)\\
            -4\\
        \end{array}
        \right].
    \end{aligned}
\end{equation}
By symmetry, agent $2$ updates its orientation function parameters via
\begin{equation}\label{eq:svo-update2-a}
    \begin{aligned}
        w^{2} & \leftarrow w^{2}+\beta \nabla_{w^{2}} \frac{1}{1-\gamma} \hat{p}^{T} r^{\text{env}}_2 \\
        =& w^{2}+\frac{\beta}{1-\gamma} \nabla_{w^{2}}\left[-\left(\theta_{1}+\Delta_{1}\right)\left(\theta_{2}+\Delta_{2}\right)-3\left(1-\theta_{1}-\Delta_{1}\right)\left(\theta_{2}+\Delta_{2}\right)\right.\\
        &\left.-2\left(1-\theta_{1}-\Delta_{1}\right)\left(1-\theta_{2}-\Delta_{2}\right)\right] \\
        =& w^{2}+\frac{\beta}{1-\gamma} \nabla_{w^{2}}\left[-\theta_2-\Delta_2-2+2\theta_1+2\Delta_1\right] \\
        =& w^{2}+\frac{\beta}{1-\gamma} \left[
        \begin{array}{l}
            \nabla_{w_{C}^{21}}\left(2\Delta_1\right)\\
            \nabla_{w_{C}^{22}}\left(-\Delta_2\right)\\
            \nabla_{w_{D}^{21}}\left(2\Delta_1\right)\\
            \nabla_{w_{D}^{22}}\left(-\Delta_2\right)
        \end{array}
        \right]
        = w^{2}+\frac{\beta}{1-\gamma} \left[
        \begin{array}{l}
            -\frac{2\zeta}{1-\gamma}\theta_2\\
            \frac{\zeta}{1-\gamma}\theta_1\\
            -\frac{4\zeta}{1-\gamma}\\
            -\frac{\zeta}{1-\gamma}(3\theta_1-2)\\
        \end{array}
        \right] \\
        =& w^{2}+\frac{\zeta \beta}{(1-\gamma)^{2}} \left[
        \begin{array}{l}
            -2\theta_2\\
            \theta_1\\
            -4\\
            -(3\theta_1-2)\\
        \end{array}
        \right].
    \end{aligned}
\end{equation}

It can be seen from (\ref{eq:svo-update1-a}) and (\ref{eq:svo-update2-a}) that in the Prisoner's Dilemma shown in Table~\ref{tab:pd}, except for $w_{C}^{11}$ and $w_{C}^{22}$, which are constantly increasing, other terms are constantly decreasing. 
This ensures that $\theta_1$ and $\theta_2$ in (\ref{eq:p-update1-a}) and (\ref{eq:p-update2-a}) will eventually continue to increase, that is, the two agents converge to mutual cooperation. 
Of course, in the above derivation, we ignore the low-rank and mutual information constraints on $w^1$ and $w^2$. 
We believe that as long as the parameters are adjusted reasonably so that the direction of $w^1, w^2$ update is consistent with the above derivation process, the algorithm can be guaranteed to converge to mutual cooperation.

\end{proof}

\begin{figure}[htb!]
    \centering
    \begin{subfigure}[b]{0.45\textwidth}
        \includegraphics[width=\textwidth]{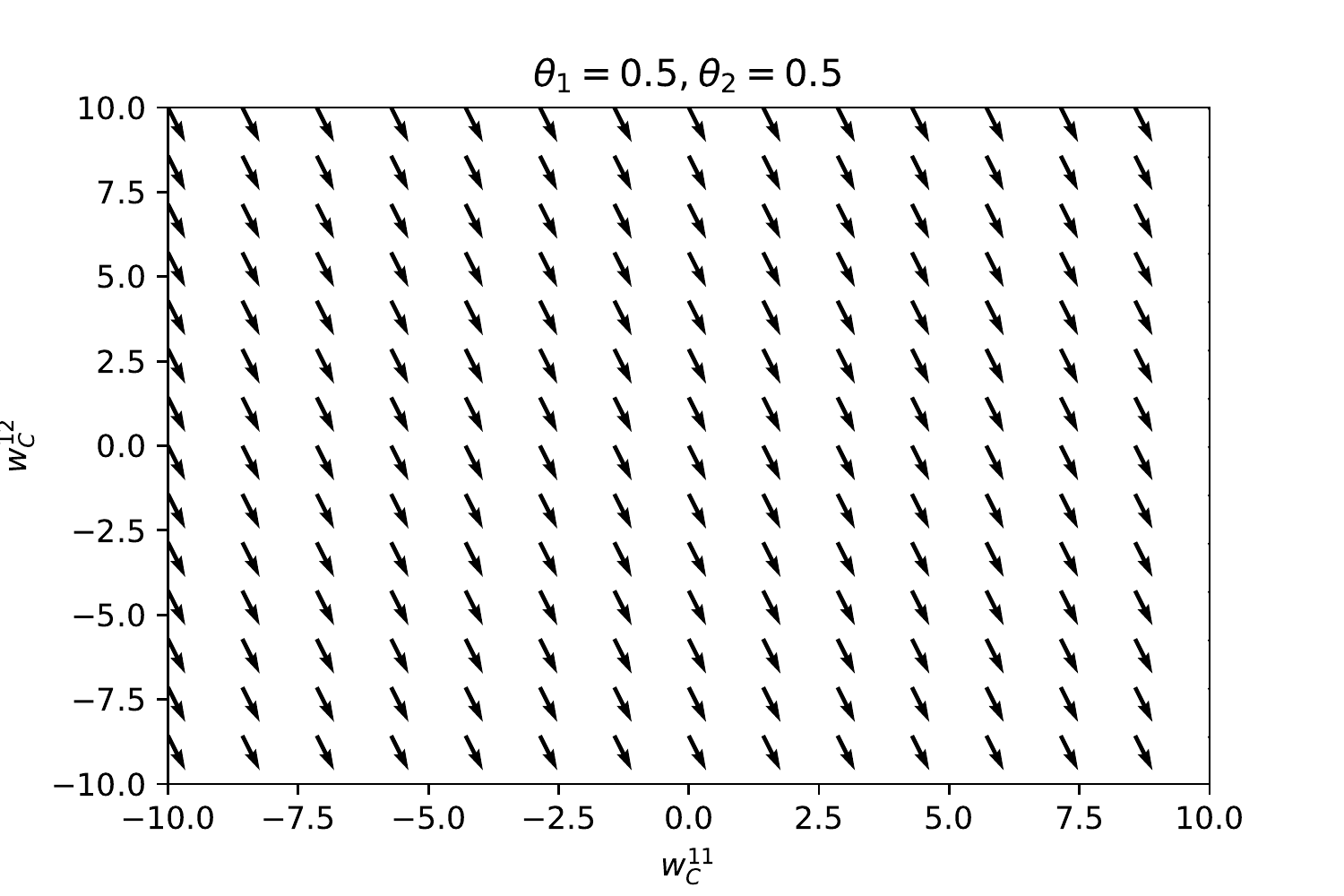}
        \caption{The vector fields of $w_{C}^{11}$ and $w_{C}^{12}$.}
        \label{fig:wc1-0505}
    \end{subfigure}
    \begin{subfigure}[b]{0.45\textwidth}
        \includegraphics[width=\textwidth]{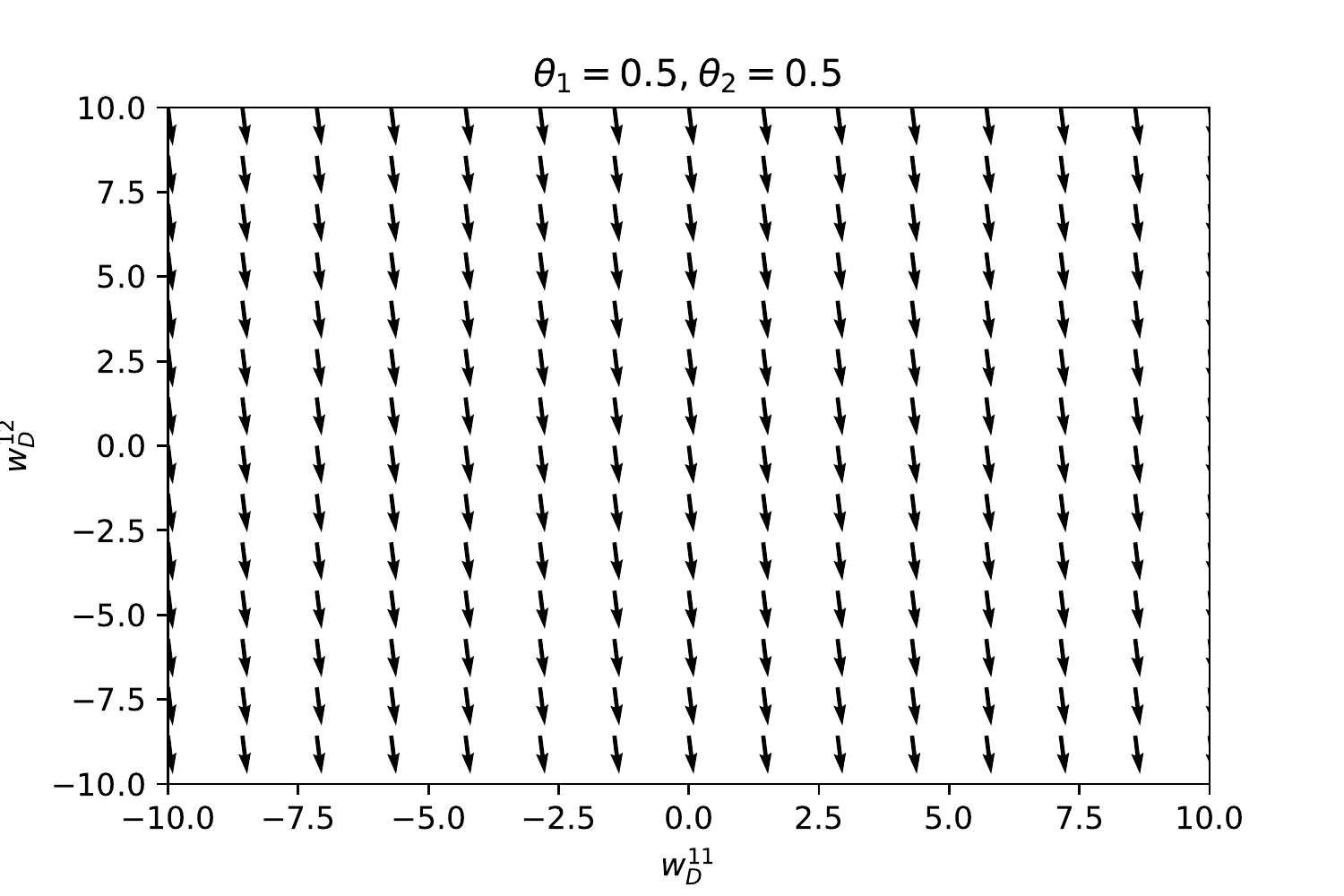}
        \caption{The vector fields of $w_{D}^{11}$ and $w_{D}^{12}$.}
        \label{fig:wd1-0505}
    \end{subfigure}
    \begin{subfigure}[b]{0.45\textwidth}
        \includegraphics[width=\textwidth]{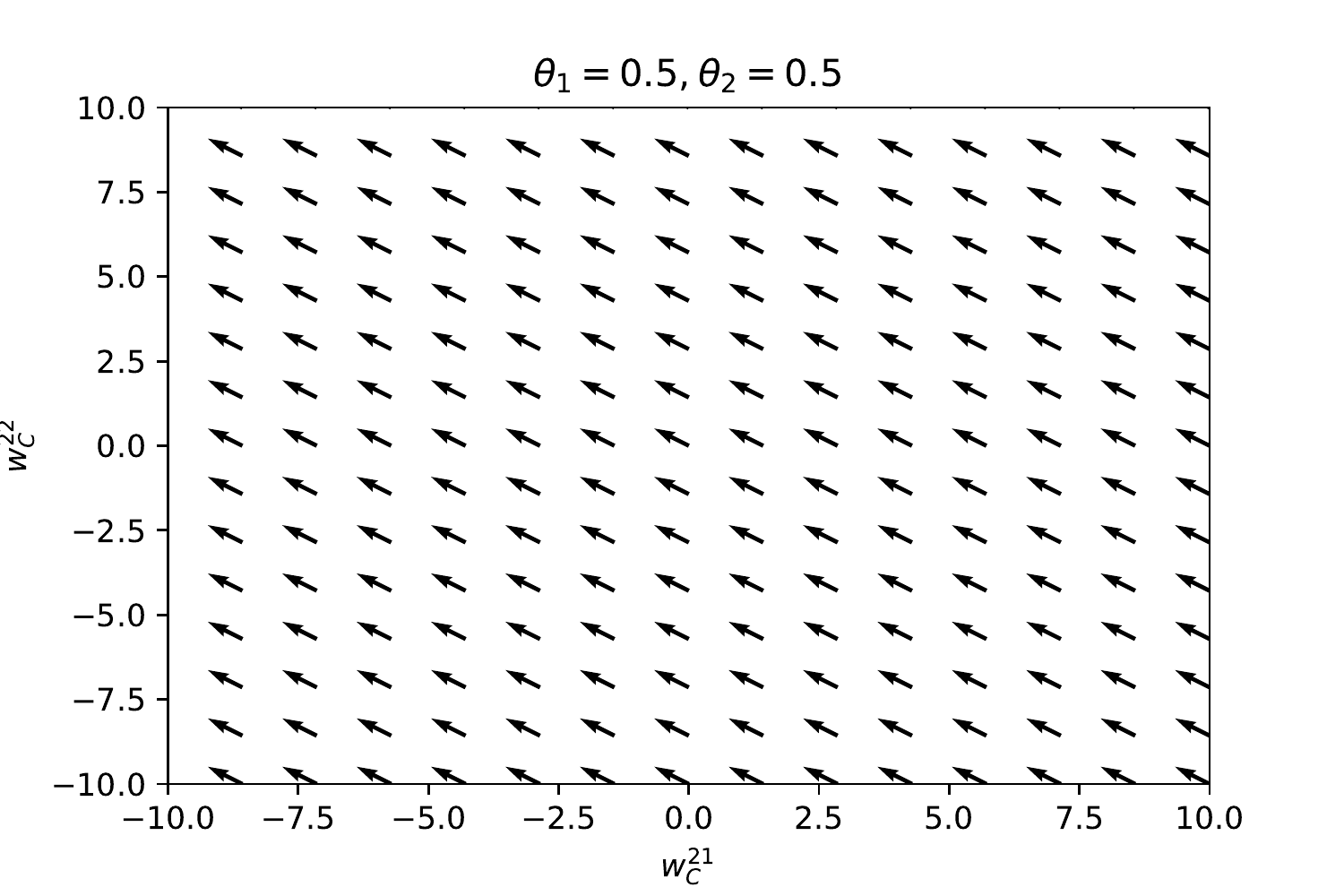}
        \caption{The vector fields of $w_{C}^{21}$ and $w_{C}^{22}$.}
        \label{fig:wc2-0505}
    \end{subfigure}
    \begin{subfigure}[b]{0.45\textwidth}
        \includegraphics[width=\textwidth]{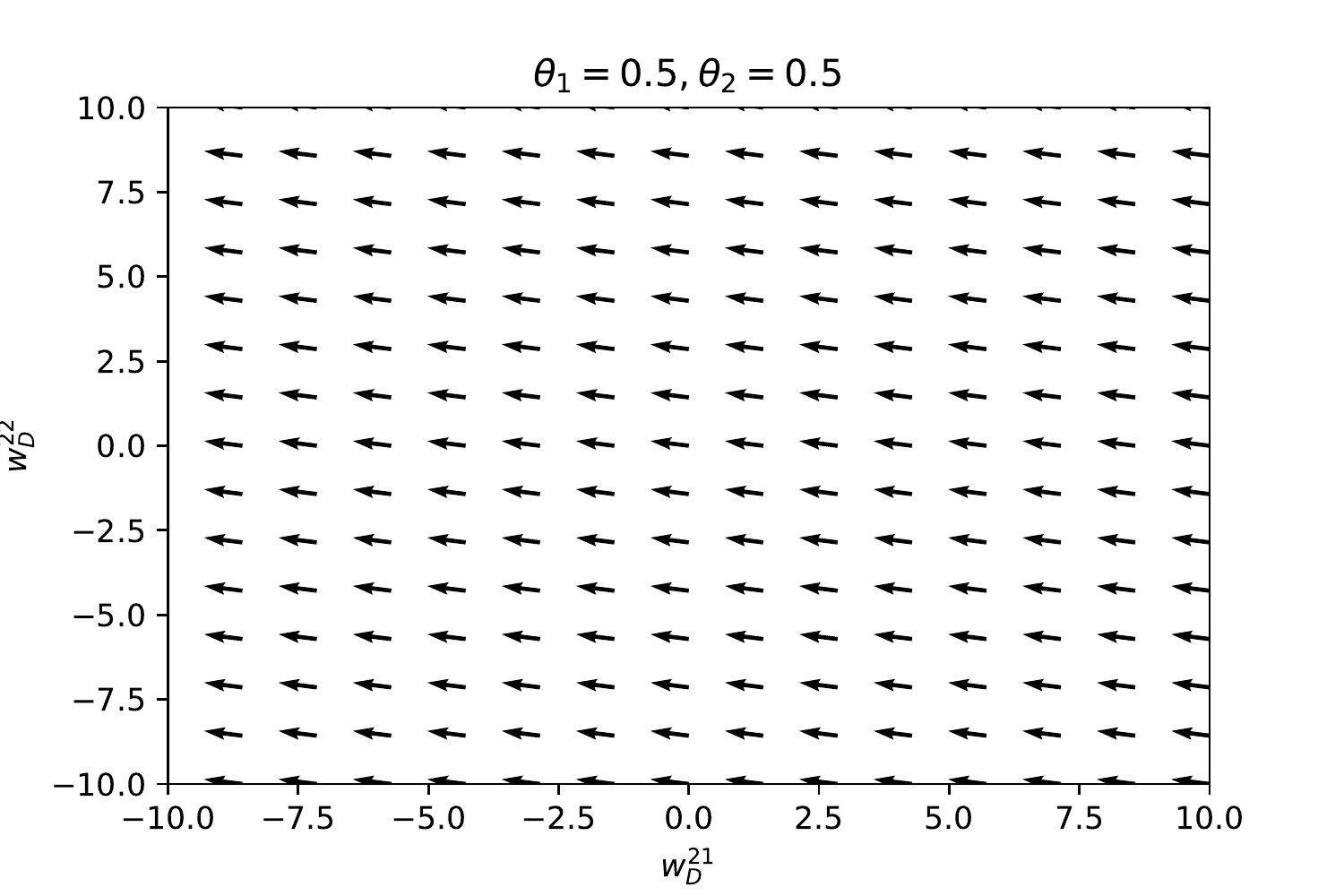}
        \caption{The vector fields of $w_{D}^{21}$ and $w_{D}^{22}$.}
        \label{fig:wd2-0505}
    \end{subfigure}
    \caption{Vector fields of orientation parameters of each agent with $\theta_1=0.5, \theta_2=0.5$.}
    \label{fig:w-change-1-a}
\end{figure}

Below we visualize the updating direction of each parameter in the form of the vector field.
We initialize $\theta_1=0.5, \theta_2=0.5, w^1=[1, 0, 1, 0], w^2=[0, 1, 0, 1]$.
After fixing the values of $\theta_1, \theta_2$, the change of the orientation parameters of each agent is shown in Figure~\ref{fig:w-change-1-a}.

\begin{figure}[htb!]
    \centering
    \begin{subfigure}[b]{0.45\textwidth}
        \includegraphics[width=\textwidth]{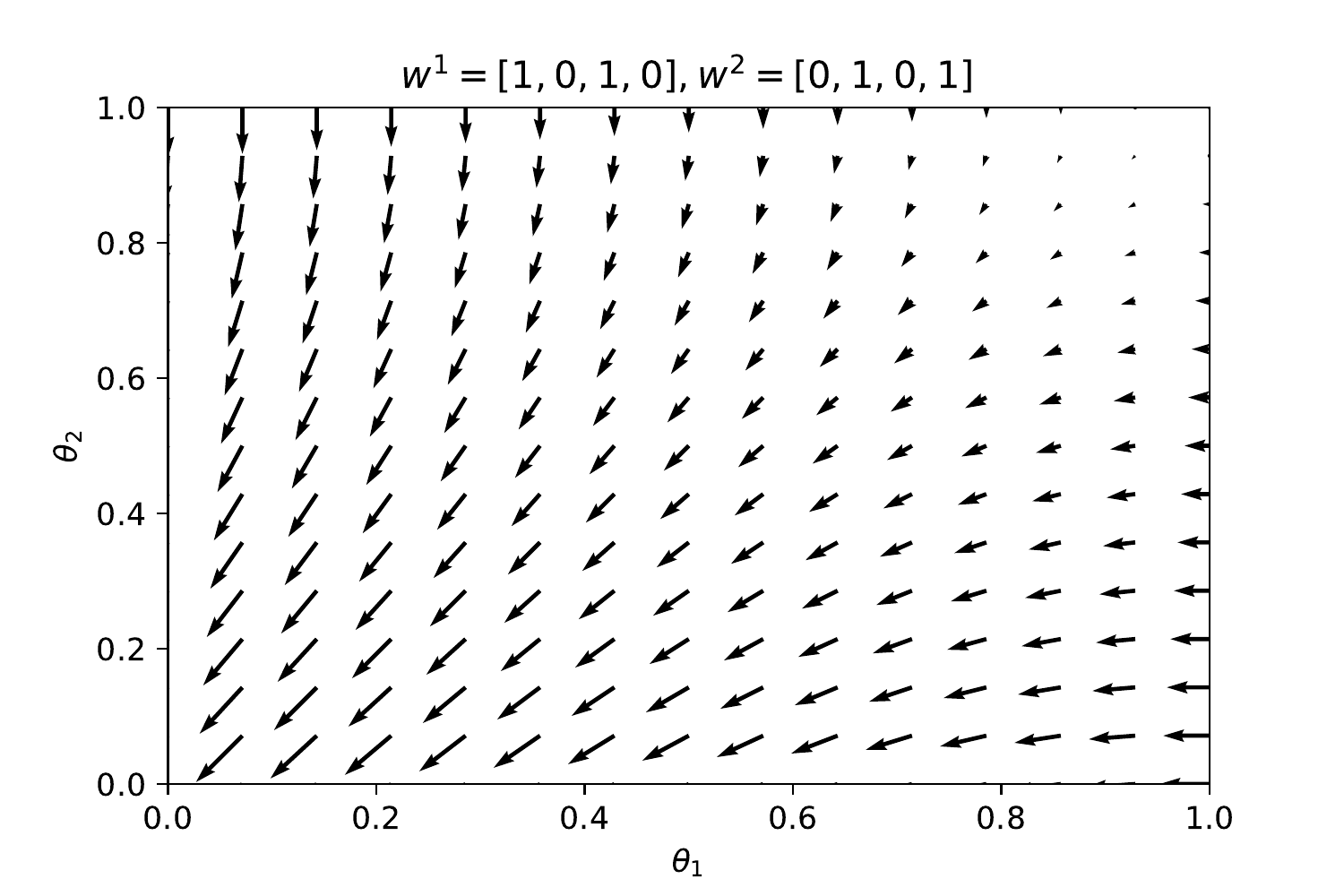}
        \caption{Before the orientation parameters update.}
        \label{fig:p0}
    \end{subfigure}
    \begin{subfigure}[b]{0.45\textwidth}
        \includegraphics[width=\textwidth]{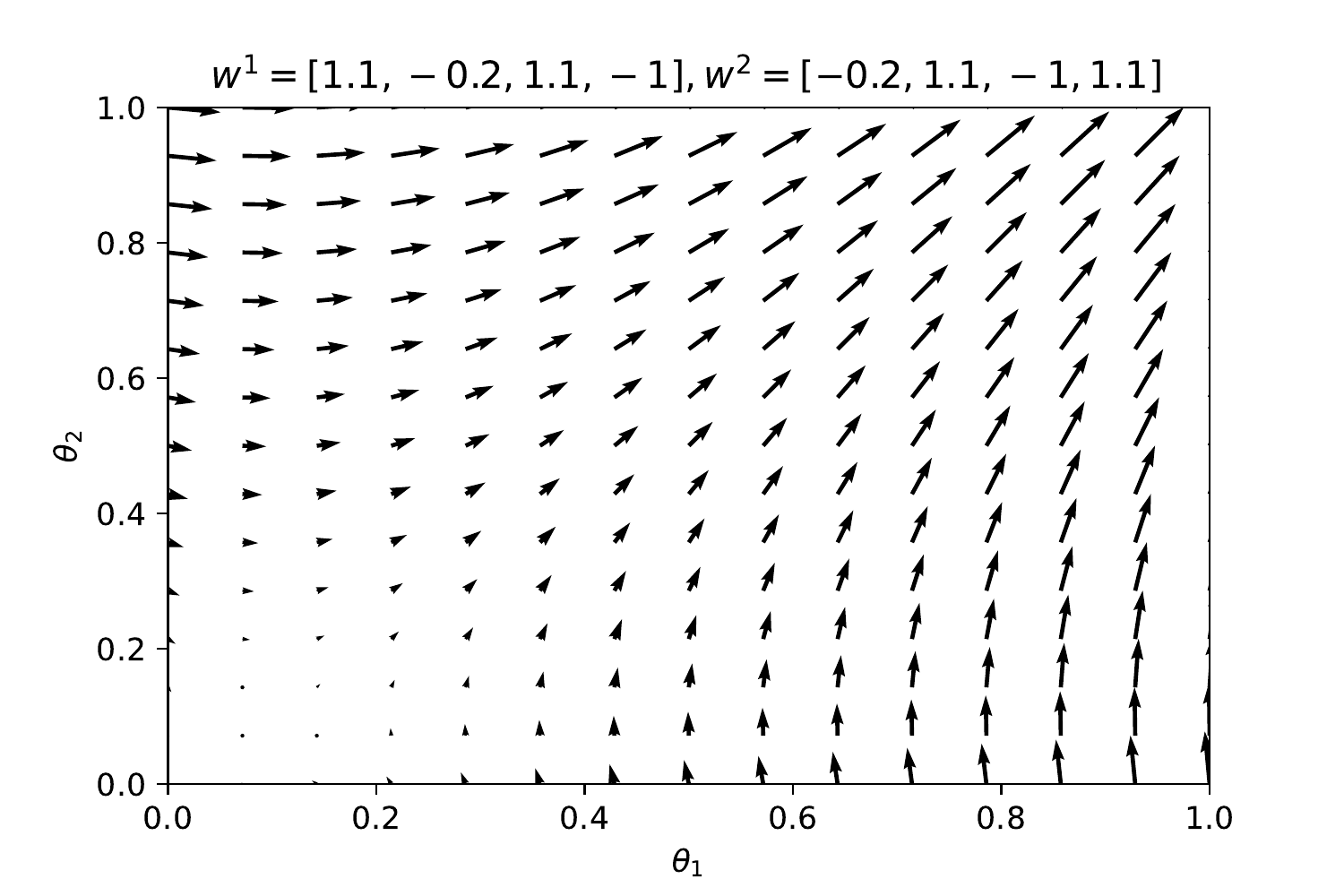}
        \caption{After the orientation parameters update.}
        \label{fig:p1}
    \end{subfigure}
    \caption{Vector fields of the probability of cooperation of each agent before and after the orientation parameters update $w^{1}: [1, 0, 1, 0]\rightarrow[1.1,-0.2,1.1,-1], w^{2}: [0, 1, 0, 1]\rightarrow[-0.2,1.1,-1,1.1]$.}
    \label{fig:p-change-1-a}
\end{figure}

From Figure~\ref{fig:w-change-1-a}., we can see that different parameters update at different speeds.
According to the speed of parameters change in Figure~\ref{fig:w-change-1-a}., we update $w^{1}: [1, 0, 1, 0]\rightarrow[1.1,-0.2,1.1,-1]$ and $w^{2}: [0, 1, 0, 1]\rightarrow[-0.2,1.1,-1,1.1]$.
Correspondingly, we show the vector fields of $\theta_1$ and $\theta_2$ before and after the orientation parameters update are shown in Figure~\ref{fig:p-change-1-a}.

\begin{figure}[htb!]
    \centering
    \begin{subfigure}[b]{0.45\textwidth}
        \includegraphics[width=\textwidth]{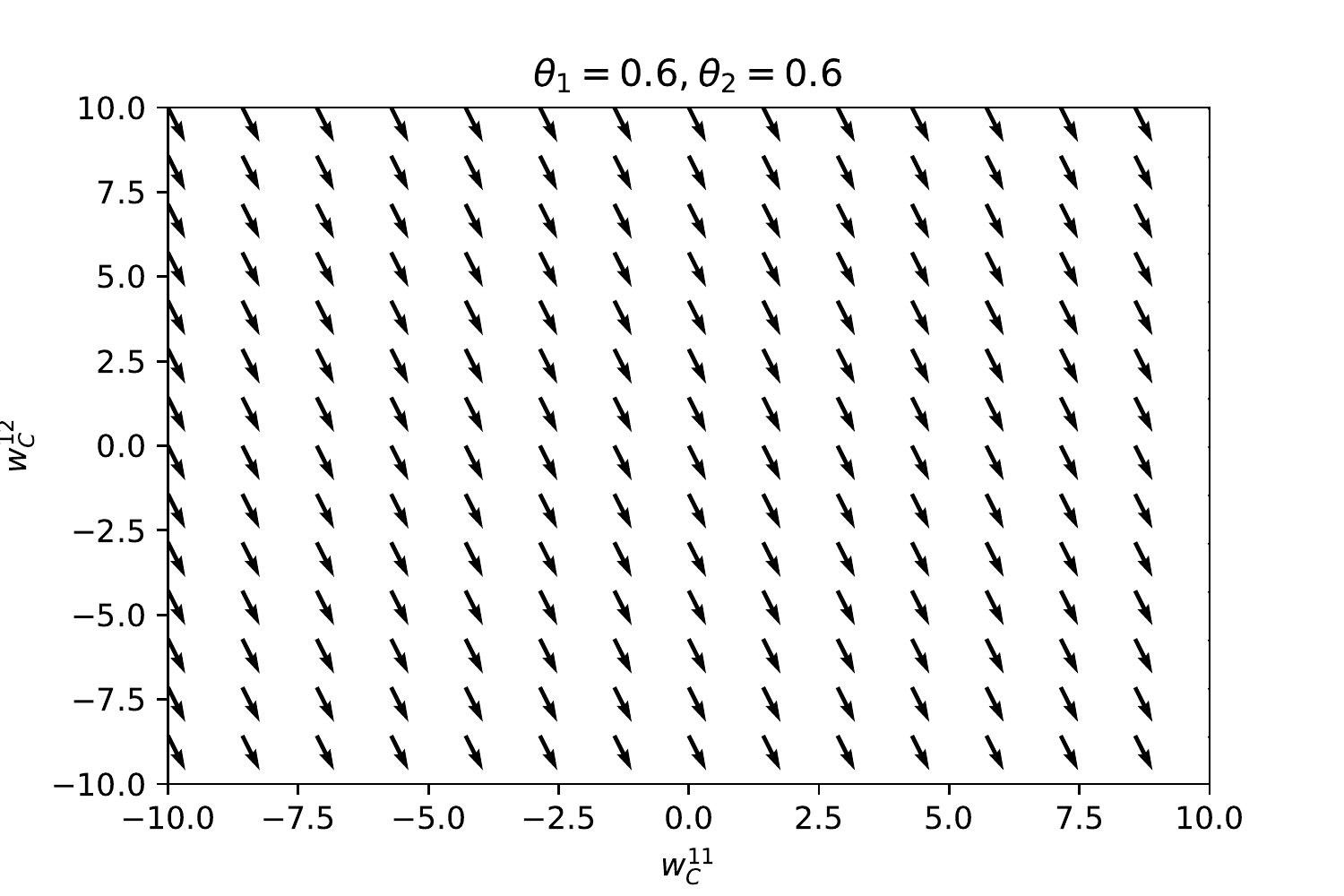}
        \caption{The vector fields of $w_{C}^{11}$ and $w_{C}^{12}$.}
        \label{fig:wc1-0606}
    \end{subfigure}
    \begin{subfigure}[b]{0.45\textwidth}
        \includegraphics[width=\textwidth]{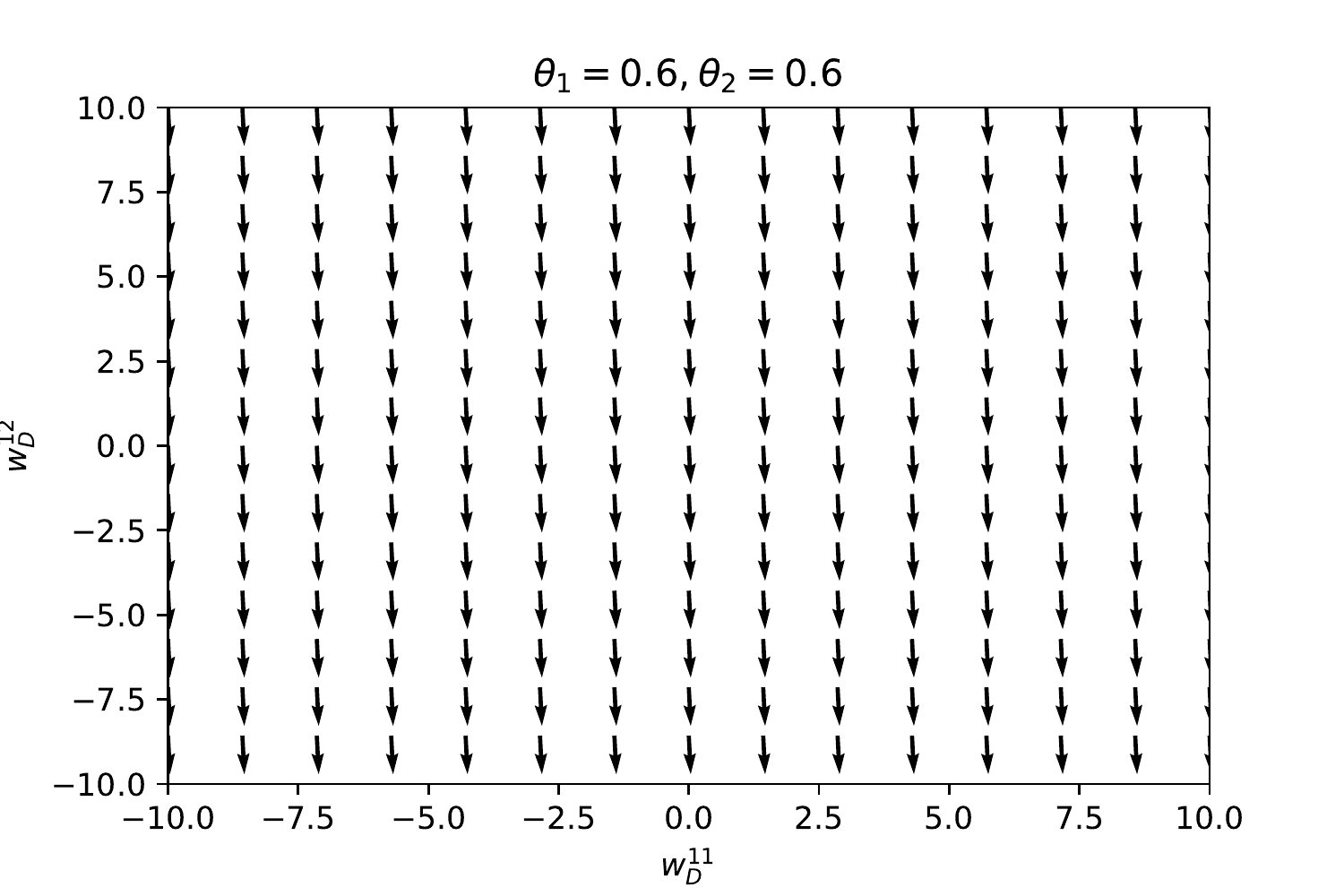}
        \caption{The vector fields of $w_{D}^{11}$ and $w_{D}^{12}$.}
        \label{fig:wd1-0606}
    \end{subfigure}
    \begin{subfigure}[b]{0.45\textwidth}
        \includegraphics[width=\textwidth]{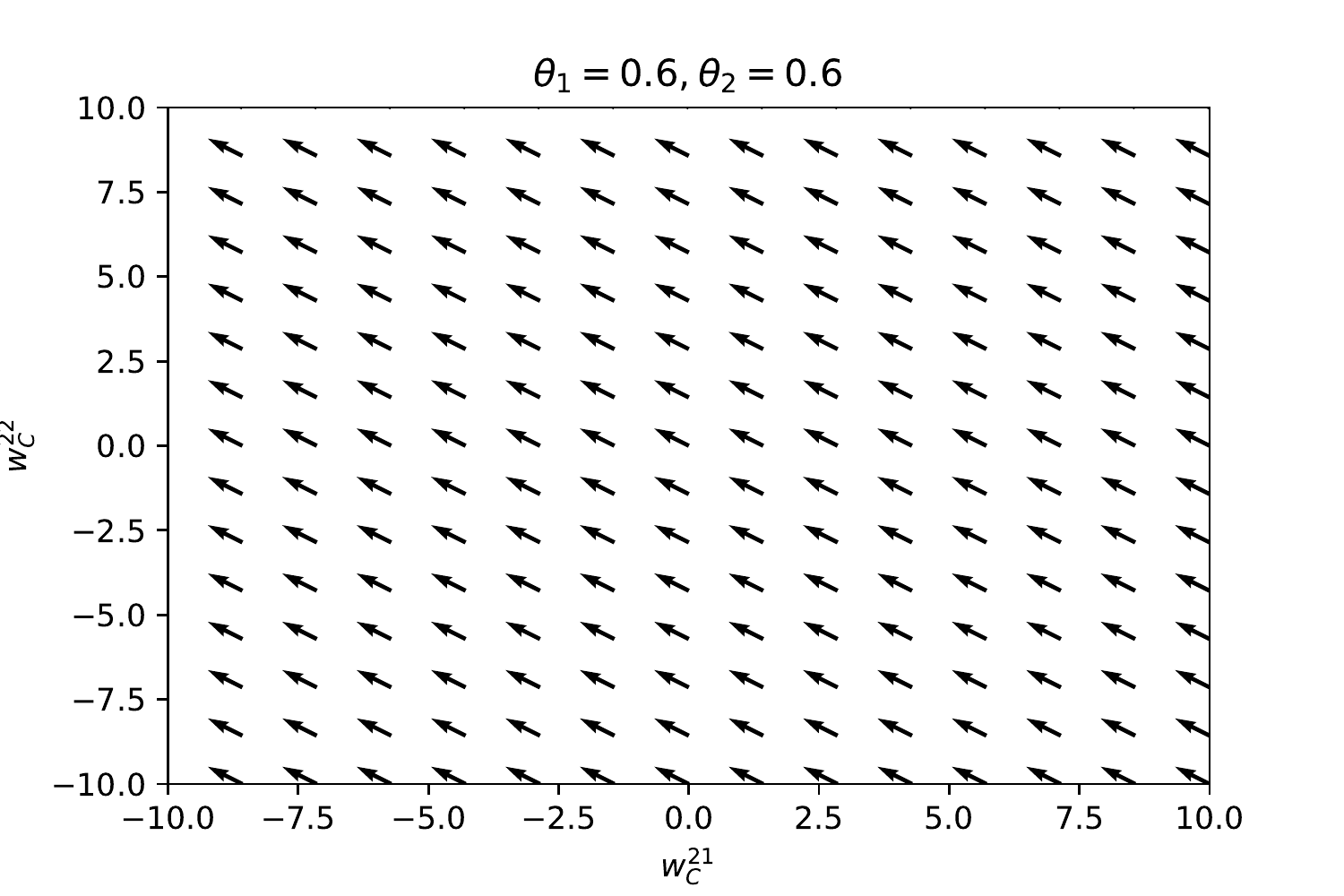}
        \caption{The vector fields of $w_{C}^{21}$ and $w_{C}^{22}$.}
        \label{fig:wc2-0606}
    \end{subfigure}
    \begin{subfigure}[b]{0.45\textwidth}
        \includegraphics[width=\textwidth]{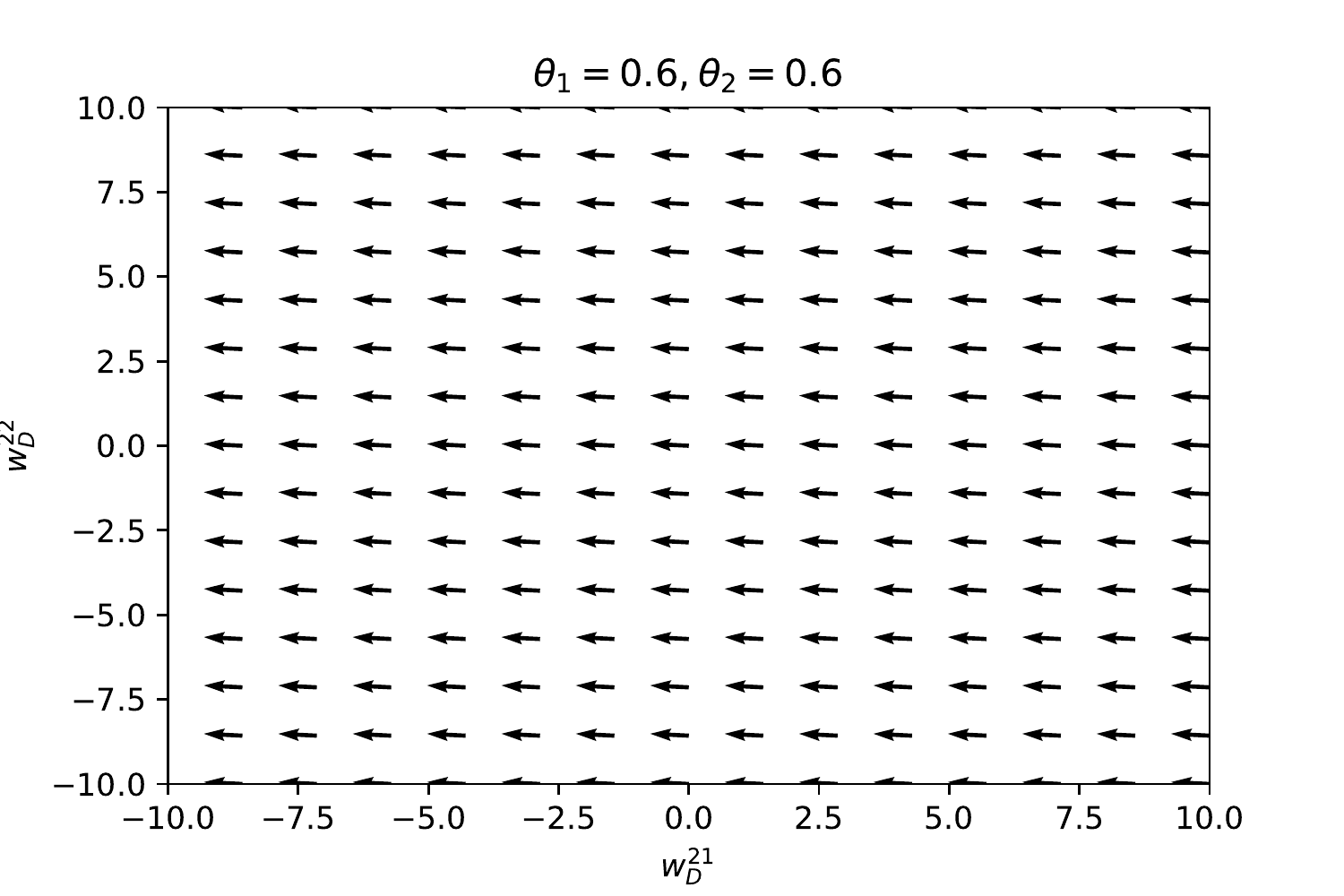}
        \caption{The vector fields of $w_{D}^{21}$ and $w_{D}^{22}$.}
        \label{fig:wd2-0606}
    \end{subfigure}
    \caption{Vector fields of orientation parameters of each agent with $\theta_1=0.6, \theta_2=0.6$.}
    \label{fig:w-change-2-a}
\end{figure}

\begin{figure}[htb!]
    \centering
    \begin{subfigure}[b]{0.45\textwidth}
        \includegraphics[width=\textwidth]{figures/vectorfields/p1.pdf}
        \caption{Before the orientation parameters update.}
        \label{fig:p1}
    \end{subfigure}
    \begin{subfigure}[b]{0.45\textwidth}
        \includegraphics[width=\textwidth]{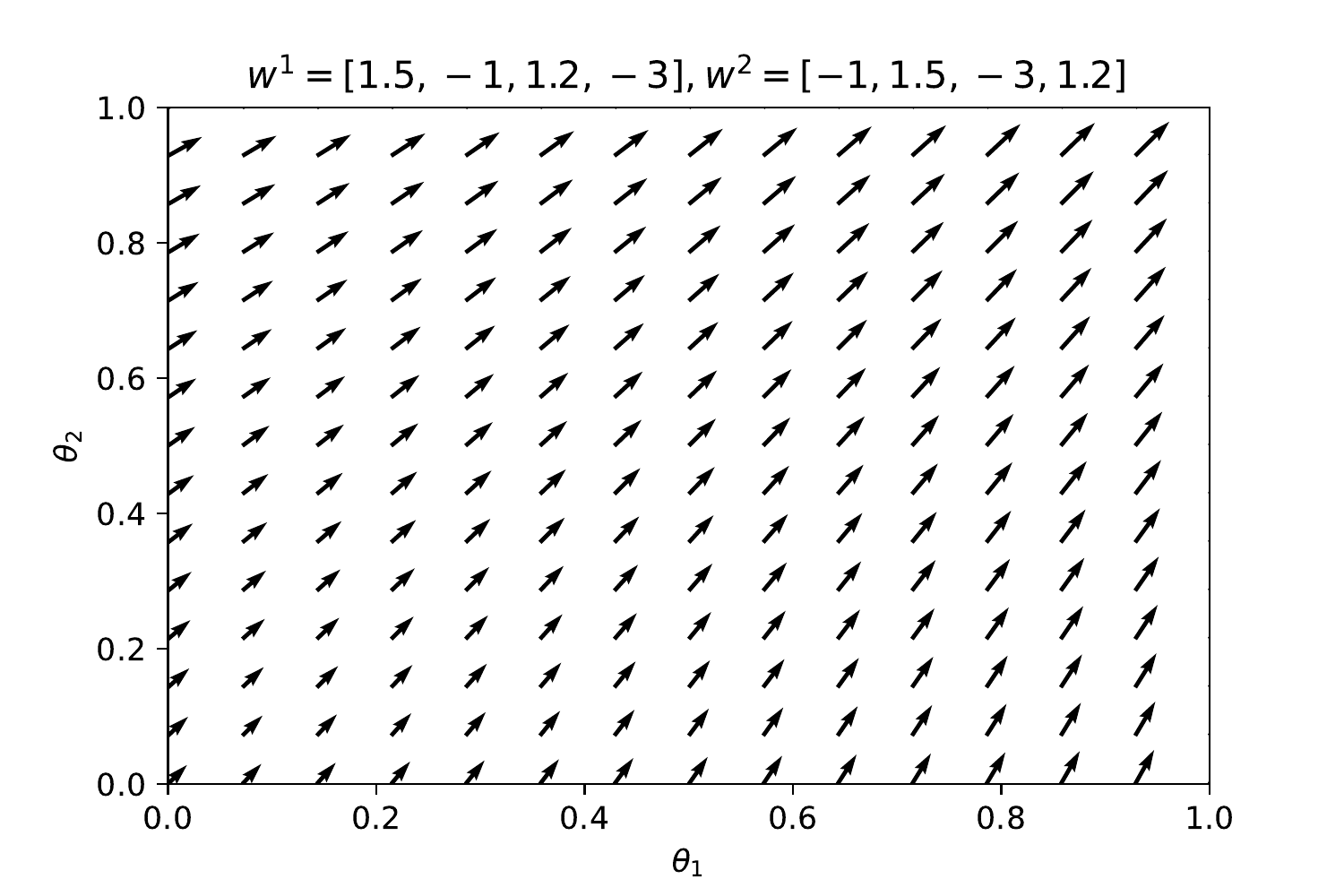}
        \caption{After the orientation parameters update.}
        \label{fig:p2}
    \end{subfigure}
    \caption{Vector fields of the probability of cooperation of each agent before and after the orientation parameters update $w^{1}: [1.1,-0.2,1.1,-1]\rightarrow[1.5,-1,1.2,-3], w^{2}: [-0.2,1.1,-1,1.1]\rightarrow[-1,1.5,-3,1.2]$.}
    \label{fig:p-change-2-a}
\end{figure}

\begin{figure}[htb!]
    \centering
    \begin{subfigure}[b]{0.45\textwidth}
        \includegraphics[width=\textwidth]{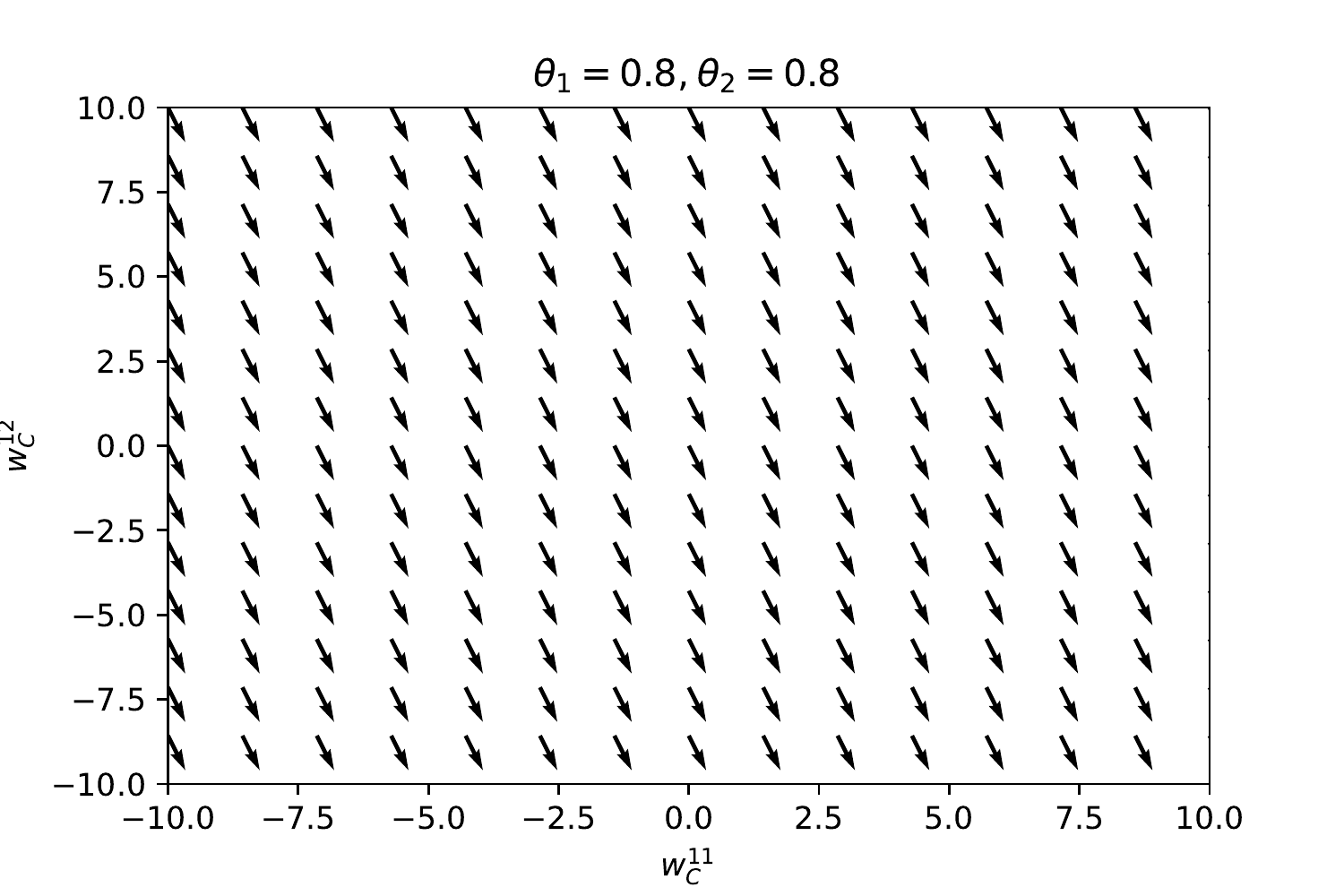}
        \caption{The vector fields of $w_{C}^{11}$ and $w_{C}^{12}$.}
        \label{fig:wc1-0808}
    \end{subfigure}
    \begin{subfigure}[b]{0.45\textwidth}
        \includegraphics[width=\textwidth]{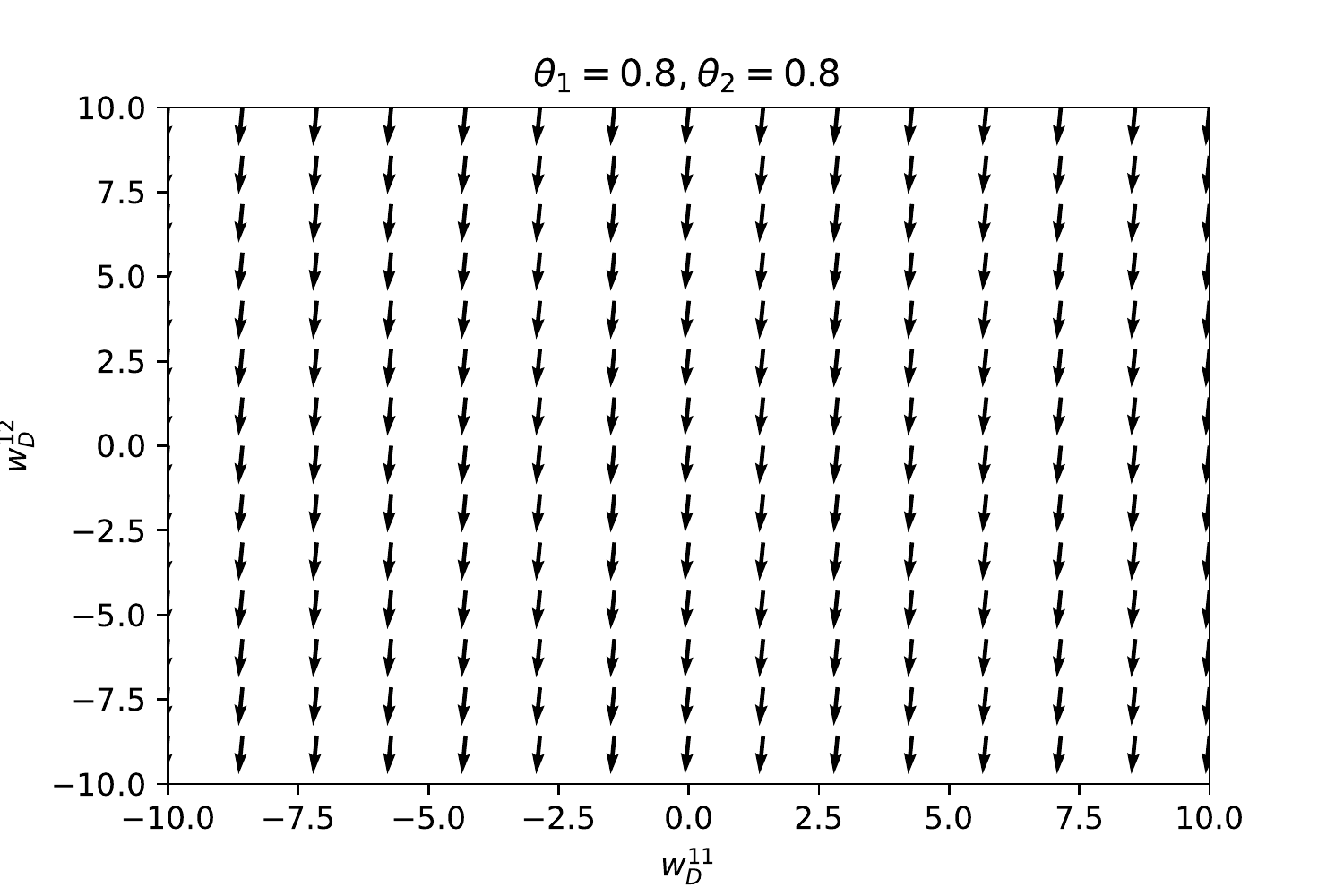}
        \caption{The vector fields of $w_{D}^{11}$ and $w_{D}^{12}$.}
        \label{fig:wd1-0808}
    \end{subfigure}
    \begin{subfigure}[b]{0.45\textwidth}
        \includegraphics[width=\textwidth]{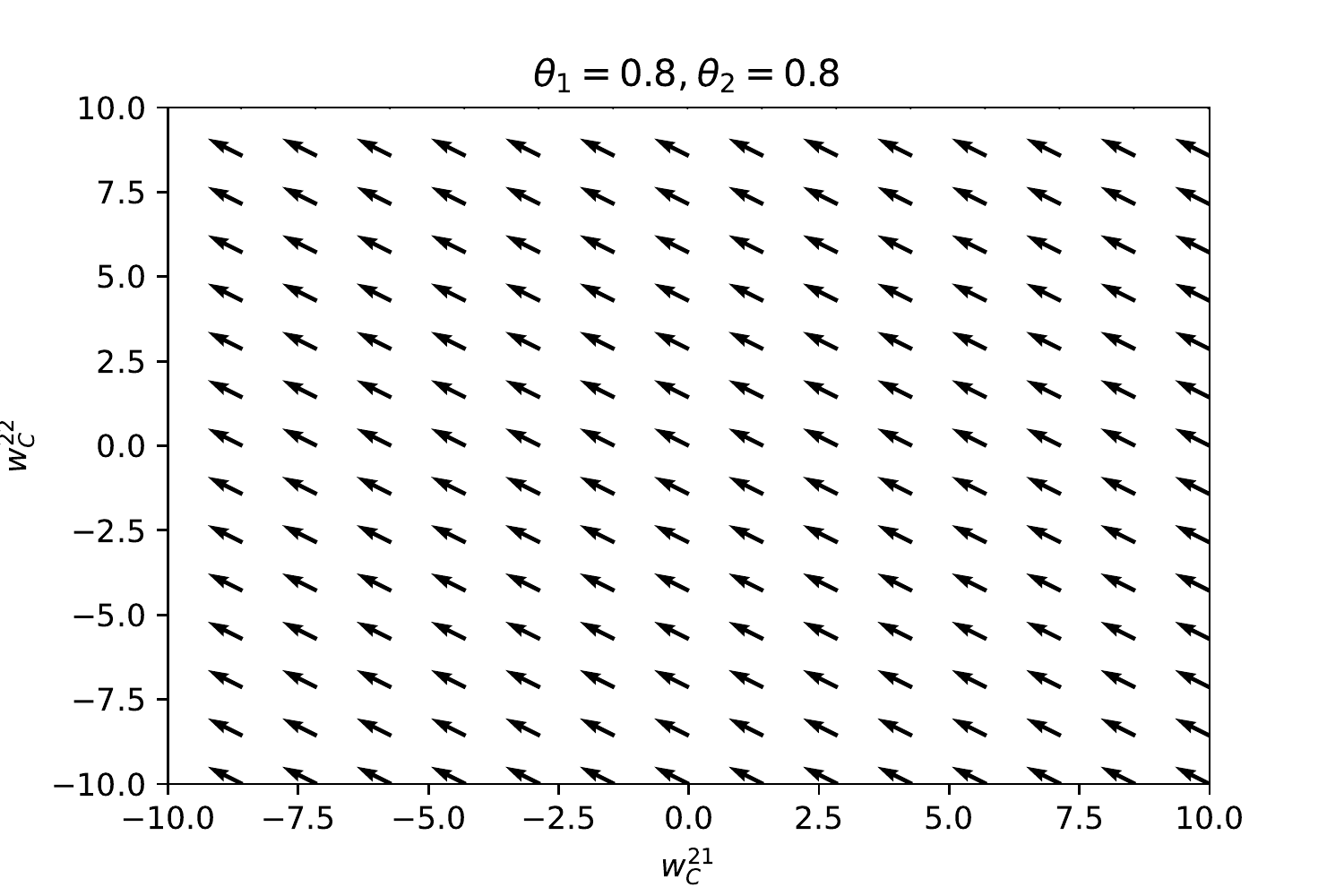}
        \caption{The vector fields of $w_{C}^{21}$ and $w_{C}^{22}$.}
        \label{fig:wc2-0808}
    \end{subfigure}
    \begin{subfigure}[b]{0.45\textwidth}
        \includegraphics[width=\textwidth]{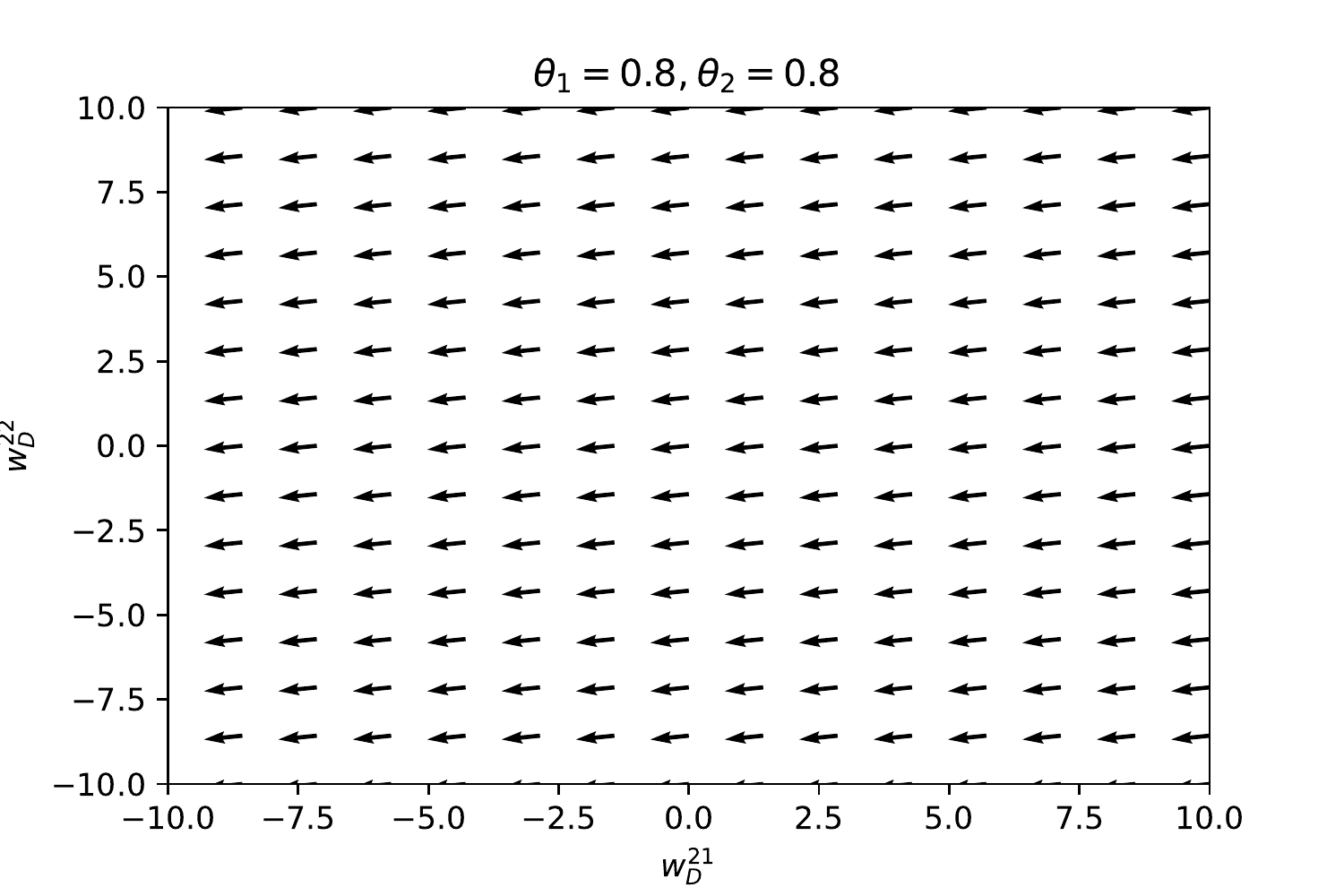}
        \caption{The vector fields of $w_{D}^{21}$ and $w_{D}^{22}$.}
        \label{fig:wd2-0808}
    \end{subfigure}
    \caption{Vector fields of orientation parameters of each agent with $\theta_1=0.8, \theta_2=0.8$.}
    \label{fig:w-change-3-a}
\end{figure}

\begin{figure}[htb!]
    \centering
    \begin{subfigure}[b]{0.45\textwidth}
        \includegraphics[width=\textwidth]{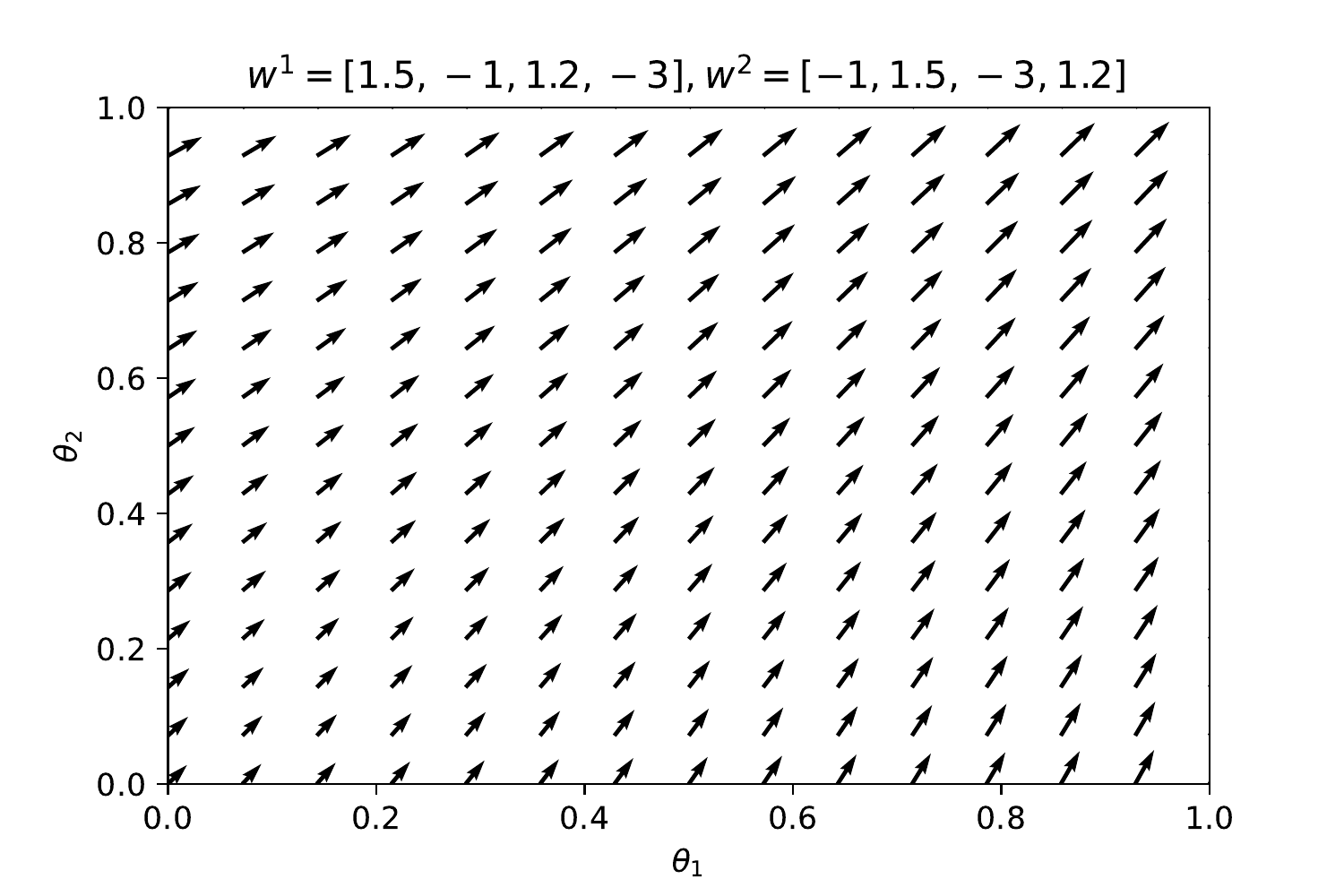}
        \caption{Before the orientation parameters update.}
        \label{fig:p2}
    \end{subfigure}
    \begin{subfigure}[b]{0.45\textwidth}
        \includegraphics[width=\textwidth]{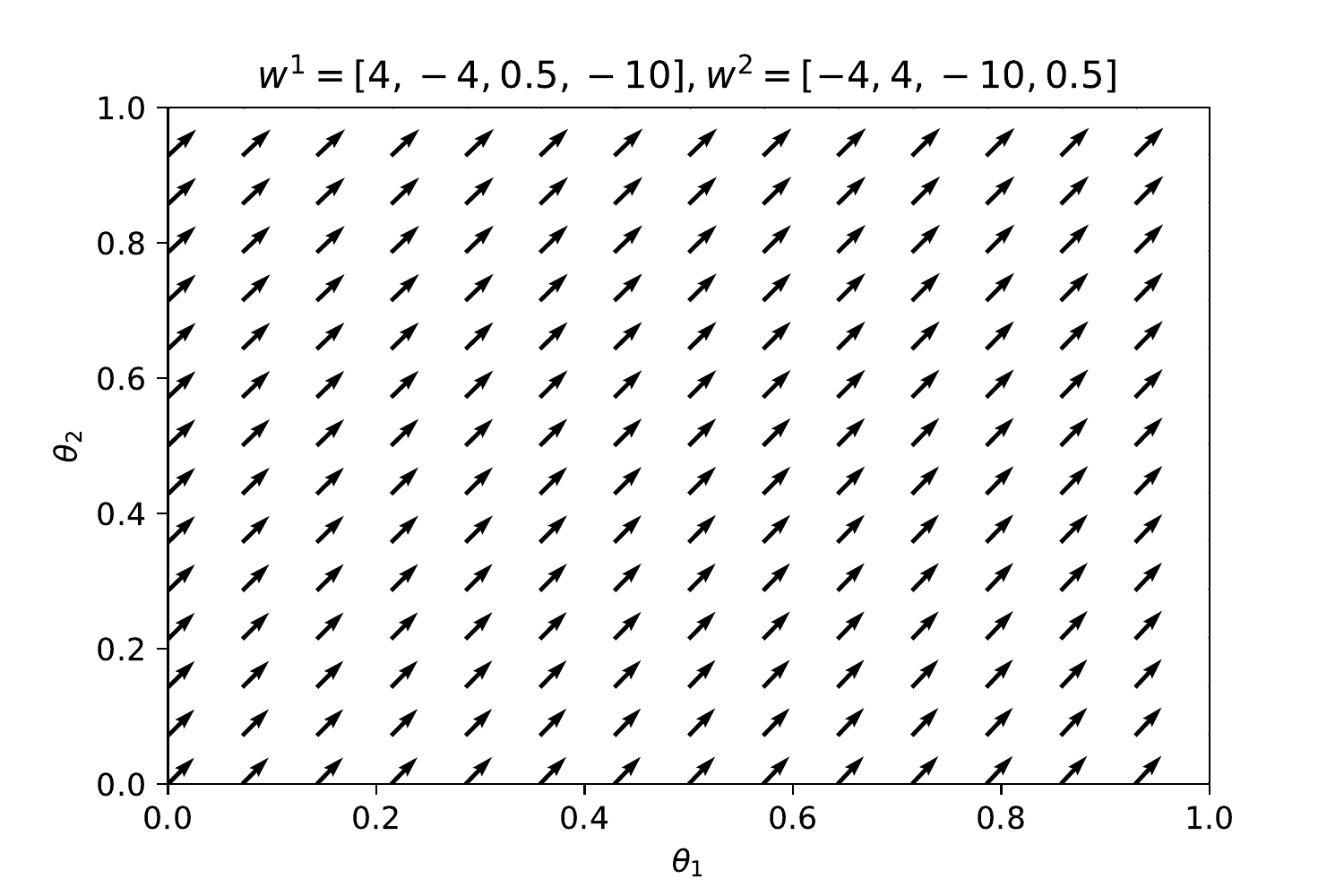}
        \caption{After the orientation parameters update.}
        \label{fig:p3}
    \end{subfigure}
    \caption{Vector fields of the probability of cooperation of each agent before and after the orientation parameters update $w^{1}: [1.5,-1,1.2,-3]\rightarrow[4,-4,0.5,-10], w^{2}: [-1,1.5,-3,1.2]\rightarrow[-4,4,-10,0.5]$.}
    \label{fig:p-change-3-a}
\end{figure}

It can be seen from Figure~\ref{fig:p-change-1-a} that after a round of parameter updates, the policies of the two agents change from mutual defection to mutual cooperation.
As the agent orientation parameters and the cooperation probabilities continue to change (Figure~\ref{fig:w-change-2-a}: $\theta_1~[0.5\rightarrow 0.6], \theta_2~[0.5\rightarrow 0.6]$; Figure~\ref{fig:p-change-2-a}: $w^{1}~[1.1,-0.2,1.1,-1]\rightarrow[1.5,-1,1.2,-3], w^{2}~[-0.2,1.1,-1,1.1]\rightarrow[-1,1.5,-3,1.2]$; Figure~\ref{fig:w-change-3-a}: $\theta_1~[0.6\rightarrow 0.8], \theta_2~[0.6\rightarrow 0.8]$; Figure~\ref{fig:p-change-3-a}: $w^{1}~[1.5,-1,1.2,-3]\rightarrow[4,-4,0.5,-10], w^{2}~[-1,1.5,-3,1.2]\rightarrow[-4,4,-10,0.5]$), we can see that the agents eventually converge to stable mutual cooperation.

\end{document}